\documentclass[sigplan,screen]{acmart}
\makeatletter
\@ifundefined{lhs2tex.lhs2tex.sty.read}%
  {\@namedef{lhs2tex.lhs2tex.sty.read}{}%
   \newcommand\SkipToFmtEnd{}%
   \newcommand\EndFmtInput{}%
   \long\def\SkipToFmtEnd#1\EndFmtInput{}%
  }\SkipToFmtEnd

\newcommand\ReadOnlyOnce[1]{\@ifundefined{#1}{\@namedef{#1}{}}\SkipToFmtEnd}
\usepackage{amstext}
\usepackage{amssymb}
\usepackage{stmaryrd}
\DeclareFontFamily{OT1}{cmtex}{}
\DeclareFontShape{OT1}{cmtex}{m}{n}
  {<5><6><7><8>cmtex8
   <9>cmtex9
   <10><10.95><12><14.4><17.28><20.74><24.88>cmtex10}{}
\DeclareFontShape{OT1}{cmtex}{m}{it}
  {<-> ssub * cmtt/m/it}{}
\newcommand{\texfamily}{\fontfamily{cmtex}\selectfont}
\DeclareFontShape{OT1}{cmtt}{bx}{n}
  {<5><6><7><8>cmtt8
   <9>cmbtt9
   <10><10.95><12><14.4><17.28><20.74><24.88>cmbtt10}{}
\DeclareFontShape{OT1}{cmtex}{bx}{n}
  {<-> ssub * cmtt/bx/n}{}

\newcommand{\Conid}[1]{\mathit{#1}}
\newcommand{\Varid}[1]{\mathit{#1}}
\newcommand{\anonymous}{\kern0.06em \vbox{\hrule\@width.5em}}

\newcommand{\bind}{\mathbin{>\!\!\!>\mkern-6.7mu=}}

\usepackage{polytable}

\@ifundefined{mathindent}%
  {\newdimen\mathindent\mathindent\leftmargini}%
  {}%

\def\resethooks{%
  \global\let\SaveRestoreHook\empty
  \global\let\ColumnHook\empty}
\newcommand*{\savecolumns}[1][default]%
  {\g@addto@macro\SaveRestoreHook{\savecolumns[#1]}}
\newcommand*{\restorecolumns}[1][default]%
  {\g@addto@macro\SaveRestoreHook{\restorecolumns[#1]}}
\newcommand*{\aligncolumn}[2]%
  {\g@addto@macro\ColumnHook{\column{#1}{#2}}}

\resethooks

\newcommand{\onelinecommentchars}{\quad-{}- }
\newcommand{\commentbeginchars}{\enskip\{-}
\newcommand{\commentendchars}{-\}\enskip}

\newcommand{\visiblecomments}{%
  \let\onelinecomment=\onelinecommentchars
  \let\commentbegin=\commentbeginchars
  \let\commentend=\commentendchars}

\newcommand{\invisiblecomments}{%
  \let\onelinecomment=\empty
  \let\commentbegin=\empty
  \let\commentend=\empty}

\visiblecomments

\newlength{\blanklineskip}
\newcommand{\hsindent}[1]{\quad}
\let\hspre\empty
\let\hspost\empty

\EndFmtInput
\makeatother
\ReadOnlyOnce{polycode.fmt}%
\makeatletter

\newcommand{\hsnewpar}[1]%
  {{\parskip=0pt\parindent=0pt\par\vskip #1\noindent}}

\newcommand{\hscodestyle}{}

\newcommand{\sethscode}[1]%
  {\expandafter\let\expandafter\hscode\csname #1\endcsname
   \expandafter\let\expandafter\endhscode\csname end#1\endcsname}

  {\par\noindent
   \advance\leftskip\mathindent
   \hscodestyle
   \let\\=\@normalcr
   \let\hspre\(\let\hspost\)%
   \pboxed}%
  {\endpboxed\)%
   \par\noindent
   \ignorespacesafterend}

  {\hsnewpar\abovedisplayskip
   \advance\leftskip\mathindent
   \hscodestyle
   \let\hspre\(\let\hspost\)%
   \pboxed}%
  {\endpboxed%
   \hsnewpar\belowdisplayskip
   \ignorespacesafterend}

  {\hsnewpar\abovedisplayskip
   \advance\leftskip\mathindent
   \hscodestyle
   \let\\=\@normalcr
   \(\pboxed}%
  {\endpboxed\)%
   \hsnewpar\belowdisplayskip
   \ignorespacesafterend}

\newcommand{\plainhs}{\sethscode{plainhscode}}

\plainhs

  {\hsnewpar\abovedisplayskip
   \advance\leftskip\mathindent
   \hscodestyle
   \let\\=\@normalcr
   \(\parray}%
  {\endparray\)%
   \hsnewpar\belowdisplayskip
   \ignorespacesafterend}

  {\parray}{\endparray}

  {\(\parray}{\endparray\)}

\def\codeframewidth{\arrayrulewidth}
\RequirePackage{calc}

  {\parskip=\abovedisplayskip\par\noindent
   \hscodestyle
   \arrayrulewidth=\codeframewidth
   \tabular{@{}|p{\linewidth-2\arraycolsep-2\arrayrulewidth-2pt}|@{}}%
   \hline\framedhslinecorrect\\{-1.5ex}%
   \let\endoflinesave=\\
   \let\\=\@normalcr
   \(\pboxed}%
  {\endpboxed\)%
   \framedhslinecorrect\endoflinesave{.5ex}\hline
   \endtabular
   \parskip=\belowdisplayskip\par\noindent
   \ignorespacesafterend}

\newcommand{\framedhslinecorrect}[2]%
  {#1[#2]}

  {\(\def\column##1##2{}%
   \let\>\undefined\let\<\undefined\let\\\undefined
   \newcommand\>[1][]{}\newcommand\<[1][]{}\newcommand\\[1][]{}%
   \def\fromto##1##2##3{##3}%
   }{\) }%

  {\let\orighscode=\hscode
   \let\origendhscode=\endhscode
   \def\endhscode{\def\hscode{\endgroup\def\@currenvir{hscode}\\}\begingroup}
   \orighscode\def\hscode{\endgroup\def\@currenvir{hscode}}}%
  {\origendhscode
   \global\let\hscode=\orighscode
   \global\let\endhscode=\origendhscode}%

\makeatother
\EndFmtInput
\ReadOnlyOnce{linear.fmt}
\makeatletter

\let\HaskellResetHook\empty
\newcommand*{\AtHaskellReset}[1]{%
  \g@addto@macro\HaskellResetHook{#1}}
\newcommand*{\HaskellReset}{\HaskellResetHook}

\newcommand*\hsUnrestrictedArrow[3]{#1}
\newcommand*\hsLinearArrow[3]{#2\global\let\hsArrow=\hsUnrestrictedArrow}
\newcommand*\hsLambdaPeriod[3]{#3\global\let\hsArrow=\hsUnrestrictedArrow}
\newcommand\hsPercent{%
  \global\let\hsOne=\hsOneAfterPercentCmd}
\newcommand*\hsOneAfterPercentCmd{%
  \global\let\hsOne=1
  \global\let\hsArrow=\hsLinearArrow}
\newcommand\hsLambdaCmd{\global\let\hsArrow=\hsLambdaPeriod}
\newcommand\hsForallCmd{\global\let\hsDot=\hsPeriod}
\newcommand\hsPeriod[2]{#2\global\let\hsDot=\hsCompose}
\newcommand\hsCompose[2]{#1}

\AtHaskellReset{%
  \global\let\hsOne=1
  \global\let\hsArrow=\hsUnrestrictedArrow
	\global\let\hsDot=\hsCompose}

\HaskellReset

\makeatother
\EndFmtInput
\usepackage[english]{babel} 
\usepackage[all]{foreign} 
\usepackage{textgreek} 
\usepackage[inline]{enumitem} 
\usepackage[htt]{hyphenat} 
\usepackage{stmaryrd,cmll} 
\usepackage{mdframed} 
\usepackage{mathpartir} 
\let\:\undefined 
\usepackage{namespc} 
\usepackage{xcolor} 
\usepackage{tikz} 
\usetikzlibrary{positioning}
\def\kinda{\tikz\fill[scale=0.4, opacity=0.4](0,.35) -- (.25,0) -- (1,.7) -- (.25,.15) -- cycle;}
\def\deffo{\tikz\fill[scale=0.4](0,.35) -- (.25,0) -- (1,.7) -- (.25,.15) -- cycle;}

\usepackage{todonotes,regexpatch}
\makeatletter
\xpatchcmd{\@todo}{\setkeys{todonotes}{#1}}{\setkeys{todonotes}{inline,#1}}{}{}
\makeatother

\allowdisplaybreaks

\newcommand{\header}[2][]{%
  \textbf{#2.}\hfill
  \if\relax\detokenize{#1}\relax
  \else
  \framebox{#1}
  \fi}

\newcommand{\tosesh}[1]{%
  \llbracket#1\rrbracket}

\newcommand{\tm}[1]{{\color[HTML]{a40038}#1}}
\newcommand{\ty}[1]{{\color[HTML]{00007a}#1}}
\newcommand{\cs}[1]{{\color[HTML]{009180}#1}}
\newcommand{\tmty}[2]{\ensuremath{\tm{#1}:\ty{#2}}}
\newcommand{\emptyenv}{\varnothing}

\newcommand{\ibind}{\mathbin{>\!\!\!>\!\!\!>\mkern-6.7mu=}}

\newcommand{\fmap}[0]{\mathbin{\langle\$\rangle}}

\usepackage{varioref}
\usepackage{hyperref}
\hypersetup{breaklinks=true}
\usepackage{breakurl}
\usepackage{cleveref}
\usepackage{doi}

\namespace*{pgv}{
  \providecommand{\tyunit}[0]{\ensuremath{\mathbf{1}}}
  \providecommand{\tyvoid}[0]{\ensuremath{\mathbf{0}}}
  \providecommand{\typrod}[2]{\ensuremath{#1\times#2}}
  \providecommand{\tysum}[2]{\ensuremath{#1+#2}}
  \providecommand{\tylolli}[4]{\ensuremath{#3\multimap_{\cs{#1}}^{\cs{#2}}#4}}
  \providecommand{\co}[1]{\ensuremath{\overline{#1}}}
  \providecommand{\tysend}[3][]{\ensuremath{!^{\cs{#1}}{#2}.{#3}}}
  \providecommand{\tyrecv}[3][]{\ensuremath{?^{\cs{#1}}{#2}.{#3}}}
  \providecommand{\tyend}[1][]{\ensuremath{\mathbf{end}^{\cs{#1}}}}
  \providecommand{\tyends}[1][]{\ensuremath{\mathbf{end}_{!}^{\cs{#1}}}}
  \providecommand{\tyendr}[1][]{\ensuremath{\mathbf{end}_{?}^{\cs{#1}}}}

  \providecommand{\subst}[4][]{%
    \if\relax\detokenize{#1}\relax
    \ensuremath{#2\{#3/#4\}}
    \else
    \ensuremath{#2(\{#3/#4\}\cup#1)}
    \fi}

  \providecommand{\andthen}[2]{\ensuremath{#1;#2}}
  \providecommand{\letbind}[3]{\ensuremath{\mathbf{let}\;#1\mathbin{=}#2\;\mathbf{in}\;#3}}
  \providecommand{\pair}[2]{\ensuremath{(#1,#2)}}
  \providecommand{\letpair}[4]{\ensuremath{\letbind{\pair{#1}{#2}}{#3}{#4}}}
  \providecommand{\labinl}[0]{\ensuremath{\mathbf{inl}}}
  \providecommand{\labinr}[0]{\ensuremath{\mathbf{inr}}}
  \providecommand{\inl}[1]{\ensuremath{\labinl\;#1}}
  \providecommand{\inr}[1]{\ensuremath{\labinr\;#1}}
  \providecommand{\casesum}[5]{\ensuremath{\mathbf{case}\;#1\;\left\{\inl{#2}\mapsto{#3};\;\inr{#4}\mapsto{#5}\right\}}}
  \providecommand{\unit}[0]{\ensuremath{()}}
  \providecommand{\letunit}[2]{\ensuremath{\letbind{\unit}{#1}{#2}}}
  \providecommand{\absurd}[1]{\ensuremath{\mathbf{absurd}\;#1}}
  
  \providecommand{\new}[0]{\ensuremath{\mathbf{new}}}

  \providecommand{\send}[0]{\ensuremath{\mathbf{send}}}
  \providecommand{\recv}[0]{\ensuremath{\mathbf{recv}}}
  \providecommand{\fork}[0]{\ensuremath{\mathbf{fork}}}
  
  \providecommand{\close}[0]{\ensuremath{\mathbf{close}}}
  \providecommand{\cancel}[0]{\ensuremath{\mathbf{cancel}}}

  \providecommand{\seq}[5]{\ensuremath{#3\vdash_{\cs{#1}}^{\cs{#2}}\tmty{#4}{#5}}}
}{}
\newcommand{\pgv}[1]{\namespace*{pgv}{}{#1}}
\newcommand{\seshtymon}[2]{\ty{\Conid{Sesh}}_{\cs{#1}}^{\cs{#2}}}
\newcommand{\seshtysend}[1][]{\ty{\Conid{Send}}^{\cs{#1}}}
\newcommand{\seshtyrecv}[1][]{\ty{\Conid{Recv}}^{\cs{#1}}}
\newcommand{\seshtyend}[1][]{\ty{\Conid{End}}^{\cs{#1}}}
\newcommand{\seshtysel}[1][]{\ty{\Conid{Select}}^{\cs{#1}}}
\newcommand{\seshtyoff}[1][]{\ty{\Conid{Offer}}^{\cs{#1}}}
\newcommand{\seshtySR}[2]{{\Conid{SR}}_{\cs{#1}}^{\cs{#2}}}
\newcommand{\seshtyRS}[2]{{\Conid{RS}}_{\cs{#1}}^{\cs{#2}}}
\newcommand{\seshtySumSrv}[1][]{{\Conid{SumSrv}}^{\cs{#1}}}
\newcommand{\seshtySumCnt}[1][]{{\Conid{SumCnt}}^{\cs{#1}}}

\begin{document}
\sloppy

\title{Deadlock-Free Session Types in Linear Haskell}

\author{Wen Kokke}
\affiliation{%
  \institution{University of Edinburgh}
  \city{Edinburgh}
  \country{Scotland}}
\email{wen.kokke@ed.ac.uk}

\author{Ornela Dardha}
\affiliation{%
  \institution{University of Glasgow}
  \city{Glasgow}
  \country{Scotland}}
\email{ornela.dardha@glasgow.ac.uk}

\begin{abstract}
  \emph{Priority Sesh} is a library for session-typed communication in Linear Haskell which offers strong compile-time correctness guarantees. Priority Sesh offers two deadlock-free APIs for session-typed communication. The first guarantees deadlock freedom by restricting the process structure to trees and forests. It is simple and composeable, but rules out cyclic structures. The second guarantees deadlock freedom via priorities, which allows the programmer to safely use cyclic structures as well.

  Our library relies on Linear Haskell to guarantee linearity, which leads to easy-to-write session types and highly idiomatic code, and lets us avoid the complex encodings of linearity in the Haskell type system that made previous libraries difficult to use.
\end{abstract}



\maketitle

%
%
%
%
%
%
%
%
%
%

\section{Introduction}\label{sec:introduction}
Session types are a type formalism used to specify and verify communication protocols~\cite{honda93,takeuchihonda94,hondavasconcelos98,hondayoshida08}. They've been studied extensively in the context of the $\pi$-calculus~\cite{sangiorgiwalker01}, a process calculus for communication an concurrency, and in the context of concurrent $\lambda$-calculi, such as the GV family of languages~\cite[``Good Variation'',][]{gayvasconcelos10,wadler14,lindleymorris15,fowlerlindley19}.

Session types have been implemented in various programming languages. We give a detailed overview in \cref{sec:related}, and \citet{orchardyoshida17} provide a complete survey of session type implementations {in Haskell}.

The main difficulty when implementing session types in most programming languages is \emph{linearity}, \ie, the guarantee that each channel endpoint is used \emph{exactly once}. There are several different approaches to guaranteeing linearity, but the main distinction is between \emph{dynamic} \cite{PadFuse,ScalasY16,Scalas2017} and \emph{static} \cite{LM17:fst,pucellatov08,lindleymorris16} usage checks. With dynamic checks, using a channel endpoint more than once simply throws a runtime error. With static checks, usage is \emph{somehow} encoded into the type system of the host language usually by encoding the entire linear typing environment into the type system using a parameterised or graded monad. Such encodings are only possible if the type system of the host language is expressive enough. However, such encodings are often quite complex, and result in a trade-off between easy-to-write session types and idiomatic programs.

Moreover, these implementations only focus on the most basic features of session types and often ignore more advanced ones, such as channel delegation or deadlock freedom: \citet{neubauerthiemann04} only provide single session channels; \citet{pucellatov08} provide multiple channels, but only the building blocks for channel delegation; \citet{imaiyuen10} extend \citet{pucellatov08} and provide full delegation. None of these works address deadlock freedom.
\citet{lindleymorris16} provide an implementation of GV into Haskell building on the work of \citet{polakow15}. To the best of our knowledge, this is the only work that guarantees deadlock freedom of session types in Haskell, albeit in a simple form. In GV, all programs must have \emph{tree-shaped} process structures. The process structure of a program is an undirected graph, where nodes represent processes, and edges represent the channels connecting them. (We explore this in more detail in \cref{sec:tree-sesh}.) Therefore, deadlock freedom is guaranteed by design: session types rule out deadlocks over a single channel, and the tree-restriction rules out sharing multiple channels between two processes. While \citet{lindleymorris16} manage to implement more advanced properties, the tree restriction rules out many interesting programs which have \emph{cyclic} process structure, but are deadlock free.

Recent works by \citet{padovaninovara15} and \citet[PGV,][]{kokkedardha21} integrate \emph{priorities}~\cite{kobayashi06,padovani14} into functional languages. Priorities are natural numbers that abstractly represent the time at which a communication action happens. Priority-based type systems check that there are \emph{no} cycles in the communication graph. The communication graph is a directed graph where nodes represent dual communication actions, and directed edges represent one action \emph{must happen} before another. (We explore this in more detail in \cref{sec:priority-sesh}.) Such type systems are \emph{more} expressive, as they allow programs to have \emph{cyclic} process structure, as long as they have an \emph{acyclic} communication graph.

With the above in mind, our research goals are as follows:
\begin{description}
\item[Q1]
  Can we have easy-to-write session types, easy linearity checks and idiomatic code at the same time?
\item[Q2]
  Can we address not only the main features of session types, but also advanced ones, such as full delegation, recursion, and deadlock freedom of programs with cyclic process structure?
\end{description}
Our \texttt{priority-sesh} library answers both questions \emph{mostly} positively. We sidestep the problems with encoding linearity in Haskell by using Linear Haskell~\cite{bernardyboespflug18}, which has native support for linear types. The resulting session type library presented in \cref{sec:sesh,sec:tree-sesh} has both easy-to-write session types, easy linearity checks, and idiomatic code. Moving to \textbf{Q2}, the library has full delegation, recursion, and the variant in \cref{sec:tree-sesh} even guarantees of deadlock freedom, albeit by restricting the process structure to trees and forests. In \cref{sec:priority-sesh}, we implement another variant which uses priorities to ensure deadlock freedom of programs with cyclic processes structure. The ease-of-writing suffers a little, as the programmer has to manually write priorities, though this isn't a \emph{huge} inconvenience. Unfortunately, GHC's ability to reason about type-level naturals currently is \emph{not} as powerful as to allow the programmer to easily write priority-polymorphic code, which is required for \emph{recursion}. Hence, while we address deadlock freedom for cyclic process structures, we do so only for the \emph{finite} setting.

\paragraph{Contributions}
In~\cref{sec:main}, we present Priority Sesh, an implementation of deadlock free session types in Linear Haskell which is:
\begin{itemize}
\item
  the \emph{first} implementation of session types to take advantage of Linear Haskell for linearity checking, and producing easy-to-write session types and highly idiomatic code;
\item
  the \emph{first} implementation of session types in Haskell to guarantee deadlock freedom of programs with cyclic process structure via \emph{priorities}; and
\item
  the \emph{first} embedding of priorities into an existing mainstream programming language.
\end{itemize}
In~\cref{sec:pgv}, we:
\begin{itemize}
\item
  present a variant of Priority GV~\cite{kokkedardha21}---the calculus upon which Priority Sesh is based---with asynchronous communication and session cancellation following~\citet{fowlerlindley19} and \emph{explicit} lower bounds on the sequent, rather than lower bounds inferred from the typing environment; and
\item
  show that Priority Sesh is related to Priority GV via monadic reflection.
\end{itemize}
%
%
%
%
%
%
%
%
%
%

\section{What is Priority Sesh?}\label{sec:main}

In this section we introduce Priority Sesh in three steps:
\begin{itemize}
\item in \cref{sec:one-shot}, we build a small library of \emph{linear} or \emph{one-shot channels} based on MVars~\cite{peytonjonesgordon96};
\item in \cref{sec:sesh}, we use these one-shot channels to build a small library of \emph{session-typed channels} \cite{dardhagiachino12}; and
\item in \cref{sec:priority-sesh}, we decorate these session types with \emph{priorities} to guarantee deadlock-freedom \cite{kokkedardha21}.
\end{itemize}

It is important to notice that the meaning of linearity in \emph{one-shot channels} differs from linearity in \emph{session channels}. A linear or one-shot channel originates from the linear $\pi$-calculus \cite{KPT99,sangiorgiwalker01}, where each endpoint of a channel must be used for \emph{exactly one} send or receive operation, whereas linearity in the context of session-typed channels, it means that each step in the protocol is performed \emph{exactly once}, but the channel itself is used multiple times.

Priority Sesh is written in Linear Haskell~\cite{bernardyboespflug18}. The type \ensuremath{\hsPercent \hsOne \hsArrow{\rightarrow }{\multimap }{\mathpunct{.}}} is syntactic sugar for the linear arrow \text{\ttfamily \char37{}1\char45{}\char62{}}. Familiar definitions refer to linear variants packaged with \texttt{linear-base}\footnote{\url{https://hackage.haskell.org/package/linear-base}} (\eg, \ensuremath{\Conid{IO}}, \ensuremath{\Conid{Functor}}, \ensuremath{\Conid{Bifunctor}}, \ensuremath{\Conid{Monad}}) or with Priority Sesh (\eg, \ensuremath{\Conid{MVar}}).

We colour the Haskell definitions which are a part of Sesh:
\begin{itemize*}[font=\bfseries]
\item[\tm{red}] for functions and constructors;
\item[\ty{blue}] for types and type families; and
\item[\cs{emerald}] for priorities and type families acting on priorities.
\end{itemize*}

\subsection{One-shot channels}\label{sec:one-shot}

We start by building a small library of \emph{linear} or \emph{one-shot channels}, \ie, channels that must be use \emph{exactly once} to send or receive a value.

The one-shot channels are at the core of our library, and their efficiency is crucial to the overall efficiency of Priority Sesh. However, we do not aim to present an efficient implementation here, rather we aim to present a compact implementation with the correct behaviour.

\paragraph{Channels}
A~one-shot channel has two endpoints, \ensuremath{\ty{\Conid{Send}_1}} and \ensuremath{\ty{\Conid{Recv}_1}}, which are two copies of the same \ensuremath{\Conid{MVar}}.
\begin{hscode}\SaveRestoreHook
\column{B}{@{}>{\hspre}l<{\hspost}@{}}%
\column{18}{@{}>{\hspre}l<{\hspost}@{}}%
\column{19}{@{}>{\hspre}l<{\hspost}@{}}%
\column{E}{@{}>{\hspre}l<{\hspost}@{}}%
\>[B]{}\mathbf{newtype}\;\ty{\Conid{Send}_1}\;{}\<[19]%
\>[19]{}\Varid{a}\mathrel{=}\tm{\Conid{Send}_1}\;(\Conid{MVar}\;\Varid{a}){}\<[E]%
\\
\>[B]{}\mathbf{newtype}\;\ty{\Conid{Recv}_1}\;{}\<[19]%
\>[19]{}\Varid{a}\mathrel{=}\tm{\Conid{Recv}_1}\;(\Conid{MVar}\;\Varid{a}){}\<[E]%
\\[\blanklineskip]%
\>[B]{}\tm{\Varid{new}_1}\mathbin{::}\Conid{IO}\;(\ty{\Conid{Send}_1}\;\Varid{a},\ty{\Conid{Recv}_1}\;\Varid{a}){}\<[E]%
\\
\>[B]{}\tm{\Varid{new}_1}\mathrel{=}\mathbf{do}\;{}\<[18]%
\>[18]{}(\Varid{mvar}_{s},\Varid{mvar}_{r})\leftarrow \Varid{dup2}\fmap\Varid{newEmptyMVar}{}\<[E]%
\\
\>[18]{}\Varid{return}\;(\tm{\Conid{Send}_1}\;(\Varid{unur}\;\Varid{mvar}_{s}),\tm{\Conid{Recv}_1}\;(\Varid{unur}\;\Varid{mvar}_{r})){}\<[E]%
\ColumnHook
\end{hscode}\resethooks
The \ensuremath{\Varid{newEmptyMVar}} function returns an \emph{unrestricted} \ensuremath{\Conid{MVar}}, which may be used non-linearly, \ie as many times as one wants. The \ensuremath{\Varid{dup2}} function creates two (unrestricted) copies of the \ensuremath{\Conid{MVar}}. The \ensuremath{\Varid{unur}} function casts each \emph{unrestricted} copy to a \emph{linear} copy. Thus, we end up with two copies of an \ensuremath{\Conid{MVar}}, each of which must be used \emph{exactly once}.

We implement \ensuremath{\tm{\Varid{send}_1}} and \ensuremath{\tm{\Varid{recv}_1}} as aliases for the corresponding \ensuremath{\Conid{MVar}} operations.
\begin{hscode}\SaveRestoreHook
\column{B}{@{}>{\hspre}l<{\hspost}@{}}%
\column{E}{@{}>{\hspre}l<{\hspost}@{}}%
\>[B]{}\tm{\Varid{send}_1}\mathbin{::}\ty{\Conid{Send}_1}\;\Varid{a}\hsPercent \hsOne \hsArrow{\rightarrow }{\multimap }{\mathpunct{.}}\Varid{a}\hsPercent \hsOne \hsArrow{\rightarrow }{\multimap }{\mathpunct{.}}\Conid{IO}\;(){}\<[E]%
\\
\>[B]{}\tm{\Varid{send}_1}\;(\tm{\Conid{Send}_1}\;\Varid{mvar}_{s})\;\Varid{x}\mathrel{=}\Varid{putMVar}\;\Varid{mvar}_{s}\;\Varid{x}{}\<[E]%
\\[\blanklineskip]%
\>[B]{}\tm{\Varid{recv}_1}\mathbin{::}\ty{\Conid{Recv}_1}\;\Varid{a}\hsPercent \hsOne \hsArrow{\rightarrow }{\multimap }{\mathpunct{.}}\Conid{IO}\;\Varid{a}{}\<[E]%
\\
\>[B]{}\tm{\Varid{recv}_1}\;(\tm{\Conid{Recv}_1}\;\Varid{mvar}_{r})\mathrel{=}\Varid{takeMVar}\;\Varid{mvar}_{r}{}\<[E]%
\ColumnHook
\end{hscode}\resethooks
The \ensuremath{\Conid{MVar}} operations implement the correct blocking behaviour for asynchronous one-shot channels: the \ensuremath{\tm{\Varid{send}_1}} operation is non-blocking, and the \ensuremath{\tm{\Varid{recv}_1}} operations blocks until a value becomes available.

\paragraph{Synchronisation}
We use \ensuremath{\ty{\Conid{Send}_1}} and \ensuremath{\ty{\Conid{Recv}_1}} to implement a construct for one-shot synchronisation between two processes, \ensuremath{\ty{\Conid{Sync}_1}}, which consists of two one-shot channels. To synchronise, each process sends a unit on the one channel, then waits to receive a unit on the other channel.
\begin{hscode}\SaveRestoreHook
\column{B}{@{}>{\hspre}l<{\hspost}@{}}%
\column{15}{@{}>{\hspre}l<{\hspost}@{}}%
\column{E}{@{}>{\hspre}l<{\hspost}@{}}%
\>[B]{}\mathbf{data}\;\ty{\Conid{Sync}_1}\mathrel{=}\tm{\Conid{Sync}_1}\;(\ty{\Conid{Send}_1}\;())\;(\ty{\Conid{Recv}_1}\;()){}\<[E]%
\\[\blanklineskip]%
\>[B]{}\tm{\Varid{newSync}_1}\mathbin{::}\Conid{IO}\;(\ty{\Conid{Sync}_1},\ty{\Conid{Sync}_1}){}\<[E]%
\\
\>[B]{}\tm{\Varid{newSync}_1}\mathrel{=}\mathbf{do}\;{}\<[15]%
\>[15]{}(\Varid{ch}_{\Varid{s1}},\Varid{ch}_{\Varid{r1}})\leftarrow \tm{\Varid{new}_1}{}\<[E]%
\\
\>[15]{}(\Varid{ch}_{\Varid{s2}},\Varid{ch}_{\Varid{r2}})\leftarrow \tm{\Varid{new}_1}{}\<[E]%
\\
\>[15]{}\Varid{return}\;(\tm{\Conid{Sync}_1}\;\Varid{ch}_{\Varid{s1}}\;\Varid{ch}_{\Varid{r2}},\tm{\Conid{Sync}_1}\;\Varid{ch}_{\Varid{s2}}\;\Varid{ch}_{\Varid{r1}}){}\<[E]%
\\[\blanklineskip]%
\>[B]{}\tm{\Varid{sync}_1}\mathbin{::}\ty{\Conid{Sync}_1}\hsPercent \hsOne \hsArrow{\rightarrow }{\multimap }{\mathpunct{.}}\Conid{IO}\;(){}\<[E]%
\\
\>[B]{}\tm{\Varid{sync}_1}\;(\tm{\Conid{Sync}_1}\;\Varid{ch}_{\Varid{s}}\;\Varid{ch}_{\Varid{r}})\mathrel{=}\mathbf{do}\;\tm{\Varid{send}_1}\;\Varid{ch}_{\Varid{s}}\;();\tm{\Varid{recv}_1}\;\Varid{ch}_{\Varid{r}}{}\<[E]%
\ColumnHook
\end{hscode}\resethooks

\paragraph{Cancellation}
We implement \emph{cancellation} for one-shot channels. One-shot channels are created in the linear \ensuremath{\Conid{IO}} monad, so \emph{forgetting} to use a channel results in a complaint from the type-checker. However, it is possible to \emph{explicitly} drop values whose types implement the \ensuremath{\Conid{Consumable}} class, using \ensuremath{\Varid{consume}\mathbin{::}\Varid{a}\hsPercent \hsOne \hsArrow{\rightarrow }{\multimap }{\mathpunct{.}}()}. The ability to cancel communications is important, as it allows us to safely throw an exception \emph{without violating linearity}, assuming that we cancel all open channels before doing so.

One-shot channels implement \ensuremath{\Conid{Consumable}} by simply dropping their \ensuremath{\Conid{MVar}}s. The Haskell runtime throws an exception when a ``thread is blocked on an \ensuremath{\Conid{MVar}}, but there are no other references to the \ensuremath{\Conid{MVar}} so it can't ever continue.''\footnote{\url{https://downloads.haskell.org/~ghc/9.0.1/docs/html/libraries/base-4.15.0.0/Control-Exception.html\#t:BlockedIndefinitelyOnMVar}} Practically, \ensuremath{\Varid{consumeAndRecv}} throws a \ensuremath{\Conid{BlockedIndefinitelyOnMVar}} exception, whereas \ensuremath{\Varid{consumeAndSend}} does not:
\begin{center}
\begin{minipage}{0.5\linewidth}
\begin{hscode}\SaveRestoreHook
\column{B}{@{}>{\hspre}l<{\hspost}@{}}%
\column{3}{@{}>{\hspre}l<{\hspost}@{}}%
\column{E}{@{}>{\hspre}l<{\hspost}@{}}%
\>[B]{}\Varid{consumeAndRecv}\mathrel{=}\mathbf{do}{}\<[E]%
\\
\>[B]{}\hsindent{3}{}\<[3]%
\>[3]{}(\Varid{ch}_{\Varid{s}},\Varid{ch}_{\Varid{r}})\leftarrow \tm{\Varid{new}_1}{}\<[E]%
\\
\>[B]{}\hsindent{3}{}\<[3]%
\>[3]{}\tm{\Varid{fork}}\mathbin{\$}\Varid{return}\;(\Varid{consume}\;\Varid{ch}_{\Varid{s}}){}\<[E]%
\\
\>[B]{}\hsindent{3}{}\<[3]%
\>[3]{}\tm{\Varid{recv}_1}\;\Varid{ch}_{\Varid{r}}{}\<[E]%
\ColumnHook
\end{hscode}\resethooks
\end{minipage}%
\begin{minipage}{0.5\linewidth}
\begin{hscode}\SaveRestoreHook
\column{B}{@{}>{\hspre}l<{\hspost}@{}}%
\column{3}{@{}>{\hspre}l<{\hspost}@{}}%
\column{E}{@{}>{\hspre}l<{\hspost}@{}}%
\>[B]{}\Varid{consumeAndSend}\mathrel{=}\mathbf{do}\;\Varid{u}{}\<[E]%
\\
\>[B]{}\hsindent{3}{}\<[3]%
\>[3]{}(\Varid{ch}_{\Varid{s}},\Varid{ch}_{\Varid{r}})\leftarrow \tm{\Varid{new}_1}{}\<[E]%
\\
\>[B]{}\hsindent{3}{}\<[3]%
\>[3]{}\tm{\Varid{fork}}\mathbin{\$}\Varid{return}\;(\Varid{consume}\;\Varid{ch}_{\Varid{r}}){}\<[E]%
\\
\>[B]{}\hsindent{3}{}\<[3]%
\>[3]{}\tm{\Varid{send}_1}\;\Varid{ch}_{\Varid{s}}\;(){}\<[E]%
\ColumnHook
\end{hscode}\resethooks
\end{minipage}%
\end{center}
Where \ensuremath{\tm{\Varid{fork}}} forks off a new thread using a linear \ensuremath{\Varid{forkIO}}. (In GV, this operation is called \ensuremath{\tm{\Varid{spawn}}}.)

As the \ensuremath{\Conid{BlockedIndefinitelyOnMVar}} check is performed by the runtime, it'll even happen when a channel is dropped for reasons other than consume, such as a process crashing.

\subsection{Session-typed channels}\label{sec:sesh}
We use the one-shot channels to build a small library of \emph{session-typed channels} based on the \emph{continuation-passing style} encoding of session types in linear types by \citet{dardhagiachino12,DardhaGS17} and in line with other libraries for Scala \cite{ScalasY16,Scalas2017}, OCaml~\cite{PadFuse}, and Rust~\cite{kokke19}.

\paragraph{An example}
Let's look at a simple example of a session-typed channel---a multiplication service, which receives two integers, sends back their product, and then terminates:
\begin{hscode}\SaveRestoreHook
\column{B}{@{}>{\hspre}l<{\hspost}@{}}%
\column{E}{@{}>{\hspre}l<{\hspost}@{}}%
\>[B]{}\mathbf{type}\;\Conid{MulServer}\mathrel{=}\seshtyrecv\;\Conid{Int}\;(\seshtyrecv\;\Conid{Int}\;(\seshtysend\;\Conid{Int}\;\seshtyend)){}\<[E]%
\\
\>[B]{}\mathbf{type}\;\Conid{MulClient}\mathrel{=}\seshtysend\;\Conid{Int}\;(\seshtysend\;\Conid{Int}\;(\seshtyrecv\;\Conid{Int}\;\seshtyend)){}\<[E]%
\ColumnHook
\end{hscode}\resethooks
We define \ensuremath{\Varid{mulServer}}, which acts on a channel of type \ensuremath{\Conid{MulServer}}, and \ensuremath{\Varid{mulClient}}, which acts on a channel of the \emph{dual} type:
\begin{center}
\begin{minipage}{0.475\linewidth}
\begin{hscode}\SaveRestoreHook
\column{B}{@{}>{\hspre}l<{\hspost}@{}}%
\column{3}{@{}>{\hspre}l<{\hspost}@{}}%
\column{9}{@{}>{\hspre}l<{\hspost}@{}}%
\column{E}{@{}>{\hspre}l<{\hspost}@{}}%
\>[B]{}\Varid{mulServer}\;(\Varid{s}\mathbin{::}\Conid{MulServer}){}\<[E]%
\\
\>[B]{}\hsindent{3}{}\<[3]%
\>[3]{}\mathrel{=}\mathbf{do}\;{}\<[9]%
\>[9]{}(\Varid{x},\Varid{s})\leftarrow \tm{\Varid{recv}}\;\Varid{s}{}\<[E]%
\\
\>[9]{}(\Varid{y},\Varid{s})\leftarrow \tm{\Varid{recv}}\;\Varid{s}{}\<[E]%
\\
\>[9]{}\Varid{s}\leftarrow \tm{\Varid{send}}\;(\Varid{x}\mathbin{*}\Varid{y},\Varid{s}){}\<[E]%
\\
\>[9]{}\tm{\Varid{close}}\;\Varid{s}{}\<[E]%
\\
\>[9]{}\Varid{return}\;(){}\<[E]%
\ColumnHook
\end{hscode}\resethooks
\end{minipage}%
\begin{minipage}{0.525\linewidth}
\begin{hscode}\SaveRestoreHook
\column{B}{@{}>{\hspre}l<{\hspost}@{}}%
\column{3}{@{}>{\hspre}l<{\hspost}@{}}%
\column{9}{@{}>{\hspre}l<{\hspost}@{}}%
\column{E}{@{}>{\hspre}l<{\hspost}@{}}%
\>[B]{}\Varid{mulClient}\;(\Varid{s}\mathbin{::}\Conid{MulClient}){}\<[E]%
\\
\>[B]{}\hsindent{3}{}\<[3]%
\>[3]{}\mathrel{=}\mathbf{do}\;{}\<[9]%
\>[9]{}\Varid{s}\leftarrow \tm{\Varid{send}}\;(\mathrm{32},\Varid{s}){}\<[E]%
\\
\>[9]{}\Varid{s}\leftarrow \tm{\Varid{send}}\;(\mathrm{41},\Varid{s}){}\<[E]%
\\
\>[9]{}(\Varid{z},\Varid{s})\leftarrow \tm{\Varid{recv}}\;\Varid{s}{}\<[E]%
\\
\>[9]{}\tm{\Varid{close}}\;\Varid{s}{}\<[E]%
\\
\>[9]{}\Varid{return}\;\Varid{z}{}\<[E]%
\ColumnHook
\end{hscode}\resethooks
\end{minipage}%
\end{center}
In order to encode the \emph{sequence} of a session type using one-shot types, each action on a session-typed channel returns a channel for the \emph{continuation} of the session---save for \ensuremath{\tm{\Varid{close}}}, which ends the session. This means that the sequence of actions specified by session types becomes a payload of one-shot channels, moving from actions \emph{in breadth} to actions \emph{in depth} (in a \emph{matryoshka doll} style). Furthermore, \ensuremath{\Varid{mulServer}} and \ensuremath{\Varid{mulClient}} act on endpoints with \emph{dual} types. \emph{Duality} is crucial to session types as it ensures that when one process sends, the other is ready to receive, and vice versa. This is the basis for communication safety guaranteed by a session type system.

\paragraph{Channels}
We start by defining the \ensuremath{\ty{\Conid{Session}}} type class, which has an \emph{associated type} \ensuremath{\ty{\Conid{Dual}}}. You may think of \ensuremath{\ty{\Conid{Dual}}} as a type-level function associated with the \ensuremath{\ty{\Conid{Session}}} class with \emph{one} case for each instance. We encode the various restrictions on  duality as constraints on the type class. Each session type must have a dual, which must itself be a session type---\ensuremath{\ty{\Conid{Session}}\;(\ty{\Conid{Dual}}\;\Varid{s})} means the dual of \ensuremath{\Varid{s}} must also implement \ensuremath{\ty{\Conid{Session}}}. Duality must be \emph{injective}---the annotation \ensuremath{\Varid{result}\hsArrow{\rightarrow }{\multimap }{\mathpunct{.}}\Varid{s}} means \ensuremath{\Varid{result}} must uniquely determine \ensuremath{\Varid{s}} and \emph{involutive}---\ensuremath{\ty{\Conid{Dual}}\;(\ty{\Conid{Dual}}\;\Varid{s})\sim\Varid{s}} means \ensuremath{\ty{\Conid{Dual}}\;(\ty{\Conid{Dual}}\;\Varid{s})} must equal \ensuremath{\Varid{s}}. These constraints are all captured by the \ensuremath{\ty{\Conid{Session}}} class, along with \ensuremath{\tm{\Varid{new}}} for constructing channels:
\begin{hscode}\SaveRestoreHook
\column{B}{@{}>{\hspre}l<{\hspost}@{}}%
\column{3}{@{}>{\hspre}l<{\hspost}@{}}%
\column{5}{@{}>{\hspre}l<{\hspost}@{}}%
\column{E}{@{}>{\hspre}l<{\hspost}@{}}%
\>[B]{}\mathbf{class}\;(\ty{\Conid{Session}}\;(\ty{\Conid{Dual}}\;\Varid{s}),\ty{\Conid{Dual}}\;(\ty{\Conid{Dual}}\;\Varid{s})\sim\Varid{s})\Rightarrow \ty{\Conid{Session}}\;\Varid{s}{}\<[E]%
\\
\>[B]{}\hsindent{3}{}\<[3]%
\>[3]{}\mathbf{where}{}\<[E]%
\\
\>[3]{}\hsindent{2}{}\<[5]%
\>[5]{}\mathbf{type}\;\ty{\Conid{Dual}}\;\Varid{s}\mathrel{=}\Varid{result}\mid \Varid{result}\hsArrow{\rightarrow }{\multimap }{\mathpunct{.}}\Varid{s}{}\<[E]%
\\
\>[3]{}\hsindent{2}{}\<[5]%
\>[5]{}\tm{\Varid{new}}\mathbin{::}\Conid{IO}\;(\Varid{s},\ty{\Conid{Dual}}\;\Varid{s}){}\<[E]%
\ColumnHook
\end{hscode}\resethooks
There are three primitive session types: \ensuremath{\seshtysend}, \ensuremath{\seshtyrecv}, and \ensuremath{\seshtyend}.
\begin{hscode}\SaveRestoreHook
\column{B}{@{}>{\hspre}l<{\hspost}@{}}%
\column{22}{@{}>{\hspre}l<{\hspost}@{}}%
\column{E}{@{}>{\hspre}l<{\hspost}@{}}%
\>[B]{}\mathbf{newtype}\;\seshtysend\;\Varid{a}\;\Varid{s}{}\<[22]%
\>[22]{}\mathrel{=}\tm{\Conid{Send}}\;(\ty{\Conid{Send}_1}\;(\Varid{a},\ty{\Conid{Dual}}\;\Varid{s})){}\<[E]%
\\
\>[B]{}\mathbf{newtype}\;\seshtyrecv\;\Varid{a}\;\Varid{s}{}\<[22]%
\>[22]{}\mathrel{=}\tm{\Conid{Recv}}\;(\ty{\Conid{Recv}_1}\;(\Varid{a},\Varid{s})){}\<[E]%
\\
\>[B]{}\mathbf{newtype}\;\seshtyend{}\<[22]%
\>[22]{}\mathrel{=}\tm{\Conid{End}}\;\ty{\Conid{Sync}_1}{}\<[E]%
\ColumnHook
\end{hscode}\resethooks
By following \citet {DardhaGS17}, a channel \ensuremath{\seshtysend} wraps a one-shot channel \ensuremath{\ty{\Conid{Send}_1}} over which we send some value---which is the intended value sent by the session channel, and the channel over which \emph{the communicating partner process} continues the session---it'll make more sense once you read the definition for \ensuremath{\tm{\Varid{send}}}.
A channel \ensuremath{\seshtyrecv} wraps a one-shot channel \ensuremath{\ty{\Conid{Recv}_1}} over which we receive some value and the channel over which \emph{we} continue the session.
Finally, an channel \ensuremath{\seshtyend} wraps a synchronisation.

We define duality for each session type---\ensuremath{\seshtysend} is dual to \ensuremath{\seshtyrecv}, \ensuremath{\seshtyrecv} is dual to \ensuremath{\seshtysend}, and \ensuremath{\seshtyend} is dual to itself:
\begin{hscode}\SaveRestoreHook
\column{B}{@{}>{\hspre}l<{\hspost}@{}}%
\column{3}{@{}>{\hspre}l<{\hspost}@{}}%
\column{5}{@{}>{\hspre}l<{\hspost}@{}}%
\column{15}{@{}>{\hspre}l<{\hspost}@{}}%
\column{E}{@{}>{\hspre}l<{\hspost}@{}}%
\>[B]{}\mathbf{instance}\;\ty{\Conid{Session}}\;\Varid{s}\Rightarrow \ty{\Conid{Session}}\;(\seshtysend\;\Varid{a}\;\Varid{s}){}\<[E]%
\\
\>[B]{}\hsindent{3}{}\<[3]%
\>[3]{}\mathbf{where}{}\<[E]%
\\
\>[3]{}\hsindent{2}{}\<[5]%
\>[5]{}\mathbf{type}\;\ty{\Conid{Dual}}\;(\seshtysend\;\Varid{a}\;\Varid{s})\mathrel{=}\seshtyrecv\;\Varid{a}\;(\ty{\Conid{Dual}}\;\Varid{s}){}\<[E]%
\\
\>[3]{}\hsindent{2}{}\<[5]%
\>[5]{}\tm{\Varid{new}}\mathrel{=}\mathbf{do}\;{}\<[15]%
\>[15]{}(\Varid{ch}_{\Varid{s}},\Varid{ch}_{\Varid{r}})\leftarrow \tm{\Varid{new}_1}{}\<[E]%
\\
\>[15]{}\Varid{return}\;(\tm{\Conid{Send}}\;\Varid{ch}_{\Varid{s}},\tm{\Conid{Recv}}\;\Varid{ch}_{\Varid{r}}){}\<[E]%
\\[\blanklineskip]%
\>[B]{}\mathbf{instance}\;\ty{\Conid{Session}}\;\Varid{s}\Rightarrow \ty{\Conid{Session}}\;(\seshtyrecv\;\Varid{a}\;\Varid{s}){}\<[E]%
\\
\>[B]{}\hsindent{3}{}\<[3]%
\>[3]{}\mathbf{where}{}\<[E]%
\\
\>[3]{}\hsindent{2}{}\<[5]%
\>[5]{}\mathbf{type}\;\ty{\Conid{Dual}}\;(\seshtyrecv\;\Varid{a}\;\Varid{s})\mathrel{=}\seshtysend\;\Varid{a}\;(\ty{\Conid{Dual}}\;\Varid{s}){}\<[E]%
\\
\>[3]{}\hsindent{2}{}\<[5]%
\>[5]{}\tm{\Varid{new}}\mathrel{=}\mathbf{do}\;{}\<[15]%
\>[15]{}(\Varid{ch}_{\Varid{s}},\Varid{ch}_{\Varid{r}})\leftarrow \tm{\Varid{new}_1}{}\<[E]%
\\
\>[15]{}\Varid{return}\;(\tm{\Conid{Recv}}\;\Varid{ch}_{\Varid{r}},\tm{\Conid{Send}}\;\Varid{ch}_{\Varid{s}}){}\<[E]%
\\[\blanklineskip]%
\>[B]{}\mathbf{instance}\;\ty{\Conid{Session}}\;\seshtyend{}\<[E]%
\\
\>[B]{}\hsindent{3}{}\<[3]%
\>[3]{}\mathbf{where}{}\<[E]%
\\
\>[3]{}\hsindent{2}{}\<[5]%
\>[5]{}\mathbf{type}\;\ty{\Conid{Dual}}\;\seshtyend\mathrel{=}\seshtyend{}\<[E]%
\\
\>[3]{}\hsindent{2}{}\<[5]%
\>[5]{}\tm{\Varid{new}}\mathrel{=}\mathbf{do}\;{}\<[15]%
\>[15]{}(\Varid{ch}_{\Varid{sync1}},\Varid{ch}_{\Varid{sync2}})\leftarrow \tm{\Varid{newSync}_1}{}\<[E]%
\\
\>[15]{}\Varid{return}\;(\tm{\Conid{End}}\;\Varid{ch}_{\Varid{sync1}},\tm{\Conid{End}}\;\Varid{ch}_{\Varid{sync2}}){}\<[E]%
\ColumnHook
\end{hscode}\resethooks
The \ensuremath{\tm{\Varid{send}}} operation constructs a channel for the continuation of the session, then sends one endpoint of that channel, along with the value, over its one-shot channel, and returns the other endpoint:
\begin{hscode}\SaveRestoreHook
\column{B}{@{}>{\hspre}l<{\hspost}@{}}%
\column{32}{@{}>{\hspre}l<{\hspost}@{}}%
\column{E}{@{}>{\hspre}l<{\hspost}@{}}%
\>[B]{}\tm{\Varid{send}}\mathbin{::}\ty{\Conid{Session}}\;\Varid{s}\Rightarrow (\Varid{a},\seshtysend\;\Varid{a}\;\Varid{s})\hsPercent \hsOne \hsArrow{\rightarrow }{\multimap }{\mathpunct{.}}\Conid{IO}\;\Varid{s}{}\<[E]%
\\
\>[B]{}\tm{\Varid{send}}\;(\Varid{x},\tm{\Conid{Send}}\;\Varid{ch}_{\Varid{s}})\mathrel{=}\mathbf{do}\;{}\<[32]%
\>[32]{}(\Varid{here},\Varid{there})\leftarrow \tm{\Varid{new}}{}\<[E]%
\\
\>[32]{}\tm{\Varid{send}_1}\;\Varid{ch}_{\Varid{s}}\;(\Varid{x},\Varid{there}){}\<[E]%
\\
\>[32]{}\Varid{return}\;\Varid{here}{}\<[E]%
\ColumnHook
\end{hscode}\resethooks
The \ensuremath{\tm{\Varid{recv}}} and \ensuremath{\tm{\Varid{close}}} operations simply wrap their corresponding one-shot operations:
\begin{hscode}\SaveRestoreHook
\column{B}{@{}>{\hspre}l<{\hspost}@{}}%
\column{E}{@{}>{\hspre}l<{\hspost}@{}}%
\>[B]{}\tm{\Varid{recv}}\mathbin{::}\seshtyrecv\;\Varid{a}\;\Varid{s}\hsPercent \hsOne \hsArrow{\rightarrow }{\multimap }{\mathpunct{.}}\Conid{IO}\;(\Varid{a},\Varid{s}){}\<[E]%
\\
\>[B]{}\tm{\Varid{recv}}\;(\tm{\Conid{Recv}}\;\Varid{ch}_{\Varid{r}})\mathrel{=}\tm{\Varid{recv}_1}\;\Varid{ch}_{\Varid{r}}{}\<[E]%
\\[\blanklineskip]%
\>[B]{}\tm{\Varid{close}}\mathbin{::}\seshtyend\hsPercent \hsOne \hsArrow{\rightarrow }{\multimap }{\mathpunct{.}}\Conid{IO}\;(){}\<[E]%
\\
\>[B]{}\tm{\Varid{close}}\;(\tm{\Conid{End}}\;\Varid{ch}_{\Varid{sync}})\mathrel{=}\tm{\Varid{sync}_1}\;\Varid{ch}_{\Varid{sync}}{}\<[E]%
\ColumnHook
\end{hscode}\resethooks

\paragraph{Cancellation}
We implement session \emph{cancellation} via the \ensuremath{\Conid{Consumable}} class.  For convenience, we provide the \ensuremath{\tm{\Varid{cancel}}} function:
\begin{hscode}\SaveRestoreHook
\column{B}{@{}>{\hspre}l<{\hspost}@{}}%
\column{9}{@{}>{\hspre}l<{\hspost}@{}}%
\column{E}{@{}>{\hspre}l<{\hspost}@{}}%
\>[9]{}\tm{\Varid{cancel}}\mathbin{::}\ty{\Conid{Session}}\;\Varid{s}\Rightarrow \Varid{s}\hsPercent \hsOne \hsArrow{\rightarrow }{\multimap }{\mathpunct{.}}\Conid{IO}{}\<[E]%
\\
\>[9]{}\tm{\Varid{cancel}}\;\Varid{s}\mathrel{=}\Varid{return}\;(\Varid{consume}\;\Varid{s}){}\<[E]%
\ColumnHook
\end{hscode}\resethooks
As with one-shot channels, \ensuremath{\Varid{consume}} simply drops the channel, and relies on the \ensuremath{\Conid{BlockedIndefinitelyOnMVar}} check, which means that \ensuremath{\Varid{cancelAndRecv}} throws an exception and \ensuremath{\Varid{cancelAndSend}} does not:
\begin{center}
\begin{minipage}{0.5\linewidth}
\begin{hscode}\SaveRestoreHook
\column{B}{@{}>{\hspre}l<{\hspost}@{}}%
\column{3}{@{}>{\hspre}l<{\hspost}@{}}%
\column{E}{@{}>{\hspre}l<{\hspost}@{}}%
\>[B]{}\Varid{cancelAndRecv}\mathrel{=}\mathbf{do}{}\<[E]%
\\
\>[B]{}\hsindent{3}{}\<[3]%
\>[3]{}(\Varid{ch}_{\Varid{s}},\Varid{ch}_{\Varid{r}})\leftarrow \tm{\Varid{new}}{}\<[E]%
\\
\>[B]{}\hsindent{3}{}\<[3]%
\>[3]{}\tm{\Varid{fork}}\mathbin{\$}\tm{\Varid{cancel}}\;\Varid{ch}_{\Varid{s}}{}\<[E]%
\\
\>[B]{}\hsindent{3}{}\<[3]%
\>[3]{}((),())\leftarrow \tm{\Varid{recv}}\;\Varid{ch}_{\Varid{r}}{}\<[E]%
\\
\>[B]{}\hsindent{3}{}\<[3]%
\>[3]{}\Varid{return}\;(){}\<[E]%
\ColumnHook
\end{hscode}\resethooks
\end{minipage}%
\begin{minipage}{0.5\linewidth}
\begin{hscode}\SaveRestoreHook
\column{B}{@{}>{\hspre}l<{\hspost}@{}}%
\column{3}{@{}>{\hspre}l<{\hspost}@{}}%
\column{E}{@{}>{\hspre}l<{\hspost}@{}}%
\>[B]{}\Varid{cancelAndSend}\mathrel{=}\mathbf{do}\;\Varid{u}{}\<[E]%
\\
\>[B]{}\hsindent{3}{}\<[3]%
\>[3]{}(\Varid{ch}_{\Varid{s}},\Varid{ch}_{\Varid{r}})\leftarrow \tm{\Varid{new}}{}\<[E]%
\\
\>[B]{}\hsindent{3}{}\<[3]%
\>[3]{}\tm{\Varid{fork}}\mathbin{\$}\tm{\Varid{cancel}}\;\Varid{ch}_{\Varid{r}}{}\<[E]%
\\
\>[B]{}\hsindent{3}{}\<[3]%
\>[3]{}()\leftarrow \tm{\Varid{send}}\;\Varid{ch}_{\Varid{s}}\;(){}\<[E]%
\\
\>[B]{}\hsindent{3}{}\<[3]%
\>[3]{}\Varid{return}\;(){}\<[E]%
\ColumnHook
\end{hscode}\resethooks
\end{minipage}%
\end{center}
These semantics correspond to EGV~\cite{fowlerlindley19}.

\paragraph{Asynchronous close}
We don't always \emph{want} session-end to involve synchronisation. Unfortunately, the \ensuremath{\tm{\Varid{close}}} operation is synchronous.

An advantage of defining session types via a type class is that its an \emph{open} class, and we can add new primitives whenever. Let's make the unit type, \ensuremath{()}, a session type:
\begin{hscode}\SaveRestoreHook
\column{B}{@{}>{\hspre}l<{\hspost}@{}}%
\column{3}{@{}>{\hspre}l<{\hspost}@{}}%
\column{5}{@{}>{\hspre}l<{\hspost}@{}}%
\column{E}{@{}>{\hspre}l<{\hspost}@{}}%
\>[B]{}\mathbf{instance}\;\ty{\Conid{Session}}\;\Varid{s}\Rightarrow \ty{\Conid{Session}}\;(){}\<[E]%
\\
\>[B]{}\hsindent{3}{}\<[3]%
\>[3]{}\mathbf{where}{}\<[E]%
\\
\>[3]{}\hsindent{2}{}\<[5]%
\>[5]{}\mathbf{type}\;\ty{\Conid{Dual}}\;()\mathrel{=}(){}\<[E]%
\\
\>[3]{}\hsindent{2}{}\<[5]%
\>[5]{}\tm{\Varid{new}}\mathrel{=}\Varid{return}\;((),()){}\<[E]%
\ColumnHook
\end{hscode}\resethooks
Units are naturally affine---they contain \emph{zero} information, so dropping them won't harm---and the linear \ensuremath{\Conid{Monad}} class allows you to silently drop unit results of monadic computations.
They're ideal for \emph{asynchronous} session end!

Using \ensuremath{()} allows us to recover the semantics of one-shot channels while keeping a session-typed language for idiomatic protocol specification.

\paragraph{Choice}
So far, we've only presented sending, receiving, and synchronisation. It is, however, possible to send and receive \emph{channels} as well as values, and we leverage that to implement most other session types by using these primitives only!

For instance, we can implement \emph{binary} choice by sending/receiving \ensuremath{\Conid{Either}} of two session continuations:
\begin{hscode}\SaveRestoreHook
\column{B}{@{}>{\hspre}l<{\hspost}@{}}%
\column{17}{@{}>{\hspre}l<{\hspost}@{}}%
\column{20}{@{}>{\hspre}l<{\hspost}@{}}%
\column{E}{@{}>{\hspre}l<{\hspost}@{}}%
\>[B]{}\mathbf{type}\;\seshtysel\;\Varid{s}_{1}\;\Varid{s}_{2}\mathrel{=}\seshtysend\;(\Conid{Either}\;(\ty{\Conid{Dual}}\;\Varid{s}_{1})\;(\ty{\Conid{Dual}}\;\Varid{s}_{2}))\;(){}\<[E]%
\\
\>[B]{}\mathbf{type}\;\seshtyoff\;\Varid{s}_{1}\;\Varid{s}_{2}\mathrel{=}\seshtyrecv\;(\Conid{Either}\;\Varid{s}_{1}\;\Varid{s}_{2})\;(){}\<[E]%
\\[\blanklineskip]%
\>[B]{}\tm{\Varid{selectLeft}}\mathbin{::}(\ty{\Conid{Session}}\;\Varid{s}_{1})\Rightarrow \seshtysel\;\Varid{s}_{1}\;\Varid{s}_{2}\hsPercent \hsOne \hsArrow{\rightarrow }{\multimap }{\mathpunct{.}}\Conid{IO}\;\Varid{s}_{1}{}\<[E]%
\\
\>[B]{}\tm{\Varid{selectLeft}}\;\Varid{s}\mathrel{=}\mathbf{do}\;{}\<[20]%
\>[20]{}(\Varid{here},\Varid{there})\leftarrow \tm{\Varid{new}}{}\<[E]%
\\
\>[20]{}\tm{\Varid{send}}\;(\Conid{Left}\;\Varid{there},\Varid{s}){}\<[E]%
\\
\>[20]{}\Varid{return}\;\Varid{here}{}\<[E]%
\\[\blanklineskip]%
\>[B]{}\tm{\Varid{offerEither}}\mathbin{::}{}\<[17]%
\>[17]{}\seshtyoff\;\Varid{s}_{1}\;\Varid{s}_{2}\hsPercent \hsOne \hsArrow{\rightarrow }{\multimap }{\mathpunct{.}}(\Conid{Either}\;\Varid{s}_{1}\;\Varid{s}_{2}\hsPercent \hsOne \hsArrow{\rightarrow }{\multimap }{\mathpunct{.}}\Conid{IO}\;\Varid{a})\hsPercent \hsOne \hsArrow{\rightarrow }{\multimap }{\mathpunct{.}}\Conid{IO}\;\Varid{a}{}\<[E]%
\\
\>[B]{}\tm{\Varid{offerEither}}\;\Varid{s}\;\Varid{match}\mathrel{=}\mathbf{do}\;(\Varid{e},())\leftarrow \tm{\Varid{recv}}\;\Varid{s};\Varid{match}\;\Varid{e}{}\<[E]%
\ColumnHook
\end{hscode}\resethooks
Differently from \ensuremath{()}, we don't have to implement the \ensuremath{\ty{\Conid{Session}}} class for \ensuremath{\seshtysel} and \ensuremath{\seshtyoff}. They're already session types!

\paragraph{Recursion}
We can write recursive session types by writing them as recursive Haskell types. Unfortunately, we cannot write recursive type synonyms, so we have to use a newtype. For instance, we can write the type for a recursive summation service, which receives numbers until the client indicates they're done, and then sends back the sum. We specify \emph{two} newtypes:
\begin{hscode}\SaveRestoreHook
\column{B}{@{}>{\hspre}l<{\hspost}@{}}%
\column{3}{@{}>{\hspre}l<{\hspost}@{}}%
\column{E}{@{}>{\hspre}l<{\hspost}@{}}%
\>[B]{}\mathbf{newtype}\;\seshtySumSrv{}\<[E]%
\\
\>[B]{}\hsindent{3}{}\<[3]%
\>[3]{}\mathrel{=}\Conid{SumSrv}\;(\seshtyoff\;(\seshtyrecv\;\Conid{Int}\;\seshtySumSrv)\;(\seshtysend\;\Conid{Int}\;\seshtyend)){}\<[E]%
\\
\>[B]{}\mathbf{newtype}\;\seshtySumCnt{}\<[E]%
\\
\>[B]{}\hsindent{3}{}\<[3]%
\>[3]{}\mathrel{=}\Conid{SumCnt}\;(\seshtysel\;(\seshtysend\;\Conid{Int}\;\seshtySumCnt)\;(\seshtyrecv\;\Conid{Int}\;\seshtyend)){}\<[E]%
\ColumnHook
\end{hscode}\resethooks
We implement the summation server as a recursive function:
\begin{hscode}\SaveRestoreHook
\column{B}{@{}>{\hspre}l<{\hspost}@{}}%
\column{3}{@{}>{\hspre}l<{\hspost}@{}}%
\column{10}{@{}>{\hspre}l<{\hspost}@{}}%
\column{E}{@{}>{\hspre}l<{\hspost}@{}}%
\>[B]{}\Varid{sumSrv}\mathbin{::}\Conid{Int}\hsPercent \hsOne \hsArrow{\rightarrow }{\multimap }{\mathpunct{.}}\seshtySumSrv[\cdot ]\hsPercent \hsOne \hsArrow{\rightarrow }{\multimap }{\mathpunct{.}}\Conid{IO}\;(){}\<[E]%
\\
\>[B]{}\Varid{sumSrv}\;\Varid{tot}\;(\Conid{SumSrv}\;\Varid{s})\mathrel{=}\tm{\Varid{offerEither}}\;\Varid{s}\mathbin{\$}\lambda \hsLambdaCmd \Varid{e}\hsArrow{\rightarrow }{\multimap }{\mathpunct{.}}\mathbf{case}\;\Varid{x}\;\mathbf{of}{}\<[E]%
\\
\>[B]{}\hsindent{3}{}\<[3]%
\>[3]{}\Conid{Left}\;{}\<[10]%
\>[10]{}\Varid{s}\hsArrow{\rightarrow }{\multimap }{\mathpunct{.}}\mathbf{do}\;(\Varid{x},\Varid{s})\leftarrow \tm{\Varid{recv}}\;\Varid{s};\Varid{sumSrv}\;(\Varid{tot}\mathbin{+}\Varid{x})\;\Varid{s}{}\<[E]%
\\
\>[B]{}\hsindent{3}{}\<[3]%
\>[3]{}\Conid{Right}\;{}\<[10]%
\>[10]{}\Varid{s}\hsArrow{\rightarrow }{\multimap }{\mathpunct{.}}\mathbf{do}\;\Varid{s}\leftarrow \tm{\Varid{send}}\;(\Varid{tot},\Varid{s});\tm{\Varid{close}}\;\Varid{s}{}\<[E]%
\ColumnHook
\end{hscode}\resethooks
As \ensuremath{\seshtySumSrv} and \ensuremath{\seshtySumCnt} are new types, we must provide instances of the \ensuremath{\ty{\Conid{Session}}} class for them.
\begin{hscode}\SaveRestoreHook
\column{B}{@{}>{\hspre}l<{\hspost}@{}}%
\column{3}{@{}>{\hspre}l<{\hspost}@{}}%
\column{5}{@{}>{\hspre}l<{\hspost}@{}}%
\column{15}{@{}>{\hspre}l<{\hspost}@{}}%
\column{E}{@{}>{\hspre}l<{\hspost}@{}}%
\>[B]{}\mathbf{instance}\;\ty{\Conid{Session}}\;\seshtySumSrv{}\<[E]%
\\
\>[B]{}\hsindent{3}{}\<[3]%
\>[3]{}\mathbf{where}{}\<[E]%
\\
\>[3]{}\hsindent{2}{}\<[5]%
\>[5]{}\mathbf{type}\;\ty{\Conid{Dual}}\;\seshtySumSrv\mathrel{=}\seshtySumCnt{}\<[E]%
\\
\>[3]{}\hsindent{2}{}\<[5]%
\>[5]{}\tm{\Varid{new}}\mathrel{=}\mathbf{do}\;{}\<[15]%
\>[15]{}(\Varid{ch}_{\Varid{srv}},\Varid{ch}_{\Varid{cnt}})\leftarrow \tm{\Varid{new}}{}\<[E]%
\\
\>[15]{}\Varid{return}\;(\Conid{SumSrv}\;\Varid{ch}_{\Varid{srv}},\Conid{SumCnt}\;\Varid{ch}_{\Varid{cnt}}){}\<[E]%
\ColumnHook
\end{hscode}\resethooks

\subsection{Deadlock freedom via process structure}\label{sec:tree-sesh}
The session-typed channels presented in \cref{sec:sesh} can be used to write deadlocking programs:
\begin{hscode}\SaveRestoreHook
\column{B}{@{}>{\hspre}l<{\hspost}@{}}%
\column{13}{@{}>{\hspre}l<{\hspost}@{}}%
\column{24}{@{}>{\hspre}l<{\hspost}@{}}%
\column{E}{@{}>{\hspre}l<{\hspost}@{}}%
\>[B]{}\Varid{woops}\mathbin{::}\Conid{IO}\;\Conid{Void}{}\<[E]%
\\
\>[B]{}\Varid{woops}\mathrel{=}\mathbf{do}\;{}\<[13]%
\>[13]{}(\Varid{ch}_{\Varid{s1}},\Varid{ch}_{\Varid{r1}})\leftarrow \tm{\Varid{new}}{}\<[E]%
\\
\>[13]{}(\Varid{ch}_{\Varid{s2}},\Varid{ch}_{\Varid{r2}})\leftarrow \tm{\Varid{new}}{}\<[E]%
\\
\>[13]{}\tm{\Varid{fork}}\mathbin{\$}\mathbf{do}\;{}\<[24]%
\>[24]{}(\Varid{void},())\leftarrow \tm{\Varid{recv}}\;\Varid{ch}_{\Varid{r1}}{}\<[E]%
\\
\>[24]{}\tm{\Varid{send}}\;(\Varid{void},\Varid{ch}_{\Varid{s2}}){}\<[E]%
\\
\>[13]{}(\Varid{void},())\leftarrow \tm{\Varid{recv}}\;\Varid{ch}_{\Varid{r2}}{}\<[E]%
\\
\>[13]{}\mathbf{let}\;(\Varid{void},\Varid{void}_{\Varid{copy}})\mathrel{=}\Varid{dup2}\;\Varid{void}{}\<[E]%
\\
\>[13]{}\tm{\Varid{send}}\;(\Varid{void},\Varid{ch}_{\Varid{s1}}){}\<[E]%
\\
\>[13]{}\Varid{return}\;\Varid{void}_{\Varid{copy}}{}\<[E]%
\ColumnHook
\end{hscode}\resethooks
Counter to what the type says, this program doesn't actually produce an inhabitant of the \emph{uninhabited} type \ensuremath{\Conid{Void}}. Instead, it deadlocks! We'd like to help the programmer avoid such programs.

As discussed in \cref{sec:introduction}, we can \emph{structurally} guarantee deadlock freedom by ensuring that the \emph{process structure} is always a tree or forest. The process structure of a program is an undirected graph, where nodes represent processes, and edges represent the channels connecting them. For instance, the process structure of \ensuremath{\Varid{woops}} is cyclic:
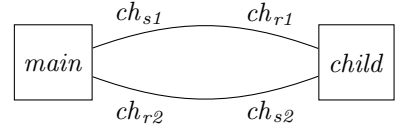
\begin{center}
\begin{tikzpicture}
  \node[draw, minimum size=1cm] (x) {$\Varid{main}$};
  \node[draw, minimum size=1cm, right=3cm of x] (y) {$\Varid{child}$};
  \path
  (x) edge[bend left=20]
      node[pos=0.2,above] {$\Varid{ch}_{\Varid{s1}}$}
      node[pos=0.8,above] {$\Varid{ch}_{\Varid{r1}}$}
  (y);
  \path
  (y) edge[bend left=20]
      node[pos=0.2,below] {$\Varid{ch}_{\Varid{s2}}$}
      node[pos=0.8,below] {$\Varid{ch}_{\Varid{r2}}$}
  (x);
\end{tikzpicture}
\end{center}
This restriction works by ensuring that between two processes there is \emph{at most} one (series of) channels over which the two can communicate. As duality rules out deadlocks on any one channel, such configurations must be deadlock free.

We can rule out cyclic process structures by hiding \ensuremath{\tm{\Varid{new}}}, and only exporting \ensuremath{\tm{\Varid{connect}}}, which creates a new channel and, \emph{crucially}, immediately passes one endpoint to a new thread:
\begin{hscode}\SaveRestoreHook
\column{B}{@{}>{\hspre}l<{\hspost}@{}}%
\column{13}{@{}>{\hspre}l<{\hspost}@{}}%
\column{E}{@{}>{\hspre}l<{\hspost}@{}}%
\>[B]{}\tm{\Varid{connect}}\mathbin{::}{}\<[13]%
\>[13]{}\ty{\Conid{Session}}\;\Varid{s}\Rightarrow {}\<[E]%
\\
\>[13]{}(\Varid{s}\hsPercent \hsOne \hsArrow{\rightarrow }{\multimap }{\mathpunct{.}}\Conid{IO}\;())\hsPercent \hsOne \hsArrow{\rightarrow }{\multimap }{\mathpunct{.}}(\ty{\Conid{Dual}}\;\Varid{s}\hsPercent \hsOne \hsArrow{\rightarrow }{\multimap }{\mathpunct{.}}\Conid{IO}\;\Varid{a})\hsPercent \hsOne \hsArrow{\rightarrow }{\multimap }{\mathpunct{.}}\Conid{IO}\;\Varid{a}{}\<[E]%
\\
\>[B]{}\tm{\Varid{connect}}\;\Varid{k}_{1}\;\Varid{k}_{2}\mathrel{=}\mathbf{do}\;(\Varid{s}_{1},\Varid{s}_{2})\leftarrow \tm{\Varid{new}};\tm{\Varid{fork}}\;(\Varid{k}_{1}\;\Varid{s}_{1});\Varid{k}_{2}\;\Varid{s}_{2}{}\<[E]%
\ColumnHook
\end{hscode}\resethooks
You can view \ensuremath{\tm{\Varid{connect}}} as the node constructor for a binary process tree. If the programmer \emph{only} uses \ensuremath{\tm{\Varid{connect}}}, their process structure is guaranteed to be a \emph{tree}. If they also use standalone \ensuremath{\tm{\Varid{fork}}}, their process structure is a \emph{forest}. Either way, their programs are guaranteed to be deadlock free.

\subsection{Deadlock freedom via priorities}\label{sec:priority-sesh}
The strategy for deadlock freedom presented in \cref{sec:tree-sesh} is simple, but \emph{very} restrictive, since it rules out \emph{all} cyclic communication structures, even the ones which don't deadlock:
\begin{hscode}\SaveRestoreHook
\column{B}{@{}>{\hspre}l<{\hspost}@{}}%
\column{3}{@{}>{\hspre}l<{\hspost}@{}}%
\column{21}{@{}>{\hspre}l<{\hspost}@{}}%
\column{32}{@{}>{\hspre}l<{\hspost}@{}}%
\column{E}{@{}>{\hspre}l<{\hspost}@{}}%
\>[3]{}\Varid{totallyFine}\mathbin{::}\Conid{IO}\;\Conid{String}{}\<[E]%
\\
\>[3]{}\Varid{totallyFine}\mathrel{=}\mathbf{do}\;{}\<[21]%
\>[21]{}(\Varid{ch}_{\Varid{s1}},\Varid{ch}_{\Varid{r1}})\leftarrow \tm{\Varid{new}}{}\<[E]%
\\
\>[21]{}(\Varid{ch}_{\Varid{s2}},\Varid{ch}_{\Varid{r2}})\leftarrow \tm{\Varid{new}}{}\<[E]%
\\
\>[21]{}\tm{\Varid{fork}}\mathbin{\$}\mathbf{do}\;{}\<[32]%
\>[32]{}(\Varid{x},())\leftarrow \tm{\Varid{recv}}\;\Varid{ch}_{\Varid{r1}}{}\<[E]%
\\
\>[32]{}\tm{\Varid{send}}\;(\Varid{x},\Varid{ch}_{\Varid{s2}}){}\<[E]%
\\
\>[21]{}\tm{\Varid{send}}\;(\text{\ttfamily \char34 Hiya!\char34},\Varid{ch}_{\Varid{s1}}){}\<[E]%
\\
\>[21]{}(\Varid{x},())\leftarrow \tm{\Varid{recv}}\;\Varid{ch}_{\Varid{r2}}{}\<[E]%
\\
\>[21]{}\Varid{return}\;\Varid{x}{}\<[E]%
\ColumnHook
\end{hscode}\resethooks
This process has \emph{exactly the same} process structure as \ensuremath{\Varid{woops}}, but it's totally fine, and returns \ensuremath{\text{\ttfamily \char34 Hiya!\char34}} as you'd expect. We'd like to enable the programmer to write such programs while still guaranteeing their programs don't deadlock.

As discussed in \cref{sec:introduction}, there is another way to rule out deadlocks---by using \emph{priorities}. Priorities are an approximation of the \emph{communication graph} of a program. The communication graph of a program is a \emph{directed graph} where nodes represent \emph{actions on channels}, and directed edges represent that one action happens before the other. Dual actions are connected with double undirected edges. (You may consider the graph contracted along these edges.) If the communication graph is cyclic, the program deadlocks. The communication graphs for \ensuremath{\Varid{woops}} and \ensuremath{\Varid{totallyFine}} are as follows:
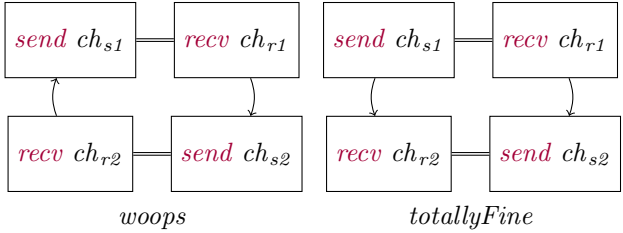
\begin{center}
  \begin{minipage}{0.5\linewidth}
    \centering
    \begin{tikzpicture}
      \node[draw, minimum size=1cm]                       (ch_s1) {\ensuremath{\tm{\Varid{send}}\;\Varid{ch}_{\Varid{s1}}}};
      \node[draw, minimum size=1cm, right=0.5cm of ch_s1] (ch_r1) {\ensuremath{\tm{\Varid{recv}}\;\Varid{ch}_{\Varid{r1}}}};
      \node[draw, minimum size=1cm, below=0.5cm of ch_r1] (ch_s2) {\ensuremath{\tm{\Varid{send}}\;\Varid{ch}_{\Varid{s2}}}};
      \node[draw, minimum size=1cm, below=0.5cm of ch_s1] (ch_r2) {\ensuremath{\tm{\Varid{recv}}\;\Varid{ch}_{\Varid{r2}}}};
      \path (ch_s1) edge[double] (ch_r1);
      \path (ch_s2) edge[double] (ch_r2);
      \path[->] (ch_r1) edge[bend left=20] (ch_s2);
      \path[->] (ch_r2) edge[bend left=20] (ch_s1);
    \end{tikzpicture}
    \\
    \ensuremath{\Varid{woops}}
  \end{minipage}%
  \begin{minipage}{0.5\linewidth}
    \centering
    \begin{tikzpicture}
      \node[draw, minimum size=1cm]                       (ch_s1) {\ensuremath{\tm{\Varid{send}}\;\Varid{ch}_{\Varid{s1}}}};
      \node[draw, minimum size=1cm, right=0.5cm of ch_s1] (ch_r1) {\ensuremath{\tm{\Varid{recv}}\;\Varid{ch}_{\Varid{r1}}}};
      \node[draw, minimum size=1cm, below=0.5cm of ch_r1] (ch_s2) {\ensuremath{\tm{\Varid{send}}\;\Varid{ch}_{\Varid{s2}}}};
      \node[draw, minimum size=1cm, below=0.5cm of ch_s1] (ch_r2) {\ensuremath{\tm{\Varid{recv}}\;\Varid{ch}_{\Varid{r2}}}};
      \path (ch_s1) edge[double] (ch_r1);
      \path (ch_s2) edge[double] (ch_r2);
      \path[->] (ch_s1) edge[bend right=20] (ch_r2);
      \path[->] (ch_r1) edge[bend left=20] (ch_s2);
    \end{tikzpicture}
    \\
    \ensuremath{\Varid{totallyFine}}
  \end{minipage}%
\end{center}
If the communication graph is acyclic, then we can assign each node a number such that directed edges only ever point to nodes with \emph{bigger} numbers. For instance, for \ensuremath{\Varid{totallyFine}} we can assign the number \ensuremath{\mathrm{0}} to \ensuremath{\tm{\Varid{send}}\;\Varid{ch}_{\Varid{s1}}} and \ensuremath{\tm{\Varid{recv}}\;\Varid{ch}_{\Varid{r2}}}, and \ensuremath{\hsOne } to \ensuremath{\tm{\Varid{recv}}\;\Varid{ch}_{\Varid{r2}}} and \ensuremath{\tm{\Varid{send}}\;\Varid{ch}_{\Varid{s2}}}. These numbers are \emph{priorities}.

In this section, we present a type system in which \emph{priorities} are used to ensure deadlock freedom, by tracking the time a process starts and finishes communicating using a graded monad~\cite{gaboardikatsumata16,orchardwadler20}. The bind operation registers the order of its actions in the type, requiring the sequentiality of their duals.

\paragraph{Priorities}
The priorities assigned to \emph{communication actions} are always natural numbers, which represent, \emph{abstractly}, at which time the action happens. When tracking the start and finish times of a program, however, we also use \ensuremath{\cs{\bot}} and \ensuremath{\cs{\top}} for programs which don't communicate. These are used as the identities for \ensuremath{\mathbin{\cs{\sqcap}}} and \ensuremath{\mathbin{\cs{\sqcup}}} in lower and upper bounds, respectively. We let \ensuremath{\Varid{o}} range over natural numbers, \ensuremath{\Varid{p}} over \emph{lower bounds}, and \ensuremath{\Varid{q}} over \emph{upper bounds}.

\begin{hscode}\SaveRestoreHook
\column{B}{@{}>{\hspre}l<{\hspost}@{}}%
\column{E}{@{}>{\hspre}l<{\hspost}@{}}%
\>[B]{}\mathbf{data}\;\ty{\Conid{Priority}}\mathrel{=}\cs{\bot}\mid \Conid{Nat}\mid \cs{\top}{}\<[E]%
\ColumnHook
\end{hscode}\resethooks

We define strict inequality ($\ensuremath{\mathrel{\cs{<}}}$), minimum (\ensuremath{\mathbin{\cs{\sqcap}}}), and maximum (\ensuremath{\mathbin{\cs{\sqcup}}}) on priorities as usual.

\paragraph{Channels}
We define \ensuremath{\seshtysend[\cs{\Varid{o}}]}, \ensuremath{\seshtyrecv[\cs{\Varid{o}}]}, and \ensuremath{\seshtyend[\cs{\Varid{o}}]}, which decorate the \emph{raw} sessions from \cref{sec:sesh} with the priority \ensuremath{\Varid{o}} of the communication action, \ie, it denoted when the communication happens. Duality (\ensuremath{\ty{\Conid{Dual}}}) preserves these priorities. These are implemented exactly as in \cref{sec:sesh}.

\paragraph{The communication monad}
We define a graded monad \ensuremath{\seshtymon{\Varid{p}}{\Varid{q}}}, which decorates \ensuremath{\Conid{IO}} with a lower bound \ensuremath{\Varid{p}} and an upper bound \ensuremath{\Varid{q}} on the priorities of its communication actions, \ie, if you run the monad, it denotes when communication begins and ends.

\begin{hscode}\SaveRestoreHook
\column{B}{@{}>{\hspre}l<{\hspost}@{}}%
\column{E}{@{}>{\hspre}l<{\hspost}@{}}%
\>[B]{}\mathbf{newtype}\;\seshtymon{\Varid{p}}{\Varid{q}}\;\Varid{a}\mathrel{=}\tm{\Conid{Sesh}}\;\{\mskip1.5mu \tm{\Varid{runSeshIO}}\mathbin{::}\Conid{IO}\;\Varid{a}\mskip1.5mu\}{}\<[E]%
\ColumnHook
\end{hscode}\resethooks

The monad operations for \ensuremath{\seshtymon{\Varid{p}}{\Varid{q}}} merely wrap those for \ensuremath{\Conid{IO}}, hence trivially obeys the monad laws.

The \ensuremath{\tm{\Varid{ireturn}}} function returns a \emph{pure} computation---the type \ensuremath{\seshtymon{\cs{\top}}{\cs{\bot}}} guarantees that all communications happen between \ensuremath{\cs{\top}} and \ensuremath{\cs{\bot}}, hence there can be no communication at all.

\begin{hscode}\SaveRestoreHook
\column{B}{@{}>{\hspre}l<{\hspost}@{}}%
\column{E}{@{}>{\hspre}l<{\hspost}@{}}%
\>[B]{}\tm{\Varid{ireturn}}\mathbin{::}\Varid{a}\hsPercent \hsOne \hsArrow{\rightarrow }{\multimap }{\mathpunct{.}}\seshtymon{\cs{\top}}{\cs{\bot}}\;\Varid{a}{}\<[E]%
\\
\>[B]{}\tm{\Varid{ireturn}}\;\Varid{x}\mathrel{=}\tm{\Conid{Sesh}}\mathbin{\$}\Varid{return}\;\Varid{x}{}\<[E]%
\ColumnHook
\end{hscode}\resethooks

The \ensuremath{\tm{\ibind}} operator sequences two actions with types \ensuremath{\seshtymon{\Varid{p}}{\Varid{q}}} and \ensuremath{\seshtymon{\Varid{p'}}{\Varid{q'}}}, and requires \ensuremath{\cs{\Varid{q}}\mathrel{\cs{<}}\cs{\Varid{p'}}}, \ie, the first action must have finished before the second starts. The resulting action has lower bound \ensuremath{\cs{\Varid{p}}\mathbin{\cs{\sqcap}}\cs{\Varid{p'}}} and upper bound \ensuremath{\cs{\Varid{q}}\mathbin{\cs{\sqcup}}\cs{\Varid{q'}}}.

\begin{hscode}\SaveRestoreHook
\column{B}{@{}>{\hspre}l<{\hspost}@{}}%
\column{E}{@{}>{\hspre}l<{\hspost}@{}}%
\>[B]{}(\tm{\ibind})\mathbin{::}(\cs{\Varid{q}}\mathrel{\cs{<}}\cs{\Varid{p'}})\Rightarrow \seshtymon{\Varid{p}}{\Varid{q}}\;\Varid{a}\hsPercent \hsOne \hsArrow{\rightarrow }{\multimap }{\mathpunct{.}}(\Varid{a}\hsPercent \hsOne \hsArrow{\rightarrow }{\multimap }{\mathpunct{.}}\seshtymon{\Varid{p'}}{\Varid{q'}}\;\Varid{b})\hsPercent \hsOne \hsArrow{\rightarrow }{\multimap }{\mathpunct{.}}\seshtymon{\cs{\Varid{p}}\mathbin{\cs{\sqcap}}\cs{\Varid{p'}}}{\cs{\Varid{q}}\mathbin{\cs{\sqcup}}\cs{\Varid{q'}}}\;\Varid{b}{}\<[E]%
\\
\>[B]{}\Varid{mx}\tm{\ibind}\Varid{mf}\mathrel{=}\tm{\Conid{Sesh}}\mathbin{\$}\tm{\Varid{runSeshIO}}\;\Varid{mx}\bind \lambda \hsLambdaCmd \Varid{x}\hsArrow{\rightarrow }{\multimap }{\mathpunct{.}}\tm{\Varid{runSeshIO}}\;(\Varid{mf}\;\Varid{x}){}\<[E]%
\ColumnHook
\end{hscode}\resethooks
In what follows, we implicitly use \ensuremath{\tm{\ibind}} with do-notation. This can be accomplished in Haskell using \texttt{RebindableSyntax}.

We define decorated variants of the concurrency and communication primitives:
\begin{itemize*}[label=\empty]
\item \ensuremath{\tm{\Varid{send}}}, \ensuremath{\tm{\Varid{recv}}}, and \ensuremath{\tm{\Varid{close}}} each perform a communication action with some priority \ensuremath{\Varid{o}}, and return a computation of type \ensuremath{\seshtymon{\Varid{o}}{\Varid{o}}}, \ie, with \emph{exact} bounds;
\item \ensuremath{\tm{\Varid{new}}}, \ensuremath{\tm{\Varid{fork}}}, and \ensuremath{\tm{\Varid{cancel}}} don't perform any communication action, and so return a \emph{pure} computation of type \ensuremath{\seshtymon{\cs{\top}}{\cs{\bot}}}.
\end{itemize*}
\begin{hscode}\SaveRestoreHook
\column{B}{@{}>{\hspre}l<{\hspost}@{}}%
\column{10}{@{}>{\hspre}l<{\hspost}@{}}%
\column{E}{@{}>{\hspre}l<{\hspost}@{}}%
\>[B]{}\tm{\Varid{new}}{}\<[10]%
\>[10]{}\mathbin{::}\ty{\Conid{Session}}\;\Varid{s}\Rightarrow \seshtymon{\cs{\top}}{\cs{\bot}}\;(\Varid{s},\ty{\Conid{Dual}}\;\Varid{s}){}\<[E]%
\\
\>[B]{}\tm{\Varid{fork}}{}\<[10]%
\>[10]{}\mathbin{::}\seshtymon{\Varid{p}}{\Varid{q}}\;()\hsPercent \hsOne \hsArrow{\rightarrow }{\multimap }{\mathpunct{.}}\seshtymon{\cs{\top}}{\cs{\bot}}\;(){}\<[E]%
\\
\>[B]{}\tm{\Varid{cancel}}{}\<[10]%
\>[10]{}\mathbin{::}\ty{\Conid{Session}}\;\Varid{s}\Rightarrow \Varid{s}\hsPercent \hsOne \hsArrow{\rightarrow }{\multimap }{\mathpunct{.}}\seshtymon{\cs{\top}}{\cs{\bot}}\;(){}\<[E]%
\\
\>[B]{}\tm{\Varid{send}}{}\<[10]%
\>[10]{}\mathbin{::}\ty{\Conid{Session}}\;\Varid{s}\Rightarrow (\Varid{a},\seshtysend[\cs{\Varid{o}}]\;\Varid{a}\;\Varid{s})\hsPercent \hsOne \hsArrow{\rightarrow }{\multimap }{\mathpunct{.}}\seshtymon{\Varid{o}}{\Varid{o}}\;\Varid{s}{}\<[E]%
\\
\>[B]{}\tm{\Varid{recv}}{}\<[10]%
\>[10]{}\mathbin{::}\seshtyrecv[\cs{\Varid{o}}]\;\Varid{a}\;\Varid{s}\hsPercent \hsOne \hsArrow{\rightarrow }{\multimap }{\mathpunct{.}}\seshtymon{\Varid{o}}{\Varid{o}}\;(\Varid{a},\Varid{s}){}\<[E]%
\\
\>[B]{}\tm{\Varid{close}}{}\<[10]%
\>[10]{}\mathbin{::}\seshtyend[\cs{\Varid{o}}]\hsPercent \hsOne \hsArrow{\rightarrow }{\multimap }{\mathpunct{.}}\seshtymon{\Varid{o}}{\Varid{o}}\;(){}\<[E]%
\ColumnHook
\end{hscode}\resethooks
From these, we derive decorated choice, as before:
\begin{hscode}\SaveRestoreHook
\column{B}{@{}>{\hspre}l<{\hspost}@{}}%
\column{14}{@{}>{\hspre}c<{\hspost}@{}}%
\column{14E}{@{}l@{}}%
\column{16}{@{}>{\hspre}l<{\hspost}@{}}%
\column{18}{@{}>{\hspre}l<{\hspost}@{}}%
\column{34}{@{}>{\hspre}l<{\hspost}@{}}%
\column{E}{@{}>{\hspre}l<{\hspost}@{}}%
\>[B]{}\mathbf{type}\;\seshtysel[\cs{\Varid{o}}]\;{}\<[16]%
\>[16]{}\Varid{s}_{1}\;\Varid{s}_{2}\mathrel{=}\seshtysend[\cs{\Varid{o}}]\;{}\<[34]%
\>[34]{}(\Conid{Either}\;(\ty{\Conid{Dual}}\;\Varid{s}_{1})\;(\ty{\Conid{Dual}}\;\Varid{s}_{2}))\;(){}\<[E]%
\\
\>[B]{}\mathbf{type}\;\seshtyoff[\cs{\Varid{o}}]\;{}\<[16]%
\>[16]{}\Varid{s}_{1}\;\Varid{s}_{2}\mathrel{=}\seshtyrecv[\cs{\Varid{o}}]\;{}\<[34]%
\>[34]{}(\Conid{Either}\;\Varid{s}_{1}\;\Varid{s}_{2})\;(){}\<[E]%
\\[\blanklineskip]%
\>[B]{}\tm{\Varid{selectLeft}}{}\<[14]%
\>[14]{}\mathbin{::}{}\<[14E]%
\>[18]{}(\ty{\Conid{Session}}\;\Varid{s}_{1})\Rightarrow \seshtysel[\cs{\Varid{o}}]\;\Varid{s}_{1}\;\Varid{s}_{2}\hsPercent \hsOne \hsArrow{\rightarrow }{\multimap }{\mathpunct{.}}\seshtymon{\Varid{o}}{\Varid{o}}\;\Varid{s}_{1}{}\<[E]%
\\
\>[B]{}\tm{\Varid{selectRight}}{}\<[14]%
\>[14]{}\mathbin{::}{}\<[14E]%
\>[18]{}(\ty{\Conid{Session}}\;\Varid{s}_{2})\Rightarrow \seshtysel[\cs{\Varid{o}}]\;\Varid{s}_{1}\;\Varid{s}_{2}\hsPercent \hsOne \hsArrow{\rightarrow }{\multimap }{\mathpunct{.}}\seshtymon{\Varid{o}}{\Varid{o}}\;\Varid{s}_{2}{}\<[E]%
\\
\>[B]{}\tm{\Varid{offerEither}}{}\<[14]%
\>[14]{}\mathbin{::}{}\<[14E]%
\>[18]{}(\cs{\Varid{o}}\mathrel{\cs{<}}\cs{\Varid{p}})\Rightarrow \seshtyoff[\cs{\Varid{o}}]\;\Varid{s}_{1}\;\Varid{s}_{2}\hsPercent \hsOne \hsArrow{\rightarrow }{\multimap }{\mathpunct{.}}{}\<[E]%
\\
\>[18]{}(\Conid{Either}\;\Varid{s}_{1}\;\Varid{s}_{2}\hsPercent \hsOne \hsArrow{\rightarrow }{\multimap }{\mathpunct{.}}\seshtymon{\Varid{p}}{\Varid{q}}\;\Varid{a})\hsPercent \hsOne \hsArrow{\rightarrow }{\multimap }{\mathpunct{.}}\seshtymon{\cs{\Varid{o}}\mathbin{\cs{\sqcap}}\cs{\Varid{p}}}{\cs{\Varid{o}}\mathbin{\cs{\sqcup}}\cs{\Varid{q}}}\;\Varid{a}{}\<[E]%
\ColumnHook
\end{hscode}\resethooks

\paragraph{Safe IO}
We can use a trick from the \ensuremath{\Conid{ST}} monad~\cite{launchburypeytonjones94} to define a ``pure'' variant of \ensuremath{\tm{\Varid{runSesh}}}, which encapsulates all use of IO within the \ensuremath{\seshtymon{\Varid{p}}{\Varid{q}}} monad. The idea is to index the \ensuremath{\seshtymon{\Varid{p}}{\Varid{q}}} and every session type constructor with an extra type parameter \ensuremath{\Varid{tok}}, which we'll call the \emph{session token}:
\begin{hscode}\SaveRestoreHook
\column{B}{@{}>{\hspre}l<{\hspost}@{}}%
\column{3}{@{}>{\hspre}l<{\hspost}@{}}%
\column{12}{@{}>{\hspre}l<{\hspost}@{}}%
\column{E}{@{}>{\hspre}l<{\hspost}@{}}%
\>[3]{}\tm{\Varid{send}}{}\<[12]%
\>[12]{}\mathbin{::}\ty{\Conid{Session}}\;\Varid{s}\Rightarrow (\Varid{a},\seshtysend[\cs{\Varid{o}}]\;\Varid{tok}\;\Varid{a}\;\Varid{s})\hsPercent \hsOne \hsArrow{\rightarrow }{\multimap }{\mathpunct{.}}\seshtymon{\Varid{o}}{\Varid{o}}\;\Varid{tok}\;\Varid{s}{}\<[E]%
\\
\>[3]{}\tm{\Varid{recv}}{}\<[12]%
\>[12]{}\mathbin{::}\seshtyrecv[\cs{\Varid{o}}]\;\Varid{tok}\;\Varid{a}\;\Varid{s}\hsPercent \hsOne \hsArrow{\rightarrow }{\multimap }{\mathpunct{.}}\seshtymon{\Varid{o}}{\Varid{o}}\;\Varid{tok}\;(\Varid{a},\Varid{s}){}\<[E]%
\\
\>[3]{}\tm{\Varid{close}}{}\<[12]%
\>[12]{}\mathbin{::}\seshtyend[\cs{\Varid{o}}]\;\Varid{tok}\hsPercent \hsOne \hsArrow{\rightarrow }{\multimap }{\mathpunct{.}}\seshtymon{\Varid{o}}{\Varid{o}}\;\Varid{tok}\;(){}\<[E]%
\ColumnHook
\end{hscode}\resethooks
The session token should never be instantiated, except by \ensuremath{\tm{\Varid{runSesh}}}, and every action under the same call to \ensuremath{\tm{\Varid{runSesh}}} should use the same type variable \ensuremath{\Varid{tok}} as its session token:
\begin{hscode}\SaveRestoreHook
\column{B}{@{}>{\hspre}l<{\hspost}@{}}%
\column{3}{@{}>{\hspre}l<{\hspost}@{}}%
\column{E}{@{}>{\hspre}l<{\hspost}@{}}%
\>[3]{}\tm{\Varid{runSesh}}\mathbin{::}(\forall \Varid{tok}\hsForallCmd \hsDot{\circ }{\mathpunct{.}}\seshtymon{\Varid{p}}{\Varid{q}}\;\Varid{tok}\;\Varid{a})\hsPercent \hsOne \hsArrow{\rightarrow }{\multimap }{\mathpunct{.}}\Varid{a}{}\<[E]%
\\
\>[3]{}\tm{\Varid{runSesh}}\;\Varid{x}\mathrel{=}\Varid{unsafePerformIO}\;(\tm{\Varid{runSeshIO}}\;\Varid{x}){}\<[E]%
\ColumnHook
\end{hscode}\resethooks
This ensures that none of the channels created in the session can escape out of the scope of \ensuremath{\tm{\Varid{runSesh}}}.

We implement this encapsulation in \texttt{priority-sesh}, though the session token is the first argument, preceding the priority bounds.

\paragraph{Recursion}
We could implement recursive session via priority-polymorphic types, or via priority-shifting~\cite{padovaninovara15}. For instance, we could give the \emph{summation service} from \cref{sec:sesh} the following type:
\begin{hscode}\SaveRestoreHook
\column{B}{@{}>{\hspre}l<{\hspost}@{}}%
\column{3}{@{}>{\hspre}l<{\hspost}@{}}%
\column{24}{@{}>{\hspre}l<{\hspost}@{}}%
\column{E}{@{}>{\hspre}l<{\hspost}@{}}%
\>[B]{}\mathbf{newtype}\;\seshtySumSrv[\Varid{o}]{}\<[E]%
\\
\>[B]{}\hsindent{3}{}\<[3]%
\>[3]{}\mathrel{=}\Conid{SumSrv}\;(\seshtyoff[\cs{\Varid{o}}]\;{}\<[24]%
\>[24]{}(\seshtyrecv[\cs{\Varid{o}\mathbin{+}\hsOne }]\;\Conid{Int}\;(\seshtySumSrv[\Varid{o}\mathbin{+}\mathrm{2}]))\;{}\<[E]%
\\
\>[24]{}(\seshtysend[\cs{\Varid{o}\mathbin{+}\hsOne }]\;\Conid{Int}\;(\seshtyend[\cs{\Varid{o}\mathbin{+}\mathrm{2}}]))){}\<[E]%
\ColumnHook
\end{hscode}\resethooks
We'd then like to assign \ensuremath{\Varid{sumSrv}} the following type:
\begin{hscode}\SaveRestoreHook
\column{B}{@{}>{\hspre}l<{\hspost}@{}}%
\column{3}{@{}>{\hspre}l<{\hspost}@{}}%
\column{10}{@{}>{\hspre}l<{\hspost}@{}}%
\column{E}{@{}>{\hspre}l<{\hspost}@{}}%
\>[B]{}\Varid{sumSrv}\mathbin{:}\Conid{Int}\hsPercent \hsOne \hsArrow{\rightarrow }{\multimap }{\mathpunct{.}}\seshtySumSrv[\Varid{o}]\hsPercent \hsOne \hsArrow{\rightarrow }{\multimap }{\mathpunct{.}}\seshtymon{\Varid{o}}{\cs{\top}}\;(){}\<[E]%
\\
\>[B]{}\Varid{sumSrv}\;\Varid{tot}\;(\Conid{SumSrv}\;\Varid{s})\mathrel{=}\tm{\Varid{offerEither}}\;\Varid{s}\mathbin{\$}\lambda \hsLambdaCmd \Varid{e}\hsArrow{\rightarrow }{\multimap }{\mathpunct{.}}\mathbf{case}\;\Varid{x}\;\mathbf{of}{}\<[E]%
\\
\>[B]{}\hsindent{3}{}\<[3]%
\>[3]{}\Conid{Left}\;{}\<[10]%
\>[10]{}\Varid{s}\hsArrow{\rightarrow }{\multimap }{\mathpunct{.}}\mathbf{do}\;(\Varid{x},\Varid{s})\leftarrow \tm{\Varid{recv}}\;\Varid{s};\Varid{sumSrv}\;(\Varid{tot}\mathbin{+}\Varid{x})\;\Varid{s}{}\<[E]%
\\
\>[B]{}\hsindent{3}{}\<[3]%
\>[3]{}\Conid{Right}\;{}\<[10]%
\>[10]{}\Varid{s}\hsArrow{\rightarrow }{\multimap }{\mathpunct{.}}\mathbf{do}\;\Varid{s}\leftarrow \tm{\Varid{send}}\;(\Varid{tot},\Varid{s});\tm{\Varid{weaken}}\;(\tm{\Varid{close}}\;\Varid{s}){}\<[E]%
\ColumnHook
\end{hscode}\resethooks
The upper bound for a recursive call should be \ensuremath{\cs{\top}}, which ensures that recursive calls are only made in \emph{tail} position~\cite{bernardidardha14,gaythiemann20}. The recursive call naturally has upper bound \ensuremath{\cs{\top}}. However, the \ensuremath{\tm{\Varid{close}}} operation happens at some \emph{concrete} priority $\cs{o + n}$, which needs to be raised to \ensuremath{\cs{\top}}, so we'd have to add a primitive \ensuremath{\tm{\Varid{weaken}}\mathbin{:}\seshtymon{\Varid{p}}{\Varid{q}}\;\Varid{a}\hsPercent \hsOne \hsArrow{\rightarrow }{\multimap }{\mathpunct{.}}\seshtymon{\Varid{p}}{\cs{\top}}\;\Varid{a}}.

Unfortunately, writing such priority-polymorphic code relies heavily on GHC's ability to reason about type-level naturals, and GHC rejects \ensuremath{\Varid{sumSrv}} complaining that it cannot verify that \ensuremath{\cs{\Varid{o}}\mathrel{\cs{<}}\cs{\Varid{o}\mathbin{+}\hsOne }}, \ensuremath{\cs{\Varid{o}\mathbin{+}\hsOne }\mathrel{\cs{<}}\cs{\Varid{o}\mathbin{+}\mathrm{2}}}, \etc. There's several possible solutions for this:
\begin{enumerate}
\item We could embrace the Hasochism~\cite{lindleymcbride13}, and provide GHC with explicit evidence, though this would make \texttt{priority-sesh} more difficult to use.
\item We could delegate \emph{some} of these problems to a GHC plugin such as \texttt{type-nat-solver}\footnote{\url{https://github.com/yav/type-nat-solver}} or \texttt{ghc-typelits-presburger}\footnote{\url{https://hackage.haskell.org/package/ghc-typelits-presburger}}. Unfortunately, \ensuremath{\mathbin{\cs{\sqcap}}} and \ensuremath{\mathbin{\cs{\sqcup}}} are beyond Presburger arithmetic, and \texttt{type-nat-solver} has not been maintained in recent years.
\item We could attempt to write type families which reduce in as many cases as possible. Unfortunately, a restriction in closed type families~\cite[\S6.1]{eisenbergvytiniotis14} prevents us from checking \emph{exactly these cases}.
\end{enumerate}
Currently, the prioritised sessions don't support recursion, and implementing one of these solutions is future work.

\paragraph{Cyclic Scheduler}
\citet{dardhagay18} and \citet{kokkedardha21} use a \emph{finite} cyclic scheduler as an example. The cyclic scheduler has the following process structure, with the flow of information indicated by the dotted arrows:
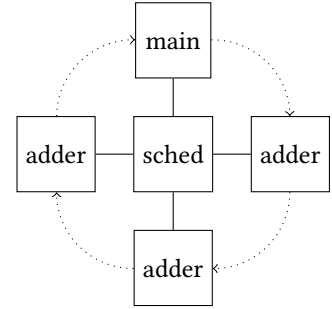
\begin{center}
\begin{tikzpicture}
  \node[draw, minimum size=1cm] (sched)  {sched};
  \node[draw, minimum size=1cm, above=0.5cm of sched] (main)   {main};
  \node[draw, minimum size=1cm, right=0.5cm of sched] (adder1) {adder};
  \node[draw, minimum size=1cm, below=0.5cm of sched] (adder2) {adder};
  \node[draw, minimum size=1cm, left =0.5cm of sched] (adder3) {adder};
  \path (sched) edge (main);
  \path (sched) edge (adder1);
  \path (sched) edge (adder2);
  \path (sched) edge (adder3);
  \path[->] (main)   edge[dotted, bend left=45] (adder1);
  \path[->] (adder1) edge[dotted, bend left=45] (adder2);
  \path[->] (adder2) edge[dotted, bend left=45] (adder3);
  \path[->] (adder3) edge[dotted, bend left=45] (main);
\end{tikzpicture}
\end{center}
We start by defining the types of the channels which connect each client process to the scheduler:
\begin{hscode}\SaveRestoreHook
\column{B}{@{}>{\hspre}l<{\hspost}@{}}%
\column{E}{@{}>{\hspre}l<{\hspost}@{}}%
\>[B]{}\mathbf{type}\;\seshtySR{\Varid{o}_{1}}{\Varid{o}_{2}}\;\Varid{a}\mathrel{=}\seshtysend[\cs{\Varid{o1}}]\;\Varid{a}\;(\seshtyrecv[\cs{\Varid{o}_{2}}]\;\Varid{a}\;()){}\<[E]%
\\
\>[B]{}\mathbf{type}\;\seshtyRS{\Varid{o}_{1}}{\Varid{o}_{2}}\;\Varid{a}\mathrel{=}\ty{\Conid{Dual}}\;(\seshtySR{\Varid{o1}}{\Varid{o}_{2}}\;\Varid{a}){}\<[E]%
\ColumnHook
\end{hscode}\resethooks
We then define the scheduler itself, which forwards messages from one process to the next in a cycle:
\begin{hscode}\SaveRestoreHook
\column{B}{@{}>{\hspre}l<{\hspost}@{}}%
\column{3}{@{}>{\hspre}l<{\hspost}@{}}%
\column{E}{@{}>{\hspre}l<{\hspost}@{}}%
\>[B]{}\Varid{sched}\mathbin{::}\seshtyRS{\mathrm{0}}{\mathrm{7}}\;\Varid{a}\hsPercent \hsOne \hsArrow{\rightarrow }{\multimap }{\mathpunct{.}}\seshtySR{\hsOne }{\mathrm{2}}\;\Varid{a}\hsPercent \hsOne \hsArrow{\rightarrow }{\multimap }{\mathpunct{.}}\seshtySR{\mathrm{3}}{\mathrm{4}}\;\Varid{a}\hsPercent \hsOne \hsArrow{\rightarrow }{\multimap }{\mathpunct{.}}\seshtySR{\mathrm{5}}{\mathrm{6}}\;\Varid{a}\hsPercent \hsOne \hsArrow{\rightarrow }{\multimap }{\mathpunct{.}}\seshtymon{\mathrm{0}}{\mathrm{7}}\;(){}\<[E]%
\\
\>[B]{}\Varid{sched}\;\Varid{s1}\;\Varid{s2}\;\Varid{s3}\;\Varid{s4}\mathrel{=}\mathbf{do}{}\<[E]%
\\
\>[B]{}\hsindent{3}{}\<[3]%
\>[3]{}(\Varid{x},\Varid{s1})\leftarrow \tm{\Varid{recv}}\;\Varid{s1}{}\<[E]%
\\
\>[B]{}\hsindent{3}{}\<[3]%
\>[3]{}\Varid{s2}\leftarrow \tm{\Varid{send}}\;(\Varid{x},\Varid{s2});(\Varid{x},())\leftarrow \tm{\Varid{recv}}\;\Varid{s2}{}\<[E]%
\\
\>[B]{}\hsindent{3}{}\<[3]%
\>[3]{}\Varid{s3}\leftarrow \tm{\Varid{send}}\;(\Varid{x},\Varid{s3});(\Varid{x},())\leftarrow \tm{\Varid{recv}}\;\Varid{s3}{}\<[E]%
\\
\>[B]{}\hsindent{3}{}\<[3]%
\>[3]{}\Varid{s4}\leftarrow \tm{\Varid{send}}\;(\Varid{x},\Varid{s4});(\Varid{x},())\leftarrow \tm{\Varid{recv}}\;\Varid{s4}{}\<[E]%
\\
\>[B]{}\hsindent{3}{}\<[3]%
\>[3]{}\tm{\Varid{send}}\;(\Varid{x},\Varid{s1}){}\<[E]%
\ColumnHook
\end{hscode}\resethooks
Finally, we define the \ensuremath{\Varid{adder}} and the \ensuremath{\Varid{main}} processes. The \ensuremath{\Varid{adder}} adds one to the value it receives, and the \ensuremath{\Varid{main}} process initiates the cycle and receives the result:
\begin{hscode}\SaveRestoreHook
\column{B}{@{}>{\hspre}l<{\hspost}@{}}%
\column{E}{@{}>{\hspre}l<{\hspost}@{}}%
\>[B]{}\Varid{adder}\mathbin{::}(\cs{\Varid{o}_{1}}\mathrel{\cs{<}}\cs{\Varid{o}_{2}})\Rightarrow \seshtyRS{\Varid{o}_{1}}{\Varid{o}_{2}}\;\Conid{Int}\hsPercent \hsOne \hsArrow{\rightarrow }{\multimap }{\mathpunct{.}}\seshtymon{\Varid{o}_{1}}{\Varid{o}_{2}}\;(){}\<[E]%
\\
\>[B]{}\Varid{adder}\;\Varid{s}\mathrel{=}\mathbf{do}\;(\Varid{x},\Varid{s})\leftarrow \tm{\Varid{recv}}\;\Varid{s};\tm{\Varid{send}}\;(\Varid{x}\mathbin{+}\hsOne ,\Varid{s}){}\<[E]%
\\[\blanklineskip]%
\>[B]{}\Varid{main}\mathbin{::}(\cs{\Varid{o}_{1}}\mathrel{\cs{<}}\cs{\Varid{o}_{2}})\Rightarrow \Conid{Int}\hsPercent \hsOne \hsArrow{\rightarrow }{\multimap }{\mathpunct{.}}\seshtySR{\Varid{o}_{1}}{\Varid{o}_{2}}\;\Conid{Int}\hsPercent \hsOne \hsArrow{\rightarrow }{\multimap }{\mathpunct{.}}\seshtymon{\Varid{o}_{1}}{\Varid{o}_{2}}\;\Conid{Int}{}\<[E]%
\\
\>[B]{}\Varid{main}\;\Varid{x}\;\Varid{s}\mathrel{=}\mathbf{do}\;;\Varid{s}\leftarrow \tm{\Varid{send}}\;(\Varid{x},\Varid{s});(\Varid{x},())\leftarrow \tm{\Varid{recv}}\;\Varid{s};\tm{\Varid{ireturn}}\;\Varid{x}{}\<[E]%
\ColumnHook
\end{hscode}\resethooks
While the process structure of the cyclic scheduler \emph{as presented} isn't cyclic, nothing prevents the user from adding communications between the various client processes, or from removing the scheduler and having the client processes communicate \emph{directly} in a ring.
%
%
%
%
%
%
%
%
%
%

\section{Relation to Priority GV}
\label{sec:pgv}
The \texttt{priority-sesh} library is based on a variant of Priority GV~\cite{kokkedardha21}, which differs in three ways:
\begin{enumerate}
\item
  it marks lower bounds \emph{explicitly} on the sequent, rather than implicitly inferring them from the typing environment;
\item
  it collapses the isomorphic types for session end, $\pgv{\tyends[o]}$ and $\pgv{\tyendr[o]}$, into $\pgv{\tyend[o]}$;
\item
  it is extended with asynchronous communication and session cancellation following \citet{fowlerlindley19}.
\end{enumerate}
These changes preserve subject reduction and progress properties, and give us \emph{tighter} bounds on priorities. To see why, note that PCP \cite{dardhagay18} and PGV \cite{kokkedardha21} use the \emph{smallest} priority in the typing environment as an approximation for the lower bound. Unfortunately, this \emph{underestimates} the lower bound in the rules \LabTirName{T-Var} and \LabTirName{T-Lam} (check \cref{fig:pgv-typing}). These rules type \emph{values}, which are pure and could have lower bound \ensuremath{\cs{\top}}, but the smallest priority in their typing environment is not necessarily \ensuremath{\cs{\top}}.

\paragraph{Priority GV}
We briefly revisit the syntax and type system of PGV, but a full discussion of PGV is out of scope for this paper. For a discussion of the \emph{synchronous} semantics for PGV, and the proofs of subject reduction, progress, and deadlock freedom, please see \citet{kokkedardha21}. For a discussion of the \emph{asynchronous} semantics and session cancellation, please see \citet{fowlerlindley19}.

As in \cref{sec:priority-sesh}, we let $\cs{o}$ range over priorities, which are natural numbers, and $\cs{p}$ and $\cs{q}$ over priority bounds, which are either natural numbers, $\cs{\top}$, or $\cs{\bot}$.

PGV is based on the standard linear $\lambda$-calculus with product types ($\pgv{\ty{\typrod{\cdot}{\cdot}}}$), sum types ($\pgv{\ty{\tysum{\cdot}{\cdot}}}$), and their units ($\pgv{\ty{\tyunit}}$ and $\pgv{\ty{\tyvoid}}$). Linear functions ($\pgv{\ty{\tylolli{p}{q}{\cdot}{\cdot}}}$) are annotated with priority bounds which tell us--when the function is applied--when communication begins and ends.

Types and session types are defined as follows:
\[
  \usingnamespace{pgv}
  \begin{array}{lcl}
    \ty{S}
    & \Coloneqq & \ty{\tysend[\cs{o}]{T}{S}}
      \mid        \ty{\tyrecv[\cs{o}]{T}{S}}
      \mid        \ty{\tyend[\cs{o}]}
    \\
    \ty{T}, \ty{U}
    & \Coloneqq & \ty{\typrod{T}{U}}
      \mid        \ty{\tyunit}
      \mid        \ty{\tysum{T}{U}}
      \mid        \ty{\tyvoid}
      \mid        \ty{\tylolli{\cs{p}}{\cs{q}}{T}{U}}
      \mid        \ty{S}
  \end{array}
\]
The types $\pgv{\ty{\tysend[\cs{o}]{T}{S}}}$ and $\pgv{\ty{\tyrecv[\cs{o}]{T}{S}}}$ mean ``send'' and ``receive'', respectively, and $\pgv{\ty{\tyend[\cs{o}]}}$ means, well, session end.

The term language is the standard linear $\lambda$-calculus extended with concurrency primitives $\pgv{\tm{K}}$:
\[
  \usingnamespace{pgv}
  \begin{array}{lcl}
    \multicolumn{3}{l}{\tm{L}, \tm{M}, \tm{N}}
    \\
    & \Coloneqq & \tm{x}
      \mid        \tm{K}
      \mid        \tm{\lambda x.M}
      \mid        \tm{M\;N} \\
    & \mid      & \tm{\unit}
      \mid        \tm{\andthen{M}{N}} \\
    & \mid      & \tm{\pair{M}{N}}
      \mid        \tm{\letpair{x}{y}{M}{N}} \\
    & \mid      & \tm{\absurd{M}} \\
    & \mid      & \tm{\inl{M}}
      \mid        \tm{\inr{M}}
      \mid        \tm{\casesum{L}{x}{M}{y}{N}}
    \\
    \tm{K}
    & \Coloneqq & \tm{\new}
      \mid        \tm{\fork}
      \mid        \tm{\send}
      \mid        \tm{\recv}
      \mid        \tm{\close}
  \end{array}
\]
The concurrency primitives are uninterpreted in the term language. Rather, they are interpreted in a configuration language based on the $\pi$-calculus, which we omit from this paper (see \citet{kokkedardha21}).

\begin{figure*}
\usingnamespace{pgv}
\header[${\seq{p}{q}{\ty{\Gamma}}{M}{T}}$]{Static Typing Rules}
\begin{mathpar}
  \inferrule*[lab=T-Var]{
  }{\seq{\top}{\bot}{\tmty{x}{T}}{x}{T}}

  \inferrule*[lab=T-Lam]{
    \seq{p}{q}{\ty{\Gamma},\tmty{x}{T}}{M}{U}
  }{\seq{\top}{\bot}{\ty{\Gamma}}{\lambda x.M}{\tylolli{p}{q}{T}{U}}}

  \inferrule*[lab=T-Const]{
  }{\seq{{\top}}{{\bot}}{\emptyenv}{K}{T}}

  \inferrule*[lab=T-App]{
    \seq{p}{q}{\ty{\Gamma}}{M}{\tylolli{p''}{q''}{T}{U}}
    \and
    \seq{p'}{q'}{\ty{\Delta}}{N}{T}
    \and
    \cs{q}<\cs{p'}
    \and
    \cs{q'}<\cs{p''}
  }{\seq
    {{p}\sqcap{p'}\sqcap{p''}}
    {{q}\sqcup{q'}\sqcup{q''}}
    {\ty{\Gamma},\ty{\Delta}}{M\;N}{U}}

  \inferrule*[lab=T-Unit]{
  }{\seq{{\top}}{{\bot}}{\emptyenv}{\unit}{\tyunit}}

  \inferrule*[lab=T-LetUnit]{
    \seq{p}{q}{\ty{\Gamma}}{M}{\tyunit}
    \and
    \seq{p'}{q'}{\ty{\Delta}}{N}{T}
    \and
    \cs{q}<\cs{p'}
  }{\seq{{p}\sqcap{p'}}{{q}\sqcup{q'}}{\ty{\Gamma},\ty{\Delta}}{\letunit{M}{N}}{T}}

  \inferrule*[lab=T-Pair]{
    \seq{p}{q}{\ty{\Gamma}}{M}{T}
    \and
    \seq{p'}{q'}{\ty{\Delta}}{N}{U}
    \and
    \cs{q}<\cs{p'}
  }{\seq{{p}\sqcap{p'}}{{q}\sqcup{q'}}{\ty{\Gamma},\ty{\Delta}}{\pair{M}{N}}{\typrod{T}{U}}}

  \inferrule*[lab=T-LetPair]{
    \seq{p}{q}{\ty{\Gamma}}{M}{\typrod{T}{T'}}
    \and
    \seq{p'}{q'}{\ty{\Delta},\tmty{x}{T},\tmty{y}{T'}}{N}{U}
    \and
    \cs{q}<\cs{p'}
  }{\seq{{p}\sqcap{p'}}{{q}\sqcup{q'}}{\ty{\Gamma},\ty{\Delta}}{\letpair{x}{y}{M}{N}}{U}}

  \inferrule*[lab=T-Inl]{
    \seq{p}{q}{\ty{\Gamma}}{M}{T}
  }{\seq{p}{q}{\ty{\Gamma}}{\inl{M}}{\tysum{T}{U}}}

  \inferrule*[lab=T-Inr]{
    \seq{p}{q}{\ty{\Gamma}}{M}{T}
  }{\seq{p}{q}{\ty{\Gamma}}{\inr{M}}{\tysum{T}{U}}}

  \inferrule*[lab=T-CaseSum]{
    \seq{p}{q}{\ty{\Gamma}}{L}{\tysum{T}{T'}}
    \and
    \seq{p'}{q'}{\ty{\Delta},\tmty{x}{T}}{M}{U}
    \and
    \seq{p'}{q'}{\ty{\Delta},\tmty{y}{T'}}{N}{U}
    \and
    \cs{q}<\cs{p'}
  }{\seq{{p}\sqcup{p'}}{{q}\sqcup{q'}}{\ty{\Gamma},\ty{\Delta}}{\casesum{L}{x}{M}{y}{N}}{U}}

  \inferrule*[lab=T-Absurd]{
    \seq{p}{q}{\ty{\Gamma}}{M}{\tyvoid}
  }{\seq{p}{q}{\ty{\Gamma}}{\absurd{M}}{T}}
\end{mathpar}
\header[${\tmty{K}{T}}$]{Type Schemas for Constants}
\begin{mathpar}
  \tmty{\new}{\tylolli{\top}{\bot}{\tyunit}{\typrod{S}{\co{S}}}}

  \tmty{\fork}{\tylolli{\top}{\bot}{(\tylolli{p}{q}{\tyunit}{\tyunit})}{\tyunit}}

  \tmty{\cancel}{\tylolli{\top}{\bot}{S}{\tyunit}}
  \\
  \tmty{\send}{\tylolli{o}{o}{\typrod{T}{\tysend[{o}]{T}{S}}}{S}}

  \tmty{\recv}{\tylolli{o}{o}{\tyrecv[{o}]{T}{S}}{\typrod{T}{S}}}

  \tmty{\close}{\tylolli{o}{o}{\tyend[{o}]}{\tyunit}}
\end{mathpar}
\caption{Typing rules for Priority GV.}
\label{fig:pgv-typing}
\end{figure*}

We present the typing rules for PGV in \cref{fig:pgv-typing}. A sequent $\pgv{\seq{p}{q}{\ty{\Gamma}}{M}{T}}$ should be read as ``$\pgv{\tm{M}}$ is well-typed PGV program with type $\pgv{\ty{T}}$ in typing environment $\pgv{\ty{\Gamma}}$, and when run it starts communicating at time $\pgv{\cs{p}}$ and stops at time $\pgv{\cs{q}}$.''

\paragraph{Monadic Reflection}
The graded monad \ensuremath{\seshtymon{\Varid{p}}{\Varid{q}}} arises from the \emph{monadic reflection}~\cite{filinski94} of the typing rules in \cref{fig:pgv-typing}. Monadic reflection is a technique for translating programs in an effectful language to \emph{monadic} programs in a pure language. For instance, \citet{filinski94} demonstrates the reflection from programs of type $\pgv{\ty{T}}$ in a language with exceptions and handlers to programs of type $\pgv{\ty{\tysum{T}{\mathbf{exn}}}}$ in a pure language where $\pgv{\ty{\mathbf{exn}}}$ is the type of exceptions.

We translate programs from PGV to Haskell programs in the \ensuremath{\seshtymon{\Varid{p}}{\Varid{q}}} monad. First, let's look at the translation of types:
\[
  \setlength{\arraycolsep}{2pt}
  \begin{array}{lcl}
    \pgv{\tosesh{\ty{\tylolli{p}{q}{T}{U}}}}
    &=
    &\ensuremath{\tosesh{\Conid{T}}\hsPercent \hsOne \hsArrow{\rightarrow }{\multimap }{\mathpunct{.}}\seshtymon{\Varid{p}}{\Varid{q}}\;\tosesh{\Conid{U}}}
    \\
    \pgv{\tosesh{\ty{\tysend[o]{T}{S}}}}
    &=
    &\ensuremath{\seshtysend[\cs{\Varid{o}}]\;\tosesh{\Conid{T}}\;\tosesh{\Conid{S}}}
    \\
    \pgv{\tosesh{\ty{\tyrecv[o]{T}{S}}}}
    &=
    &\ensuremath{\seshtyrecv[\cs{\Varid{o}}]\;\tosesh{\Conid{T}}\;\tosesh{\Conid{S}}}
    \\
    \pgv{\tosesh{\ty{\tyend[o]}}}
    &=
    &\ensuremath{\seshtyend[\cs{\Varid{o}}]}
  \end{array}
  \;
  \begin{array}{lcl}
    \pgv{\tosesh{\ty{\tyunit}}}
    &=
    &\ensuremath{()}
    \\
    \pgv{\tosesh{\ty{\typrod{T}{U}}}}
    &=
    &\ensuremath{(\tosesh{\Conid{T}},\tosesh{\Conid{U}})}
    \\
    \pgv{\tosesh{\ty{\tyvoid}}}
    &=
    &\ensuremath{\Conid{Void}}
    \\
    \pgv{\tosesh{\ty{\tysum{T}{U}}}}
    &=
    &\ensuremath{\Conid{Either}\;\tosesh{\Conid{T}}\;\tosesh{\Conid{U}}}
  \end{array}
\]
Now, let's look at the translation of terms. A term of type $\pgv{\ty{T}}$ with lower bound $\cs{p}$ and upper bound $\cs{q}$ is translated to a Haskell program of type \ensuremath{\seshtymon{\Varid{p}}{\Varid{q}}\;\tosesh{\Conid{T}}}:
\[
  \setlength{\arraycolsep}{2pt}
  \begin{array}{lcl}
    \pgv{\tosesh{\tm{x}}}
    &=
    &\ensuremath{\tm{\Varid{ireturn}}\;\Varid{x}}
    \\
    \pgv{\tosesh{\tm{\lambda x.L}}}
    &=
    &\ensuremath{\tm{\Varid{ireturn}}\;(\lambda \hsLambdaCmd \Varid{x}\hsArrow{\rightarrow }{\multimap }{\mathpunct{.}}\tosesh{\Conid{L}})}
    \\
    \pgv{\tosesh{\tm{K}}}
    &=
    &\ensuremath{\tm{\Varid{ireturn}}\;\tosesh{\Conid{K}}}
    \\
    \pgv{\tosesh{\tm{L\;M}}}
    &=
    &\ensuremath{\tosesh{\Conid{L}}\tm{\ibind}\lambda \hsLambdaCmd \Varid{f}\hsArrow{\rightarrow }{\multimap }{\mathpunct{.}}\tosesh{\Conid{M}}\bind \lambda \hsLambdaCmd \Varid{x}\hsArrow{\rightarrow }{\multimap }{\mathpunct{.}}\Varid{f}\;\Varid{x}}
    \\
    \pgv{\tosesh{\tm{\unit}}}
    &=
    &\ensuremath{\tm{\Varid{ireturn}}\;()}
    \\
    \pgv{\tosesh{\tm{\letunit{L}{M}}}}
    &=
    &\ensuremath{\tosesh{\Conid{L}}\tm{\ibind}\lambda \hsLambdaCmd ()\hsArrow{\rightarrow }{\multimap }{\mathpunct{.}}\Conid{M}}
    \\
    \pgv{\tosesh{\tm{\pair{L}{M}}}}
    &=
    &\ensuremath{\tosesh{\Conid{L}}\tm{\ibind}\lambda \hsLambdaCmd \Varid{x}\hsArrow{\rightarrow }{\multimap }{\mathpunct{.}}\tosesh{\Conid{M}}\bind \lambda \hsLambdaCmd \Varid{y}\hsArrow{\rightarrow }{\multimap }{\mathpunct{.}}\tm{\Varid{ireturn}}\;(\Varid{x},\Varid{y})}
    \\
    \pgv{\tosesh{\tm{\letpair{x}{y}{L}{M}}}}
    &=
    &\ensuremath{\tosesh{\Conid{L}}\tm{\ibind}\lambda \hsLambdaCmd (\Varid{x},\Varid{y})\hsArrow{\rightarrow }{\multimap }{\mathpunct{.}}\tosesh{\Conid{M}}}
    \\
    \pgv{\tosesh{\tm{\absurd{L}}}}
    &=
    &\ensuremath{\tosesh{\Conid{L}}\tm{\ibind}\lambda \hsLambdaCmd \Varid{x}\hsArrow{\rightarrow }{\multimap }{\mathpunct{.}}\Varid{absurd}\;\Varid{x}}
    \\
    \pgv{\tosesh{\tm{\inl{L}}}}
    &=
    &\ensuremath{\tosesh{\Conid{L}}\tm{\ibind}\lambda \hsLambdaCmd \Varid{x}\hsArrow{\rightarrow }{\multimap }{\mathpunct{.}}\tm{\Varid{ireturn}}\;(\Conid{Left}\;\Varid{x})}
    \\
    \pgv{\tosesh{\tm{\inr{L}}}}
    &=
    &\ensuremath{\tosesh{\Conid{L}}\tm{\ibind}\lambda \hsLambdaCmd \Varid{x}\hsArrow{\rightarrow }{\multimap }{\mathpunct{.}}\tm{\Varid{ireturn}}\;(\Conid{Right}\;\Varid{x})}
    \\
    \multicolumn{3}{l}{\pgv{\tosesh{\tm{\casesum{L}{x}{M}{y}{N}}}} =}
    \\
    \multicolumn{3}{r}{\ensuremath{\tosesh{\Conid{L}}\tm{\ibind}\lambda \hsLambdaCmd \Varid{x}\hsArrow{\rightarrow }{\multimap }{\mathpunct{.}}\mathbf{case}\;\Varid{x}\;\mathbf{of}\;\{\mskip1.5mu \Conid{Left}\;\Varid{x}\hsArrow{\rightarrow }{\multimap }{\mathpunct{.}}\tosesh{\Conid{M}};\Conid{Right}\;\Varid{y}\hsArrow{\rightarrow }{\multimap }{\mathpunct{.}}\tosesh{\Conid{N}}\mskip1.5mu\}}}
  \end{array}
\]
We translate the communication primitives from PGV to those with the same name in \texttt{priority-sesh}, with some minor changes in the translations of $\pgv{\tm{\new}}$ and $\pgv{\tm{\fork}}$, where PGV needs some unit arguments to create thunks in PGV, as it's call-by-value, which aren't needed in Haskell:
\[
  \begin{array}{l}
    \pgv{\tosesh{\tmty{\new}{\tylolli{\top}{\bot}{\tyunit}{\typrod{S}{\co{S}}}}}}
    \\
    \quad=\ensuremath{\lambda \hsLambdaCmd ()\hsArrow{\rightarrow }{\multimap }{\mathpunct{.}}\tm{\Varid{new}}\mathbin{::}()\hsPercent \hsOne \hsArrow{\rightarrow }{\multimap }{\mathpunct{.}}(\tosesh{\Conid{S}},\tosesh{(\ty{\Conid{Dual}}\;\Conid{S})})}
    \\
    \pgv{\tosesh{\tmty{\fork}{\tylolli{\top}{\bot}{(\tylolli{p}{q}{\tyunit}{\tyunit})}{\tyunit}}}}
    \\
    \quad=\ensuremath{\lambda \hsLambdaCmd \Varid{k}\hsArrow{\rightarrow }{\multimap }{\mathpunct{.}}\tm{\Varid{fork}}\;(\Varid{k}\;())\mathbin{::}(()\hsPercent \hsOne \hsArrow{\rightarrow }{\multimap }{\mathpunct{.}}\seshtymon{\Varid{p}}{\Varid{q}}\;())\hsPercent \hsOne \hsArrow{\rightarrow }{\multimap }{\mathpunct{.}}\seshtymon{\cs{\top}}{\cs{\bot}}\;()}
  \end{array}
\]
The rest of PGV's communication primitives line up exactly with those of \texttt{priority-sesh}:
\[
  \begin{array}{l}
    \pgv{\tosesh{\tmty{\send}{\tylolli{o}{o}{\typrod{T}{\tysend[{o}]{T}{S}}}{S}}}}
    \\
    \quad=\ensuremath{\tm{\Varid{send}}\mathbin{::}\ty{\Conid{Session}}\;\tosesh{\Conid{S}}\Rightarrow (\tosesh{\Conid{T}},\seshtysend[\cs{\Varid{o}}]\;\tosesh{\Conid{T}}\;\tosesh{\Conid{S}})\hsPercent \hsOne \hsArrow{\rightarrow }{\multimap }{\mathpunct{.}}\seshtymon{\Varid{o}}{\Varid{o}}\;\tosesh{\Conid{S}}}
    \\
    \pgv{\tosesh{\tmty{\recv}{\tylolli{o}{o}{\tyrecv[{o}]{T}{S}}{\typrod{T}{S}}}}}
    \\
    \quad=\ensuremath{\tm{\Varid{recv}}\mathbin{::}\seshtyrecv[\cs{\Varid{o}}]\;\tosesh{\Conid{T}}\;\tosesh{\Conid{S}}\hsPercent \hsOne \hsArrow{\rightarrow }{\multimap }{\mathpunct{.}}\seshtymon{\Varid{o}}{\Varid{o}}\;(\tosesh{\Conid{T}},\tosesh{\Conid{S}})}
    \\
    \pgv{\tosesh{\tmty{\close}{\tylolli{o}{o}{\tyend[{o}]}{\tyunit}}}}
    \\
    \quad=\ensuremath{\tm{\Varid{close}}\mathbin{::}\seshtyend[\cs{\Varid{o}}]\hsPercent \hsOne \hsArrow{\rightarrow }{\multimap }{\mathpunct{.}}\seshtymon{\Varid{o}}{\Varid{o}}\;()}
    \\
    \pgv{\tosesh{\tmty{\cancel}{\tylolli{\top}{\bot}{S}{\tyunit}}}}
    \\
    \quad=\ensuremath{\tm{\Varid{cancel}}\mathbin{::}\ty{\Conid{Session}}\;\tosesh{\Conid{S}}\Rightarrow \tosesh{\Conid{S}}\hsPercent \hsOne \hsArrow{\rightarrow }{\multimap }{\mathpunct{.}}\seshtymon{\cs{\top}}{\cs{\bot}}\;()}
  \end{array}
\]
These two translations, on types and terms, comprise a \emph{monadic reflection} from PGV into \texttt{priority-sesh}, which preserves typing. We state this theorem formally, using $\Gamma\vdash\ensuremath{\Varid{x}\mathbin{::}\Varid{a}}$ to mean that the Haskell program \ensuremath{\Varid{x}} has type \ensuremath{\Varid{a}} in typing environment \ensuremath{\Gamma}:
\begin{theorem}
  If $\pgv{\seq{p}{q}{\Gamma}{M}{T}}$, then $\ensuremath{\tosesh{\Gamma}}\vdash\ensuremath{\tosesh{\Conid{M}}\mathbin{::}\seshtymon{\Varid{p}}{\Varid{q}}\;\tosesh{\Conid{T}}}$.
\end{theorem}
\begin{proof}
  \Cref{fig:pgv-to-sesh-typing} presents the translation from typing derivations in PGV to abbreviated typing derivations in Haskell with \texttt{priority-sesh}.
\end{proof}

%
%
%
%
%
%
%
%
%
%
\begin{figure*}
\def\sep{\thickspace\thickspace\thickspace}%
\begin{mathpar}
  \inferrule*{
  }{\pgv{\seq{\top}{\bot}{\tmty{x}{T}}{x}{T}}}
  =
  \inferrule*{
    \ensuremath{\Varid{x}\mathbin{::}\tosesh{\Conid{T}}}\vdash\ensuremath{\Varid{x}\mathbin{::}\tosesh{\Conid{T}}}
  }{\ensuremath{\tm{\Varid{ireturn}}\;\Varid{x}\mathbin{::}\seshtymon{\cs{\top}}{\cs{\bot}}\;\tosesh{\Conid{T}}}}
  \\

  \inferrule*{
    \pgv{\seq{p}{q}{\ty{\Gamma},\tmty{x}{T}}{L}{U}}
  }{\pgv{\seq{\top}{\bot}{\ty{\Gamma}}{\lambda x.L}{\tylolli{p}{q}{T}{U}}}}
  =
  \inferrule*{
    \ensuremath{\tosesh{\Gamma},\Varid{x}\mathbin{::}\tosesh{\Conid{T}}}\vdash\ensuremath{\tosesh{\Conid{L}}\mathbin{::}\seshtymon{\Varid{p}}{\Varid{q}}\;\tosesh{\Conid{U}}}
  }{\ensuremath{\tm{\Varid{ireturn}}\;(\lambda \hsLambdaCmd \Varid{x}\hsArrow{\rightarrow }{\multimap }{\mathpunct{.}}\tosesh{\Conid{L}})\mathbin{::}\seshtymon{\cs{\top}}{\cs{\bot}}\;(\tosesh{\Conid{T}}\hsPercent \hsOne \hsArrow{\rightarrow }{\multimap }{\mathpunct{.}}\seshtymon{\Varid{p}}{\Varid{q}}\;\tosesh{\Conid{U}})}}
  \\

  \inferrule*{
  }{\pgv{\seq{{\top}}{{\bot}}{\emptyenv}{K}{T}}}
  =
  \inferrule*{
  }{\ensuremath{\tm{\Varid{ireturn}}\;\tosesh{\Conid{K}}\mathbin{::}\seshtymon{\cs{\top}}{\cs{\bot}}\;\tosesh{\Conid{T}}}}
  \\

  \inferrule*{
    \pgv{\seq{p}{q}{\ty{\Gamma}}{L}{\tylolli{p''}{q''}{T}{U}}}
    \sep
    \pgv{\seq{p'}{q'}{\ty{\Delta}}{M}{T}}
    \sep
    \pgv{\cs{q}<\cs{p'}}
    \sep
    \pgv{\cs{q'}<\cs{p''}}
  }{\pgv{\seq{{p}\sqcap{p'}\sqcap{p''}}{{q}\sqcup{q'}\sqcup{q''}}{\ty{\Gamma},\ty{\Delta}}{L\;M}{U}}}
  =
  \inferrule*{
    \ensuremath{\tosesh{\Gamma}}\vdash\ensuremath{\tosesh{\Conid{L}}\mathbin{::}\seshtymon{\Varid{p}}{\Varid{q}}\;(\tosesh{\Conid{T}}\hsPercent \hsOne \hsArrow{\rightarrow }{\multimap }{\mathpunct{.}}\seshtymon{\Varid{p''}}{\Varid{q''}}\;\tosesh{\Conid{U}})}
    \sep
    \ensuremath{\tosesh{\Delta}}\vdash\ensuremath{\tosesh{\Conid{M}}\mathbin{::}\seshtymon{\Varid{p'}}{\Varid{q'}}\;\tosesh{\Conid{T}}}
  }{\ensuremath{\tosesh{\Conid{L}}\tm{\ibind}\lambda \hsLambdaCmd \Varid{f}\hsArrow{\rightarrow }{\multimap }{\mathpunct{.}}\tosesh{\Conid{M}}\bind \lambda \hsLambdaCmd \Varid{x}\hsArrow{\rightarrow }{\multimap }{\mathpunct{.}}\Varid{f}\;\Varid{x}\mathbin{::}(\cs{\Varid{q}}\mathrel{\cs{<}}\cs{\Varid{p'}},\cs{\Varid{q'}}\mathrel{\cs{<}}\cs{\Varid{p''}})\Rightarrow \seshtymon{\Varid{p}\mathbin{\cs{\sqcap}}\Varid{p'}\mathbin{\cs{\sqcap}}\Varid{p''}}{\Varid{q}\mathbin{\cs{\sqcup}}\Varid{q'}\mathbin{\cs{\sqcup}}\Varid{q''}}\;\tosesh{\Conid{U}}}}
  \\

  \inferrule*{
  }{\pgv{\seq{{\top}}{{\bot}}{\emptyenv}{\unit}{\tyunit}}}
  =
  \inferrule*{
  }{\ensuremath{\tm{\Varid{ireturn}}\;()\mathbin{::}\seshtymon{\cs{\top}}{\cs{\bot}}\;()}}
  \\

  \inferrule*{
    \pgv{\seq{p}{q}{\ty{\Gamma}}{L}{\tyunit}}
    \sep
    \pgv{\seq{p'}{q'}{\ty{\Delta}}{M}{T}}
    \sep
    \pgv{\cs{q}<\cs{p'}}
  }{\pgv{\seq{{p}\sqcap{p'}}{{q}\sqcup{q'}}{\ty{\Gamma},\ty{\Delta}}{\letunit{L}{M}}{T}}}
  =
  \inferrule*{
    \ensuremath{\tosesh{\Gamma}}\vdash\ensuremath{\tosesh{\Conid{L}}\mathbin{::}\seshtymon{\Varid{p}}{\Varid{q}}\;()}
    \sep
    \ensuremath{\tosesh{\Delta}}\vdash\ensuremath{\tosesh{\Conid{M}}\mathbin{::}\seshtymon{\Varid{p'}}{\Varid{q'}}\;\tosesh{\Conid{T}}}
  }{\ensuremath{\tosesh{\Conid{L}}\tm{\ibind}\lambda \hsLambdaCmd ()\hsArrow{\rightarrow }{\multimap }{\mathpunct{.}}\Conid{M}\mathbin{::}(\cs{\Varid{p}}\mathrel{\cs{<}}\cs{\Varid{q'}})\Rightarrow \seshtymon{\cs{\Varid{p}}\mathbin{\cs{\sqcap}}\cs{\Varid{p'}}}{\cs{\Varid{q}}\mathbin{\cs{\sqcup}}\cs{\Varid{q'}}}\;\tosesh{\Conid{T}}}}
  \\

  \inferrule*{
    \pgv{\seq{p}{q}{\ty{\Gamma}}{L}{T}}
    \sep
    \pgv{\seq{p'}{q'}{\ty{\Delta}}{M}{U}}
    \sep
    \pgv{\cs{q}<\cs{p'}}
  }{\pgv{\seq{{p}\sqcap{p'}}{{q}\sqcup{q'}}{\ty{\Gamma},\ty{\Delta}}{\pair{L}{M}}{\typrod{T}{U}}}}
  =
  \inferrule*{
    \ensuremath{\tosesh{\Gamma}}\vdash\ensuremath{\tosesh{\Conid{L}}\mathbin{::}\seshtymon{\Varid{p}}{\Varid{q}}\;\tosesh{\Conid{T}}}
    \sep
    \ensuremath{\tosesh{\Delta}}\vdash\ensuremath{\tosesh{\Conid{M}}\mathbin{::}\seshtymon{\Varid{p'}}{\Varid{q'}}\;\tosesh{\Conid{U}}}
  }{\ensuremath{\tosesh{\Conid{L}}\tm{\ibind}\lambda \hsLambdaCmd \Varid{x}\hsArrow{\rightarrow }{\multimap }{\mathpunct{.}}\tosesh{\Conid{M}}\bind \lambda \hsLambdaCmd \Varid{y}\hsArrow{\rightarrow }{\multimap }{\mathpunct{.}}\tm{\Varid{ireturn}}\;(\Varid{x},\Varid{y})\mathbin{::}(\cs{\Varid{q}}\mathrel{\cs{<}}\cs{\Varid{p'}})\Rightarrow \seshtymon{\cs{\Varid{p}}\mathbin{\cs{\sqcap}}\cs{\Varid{p'}}}{\cs{\Varid{q}}\mathbin{\cs{\sqcup}}\cs{\Varid{q'}}}\;(\tosesh{\Conid{T}},\tosesh{\Conid{U}})}}
  \\

  \inferrule*{
    \pgv{\seq{p}{q}{\ty{\Gamma}}{L}{\typrod{T}{T'}}}
    \sep
    \pgv{\seq{p'}{q'}{\ty{\Delta},\tmty{x}{T},\tmty{y}{T'}}{M}{U}}
    \sep
    \pgv{\cs{q}<\cs{p'}}
  }{\pgv{\seq{{p}\sqcap{p'}}{{q}\sqcup{q'}}{\ty{\Gamma},\ty{\Delta}}{\letpair{x}{y}{L}{M}}{U}}}
  =
  \inferrule*{
    \ensuremath{\tosesh{\Gamma}}\vdash\ensuremath{\tosesh{\Conid{L}}\mathbin{::}\seshtymon{\Varid{p}}{\Varid{q}}\;(\tosesh{\Conid{T}},\tosesh{\Conid{T'}})}
    \sep
    \ensuremath{\tosesh{\Delta},\Varid{x}\mathbin{::}\tosesh{\Conid{T}},\Varid{y}\mathbin{::}\tosesh{\Conid{T'}}}\vdash\ensuremath{\tosesh{\Conid{M}}\mathbin{::}\seshtymon{\Varid{p'}}{\Varid{q'}}\;\tosesh{\Conid{U}}}
  }{\ensuremath{\tosesh{\Conid{L}}\tm{\ibind}\lambda \hsLambdaCmd (\Varid{x},\Varid{y})\hsArrow{\rightarrow }{\multimap }{\mathpunct{.}}\tosesh{\Conid{M}}\mathbin{::}(\cs{\Varid{q}}\mathrel{\cs{<}}\cs{\Varid{p'}})\Rightarrow \seshtymon{\cs{\Varid{p}}\mathbin{\cs{\sqcap}}\cs{\Varid{p'}}}{\cs{\Varid{q}}\mathbin{\cs{\sqcup}}\cs{\Varid{q'}}}\;\tosesh{\Conid{U}}}}
  \\

  \inferrule*{
    \pgv{\seq{p}{q}{\ty{\Gamma}}{L}{T}}
  }{\pgv{\seq{p}{q}{\ty{\Gamma}}{\inl{L}}{\tysum{T}{U}}}}
  =
  \inferrule*{
    \ensuremath{\tosesh{\Gamma}}\vdash\ensuremath{\tosesh{\Conid{L}}\mathbin{::}\seshtymon{\Varid{p}}{\Varid{q}}\;\tosesh{\Conid{T}}}
  }{\ensuremath{\tosesh{\Gamma}}\vdash\ensuremath{\tosesh{\Conid{L}}\tm{\ibind}\lambda \hsLambdaCmd \Varid{x}\hsArrow{\rightarrow }{\multimap }{\mathpunct{.}}\tm{\Varid{ireturn}}\;(\Conid{Left}\;\Varid{x})\mathbin{::}\seshtymon{\Varid{p}}{\Varid{q}}\;(\Conid{Either}\;\tosesh{\Conid{T}}\;\tosesh{\Conid{U}})}}
  \\

  \inferrule*{
    \pgv{\seq{p}{q}{\ty{\Gamma}}{L}{T}}
  }{\pgv{\seq{p}{q}{\ty{\Gamma}}{\inr{L}}{\tysum{T}{U}}}}
  =
  \inferrule*{
    \ensuremath{\tosesh{\Gamma}}\vdash\ensuremath{\tosesh{\Conid{L}}\mathbin{::}\seshtymon{\Varid{p}}{\Varid{q}}\;\tosesh{\Conid{U}}}
  }{\ensuremath{\tosesh{\Gamma}}\vdash\ensuremath{\tosesh{\Conid{L}}\tm{\ibind}\lambda \hsLambdaCmd \Varid{x}\hsArrow{\rightarrow }{\multimap }{\mathpunct{.}}\tm{\Varid{ireturn}}\;(\Conid{Right}\;\Varid{x})\mathbin{::}\seshtymon{\Varid{p}}{\Varid{q}}\;(\Conid{Either}\;\tosesh{\Conid{T}}\;\tosesh{\Conid{U}})}}
  \\

  \inferrule*{
    \pgv{\seq{p}{q}{\ty{\Gamma}}{L}{\tysum{T}{T'}}}
    \sep
    \pgv{\seq{p'}{q'}{\ty{\Delta},\tmty{x}{T}}{M}{U}}
    \sep
    \pgv{\seq{p'}{q'}{\ty{\Delta},\tmty{y}{T'}}{N}{U}}
    \sep
    \pgv{\cs{q}<\cs{p'}}
  }{\pgv{\seq{{p}\sqcup{p'}}{{q}\sqcup{q'}}{\ty{\Gamma},\ty{\Delta}}{\casesum{L}{x}{M}{y}{N}}{U}}}
  =
  \inferrule*{
    \ensuremath{\tosesh{\Gamma}}\vdash\ensuremath{\tosesh{\Conid{L}}\mathbin{::}\seshtymon{\Varid{p}}{\Varid{q}}\;(\Conid{Either}\;\tosesh{\Conid{T}}\;\tosesh{\Conid{T'}})}
    \sep
    \ensuremath{\tosesh{\Delta}}\vdash\ensuremath{\Varid{x}\mathbin{::}\tosesh{\Conid{T}}}\vdash\ensuremath{\tosesh{\Conid{M}}\mathbin{::}\seshtymon{\Varid{p'}}{\Varid{q'}}\;\tosesh{\Conid{U}}}
    \sep
    \ensuremath{\tosesh{\Delta}}\vdash\ensuremath{\Varid{y}\mathbin{::}\tosesh{\Conid{T'}}}\vdash\ensuremath{\tosesh{\Conid{N}}\mathbin{::}\seshtymon{\Varid{p'}}{\Varid{q'}}\;\tosesh{\Conid{U}}}
  }{\ensuremath{\tosesh{\Gamma},\tosesh{\Delta}}\vdash\ensuremath{\tosesh{\Conid{L}}\tm{\ibind}\lambda \hsLambdaCmd \Varid{x}\hsArrow{\rightarrow }{\multimap }{\mathpunct{.}}\mathbf{case}\;\Varid{x}\;\{\mskip1.5mu \Conid{Left}\;\Varid{x}\hsArrow{\rightarrow }{\multimap }{\mathpunct{.}}\tosesh{\Conid{M}};\Conid{Right}\;\Varid{y}\hsArrow{\rightarrow }{\multimap }{\mathpunct{.}}\tosesh{\Conid{N}}\mskip1.5mu\}\mathbin{::}\seshtymon{\cs{\Varid{p}}\mathbin{\cs{\sqcap}}\cs{\Varid{p'}}}{\cs{\Varid{q}}\mathbin{\cs{\sqcup}}\cs{\Varid{q'}}}\;\tosesh{\Conid{U}}}}
  \\

  \inferrule*{
    \pgv{\seq{p}{q}{\ty{\Gamma}}{L}{\tyvoid}}
  }{\pgv{\seq{p}{q}{\ty{\Gamma}}{\absurd{L}}{T}}}
  =
  \inferrule*{
    \ensuremath{\tosesh{\Gamma}}\vdash\ensuremath{\tosesh{\Conid{L}}\mathbin{::}\seshtymon{\Varid{p}}{\Varid{q}}\;\Conid{Void}}
  }{\ensuremath{\tosesh{\Gamma}}\vdash\ensuremath{\tosesh{\Conid{L}}\tm{\ibind}\lambda \hsLambdaCmd \Varid{x}\hsArrow{\rightarrow }{\multimap }{\mathpunct{.}}\Varid{absurd}\;\Varid{x}\mathbin{::}\seshtymon{\Varid{p}}{\Varid{q}}\;\tosesh{\Conid{T}}}}
\end{mathpar}
\caption{Translation from Priority GV to Sesh preserves types.}
\label{fig:pgv-to-sesh-typing}
\end{figure*}
%
%
%
%
%
%
%
%
%
%

\section{Related work}\label{sec:related}
\paragraph{Session types in Haskell}
\begin{figure*}
  \centering
  \begin{tabular}{l | l l l l l l | l l l}
      \hline
    &
    &
    &
    &
    &
    &
    & \multicolumn{3}{c}{\texttt{priority-sesh}}
    \\
    & NT04
    & PT08
    & SE08
    & IYA10
    & OY16
    & LM16
    & \cref{sec:sesh}
    & \cref{sec:tree-sesh}
    & \cref{sec:priority-sesh}
    \\ \hline
    Recursion
    & \deffo
    & \deffo
    & \deffo
    & \deffo
    & \deffo
    &
    & \deffo
    & \deffo
    &
    \\
    Delegation
    &
    & \kinda
    & \deffo
    & \deffo
    & \deffo
    & \deffo
    & \deffo
    & \deffo
    & \deffo
    \\
    Multiple channels
    &
    & \deffo
    & \deffo
    & \deffo
    & \deffo
    & \deffo
    & \deffo
    & \deffo
    & \deffo
    \\
    Idiomatic code
    & \kinda
    & \kinda
    &
    & \deffo
    & \deffo
    & \kinda
    & \deffo
    & \deffo
    & \deffo
    \\
    Easy-to-write session types
    & \deffo
    & \deffo
    & \deffo
    &
    & \deffo
    & \deffo
    & \deffo
    & \deffo
    & \deffo
    \\
    Deadlock freedom
    & \kinda
    &
    &
    &
    &
    & \kinda
    &
    & \kinda
    & \deffo
    \\
    \emph{via process structure}
    & \kinda
    &
    &
    &
    &
    & \deffo
    &
    & \deffo
    &
    \\
    \emph{via priorities}
    &
    &
    &
    &
    &
    &
    &
    &
    & \deffo
  \end{tabular}
  \caption{Capabilities of various implementations of session types in Haskell~\cite[adapted from][]{orchardyoshida17}.}
  \label{fig:table}
\end{figure*}

\citet{orchardyoshida17} discuss various approaches to implementing session types in Haskell. Their overview is reproduced below:
\begin{itemize}
\item
  \citet{neubauerthiemann04} give an encoding of first-order single-channel session-types with recursion;
\item
  Using \emph{parameterised monads}, \citet{pucellatov08} provide multiple channels, recursion, and some building blocks for delegation, but require manual manipulation of a session typing environment;
\item
  \citet{sackmaneisenbach08} provide an alternate approach where session types are constructed via a value-level witnesses;
\item
  \citet{imaiyuen10} extend \citet{pucellatov08} with delegation and a more user-friendly approach to handling multiple channels;
\item
  \citet{orchardyoshida16} use an embedding of effect systems into Haskell via graded monads based on a formal encoding of session-typed \textpi-calculus into PCF with an effect system;
\item
  \citet{lindleymorris16} provide a \emph{finally tagless} embedding of the GV session-typed functional calculus into Haskell, building on a linear \textlambda-calculus embedding due to \citet{polakow15}.
\end{itemize}
With respect to linearity, all works above---except \citet{neubauerthiemann04}---guarantee linearity by encoding a linear typing environment in the Haskell type system, which leads to a trade-off between having easy-to-write session types and having idiomatic programs.
We side-step this trade-off by relying on Linear Haskell to check linearity. Furthermore, our implementation supports all relevant features, including multiple channels, full delegation, recursion, and highly idiomatic code.

With respect to deadlock freedom, none of the works above---except \citet{lindleymorris16}---guarantee deadlock freedom. However, \citet{lindleymorris16} guarantee deadlock freedom \emph{structurally}, by implementing GV. As discussed in~\cref{sec:introduction}, structure-based deadlock freedom is more restrictive than priority-based deadlock freedom, as it restricts communication graphs to \emph{trees}, whereas the priority-based approach allows programs to have \emph{cyclic} process structures.

\citet{orchardyoshida17} summarise the capabilities of the various implementations of session types in Haskell in a table, which we adapted in \cref{fig:table} by adding columns for the various versions of \texttt{priority-sesh}. In general, you may read \kinda as ``Kinda'' and \deffo as a resounding ``Yes!'' For instance, \citet{pucellatov08} only provide \emph{partial} delegation, \citet{neubauerthiemann04}, \citet{pucellatov08}, and \citet{lindleymorris16} still need to use combinators instead of standard Haskell application, abstraction, or variables in \emph{some} places, and \citet{neubauerthiemann04} is only deadlock free on the technicality that they don't support multiple channels.

\paragraph{Session types in other programming languages}
Session types have been integrated in other programming language paradigms.
\citet{JML15,ScalasY16,PadFuse} integrate \emph{binary} session types in the \emph{native} host language, without language extensions; this to avoid hindering session types use in practice.
To obtain this integration of session types without extensions \citet{ScalasY16,PadFuse}) combine \emph{static} typing of input and output actions with \emph{runtime} checking of linearity of channel usage.

Implementations of \emph{multiparty} session types (MPST) are less common than binary implementations.
\citet{Scalas2017} integrate MPST in Scala building upon \citet{ScalasY16} and a continuation-passing style encoding of session types into linear types \citet{dardhagiachino12}.
There are several works on implementations of MPST in Java:
\citet{SivaramakrishnanNZE10} implement MPST leveraging an extension of Java with session primitives;
\citet{HY16} develops a MPST-based API generation for Java leveraging CFSMs by \citet{Brand1983CFM}; and \citet{KDPG16} implement session types in the form of \emph{typestates} in Java.
\citet{DHHNY2015} implement MPST in Python and \citet{Fowler16,NY2017} in Erlang, focusing on {purely dynamic} MPST verification via runtime monitoring.
\citet{NY2017A,NBY2017} extend the work by
\citet{DHHNY2015} with actors and timed specifications.
\citet{LMMNSVY2015} adopt a dependently-typed MPST theory to verify MPI programs.

\paragraph{Session types, linear logic and deadlock freedom}
The main line of work regarding deadlock freedom in session-typed systems is that of Curry-Howard correspondences with linear logic \cite{girard87}.
\citet{CP10} defined a correspondence between session types and dual intuitionistic linear logic and \citet{wadler14} between session types and classical linear logic.
These works guarantee deadlock freedom \emph{by design} as the communication structures are restricted to trees and due to the \emph{cut} rule, processes share \emph{only} one channel between them.
\citet{dardhagay18} extend \citet{wadler14} with \emph{priorities} following \citet{kobayashi06,padovani14}, thus allowing processes to share more than one channel in parallel, while guaranteeing deadlock freedom.
\citet{balzertoninho19} introduce \emph{sharing} and guarantee deadlock freedom via priorities.
All the above works deal with deadlock freedom in a session-typed $\pi$-calculus.
With regards to function languages, the original works on GV \cite{gayvasconcelos10,gayvasconcelos12} did not guarantee deadlock freedom. This was later addressed by \citet{lindleymorris15,wadler15} via syntactic restrictions where communication once again follows a tree structure. \citet{kokkedardha21} introduce PGV--Priority GV, by following \citet{dardhagay18} and allowing for more flexible programming in GV.

Other works on deadlock freedom in session-typed systems include the works by \citet{dezani-ciancaglinimostrous06}, where deadlock freedom is guaranteed by allowing only one active session at a time and by \citet{dezani-ciancagliniliguoro09progress}, where priorities are used for correct interleaving of channels.
\citet{hondayoshida08} guarantee deadlock freedom \emph{within a single} session of MPST, but not for session interleaving.
\citet{kokke19} guarantees deadlock freedom of session types in Rust by enforcing a tree structure of communication actions.

\section{Discussion and future work}
We presented \texttt{priority-sesh}, an implementation of deadlock-free session types in Linear Haskell. Using Linear Haskell allows us to check linearity---or more accurately, have linearity guaranteed for us---without relying on complex type-level machinery. Consequently, we have easy-to-write session types and idiomatic code---in fact, probably \emph{the most} idiomatic code when compared with previous work, though in fairness, all previous work predates Linear Haskell. Unfortunately, there are some drawbacks to using Linear Haskell. Most importantly, Linear Haskell is not very mature at this stage. For instance:
\begin{itemize}
\item
  Anonymous functions are assumed to be unrestricted rather than linear, meaning anonymous functions must be factored out into a let-binding or where-clause with \emph{at least} a minimal type signature such as \ensuremath{\anonymous \hsPercent \hsOne \hsArrow{\rightarrow }{\multimap }{\mathpunct{.}}\anonymous }.
\item
  There is no integration with \texttt{base} or popular Haskell packages, and given that \texttt{LinearTypes} is an extension, there likely won't be for quite a while. There's \texttt{linear-base}, which provides linear variants of many of the constructs in \texttt{base}. However, \texttt{linear-base} relies heavily on \text{\ttfamily unsafeCoerce}, which, \emph{ironically}, may affect Haskell's performance.
\item
  Generally, there is little integration with the Haskell ecosystem, \eg, one other contribution we made are the formatting directives for Linear Haskell in lhs2\TeX\footnote{\url{https://hackage.haskell.org/package/lhs2tex}}.
\end{itemize}
However, we believe that many of these drawbacks will disappear as the Linear Haskell ecosystem matures.

Our work also provides a library which guarantees deadlock freedom via \emph{priorities}, which allows for more flexible typing than previous work on deadlock freedom via a \emph{tree process structure}.

In the future, we plan to address the issue of priority-polymorphic code and recursion session types in our implementation. (While the versions of our library in~\cref{sec:sesh,sec:tree-sesh} support recursion, that is not yet the case for the priority-based version in~\cref{sec:priority-sesh}.) This is a challenging task, as it requires complex reasoning about type-level naturals. We outlined various approaches in \cref{sec:priority-sesh}. However, an alternative we would like to investigate, would be to implement \texttt{priority-sesh} in Idris2~\cite{brady13,brady17}, which supports \emph{both} linear types \emph{and} complex type-level reasoning.

\begin{acks}
  We thank Simon Fowler and April Gon\c{c}alves for comments on the manuscript.
\end{acks}

\bibliographystyle{ACM-Reference-Format}
\bibliography{main}


\begin{thebibliography}{61}


\ifx \showCODEN    \undefined \def \showCODEN     #1{\unskip}     \fi
\ifx \showDOI      \undefined \def \showDOI       #1{#1}\fi
\ifx \showISBNx    \undefined \def \showISBNx     #1{\unskip}     \fi
\ifx \showISBNxiii \undefined \def \showISBNxiii  #1{\unskip}     \fi
\ifx \showISSN     \undefined \def \showISSN      #1{\unskip}     \fi
\ifx \showLCCN     \undefined \def \showLCCN      #1{\unskip}     \fi
\ifx \shownote     \undefined \def \shownote      #1{#1}          \fi
\ifx \showarticletitle \undefined \def \showarticletitle #1{#1}   \fi
\ifx \showURL      \undefined \def \showURL       {\relax}        \fi
\providecommand\bibfield[2]{#2}
\providecommand\bibinfo[2]{#2}
\providecommand\natexlab[1]{#1}
\providecommand\showeprint[2][]{arXiv:#2}

\bibitem[\protect\citeauthoryear{Balzer, Toninho, and Pfenning}{Balzer
  et~al\mbox{.}}{2019}]%
        {balzertoninho19}
\bibfield{author}{\bibinfo{person}{Stephanie Balzer}, \bibinfo{person}{Bernardo
  Toninho}, {and} \bibinfo{person}{Frank Pfenning}.}
  \bibinfo{year}{2019}\natexlab{}.
\newblock \showarticletitle{Manifest Deadlock-Freedom for Shared Session
  Types}. In \bibinfo{booktitle}{\emph{Proc.\ of {ESOP}}}
  \emph{(\bibinfo{series}{Lecture Notes in Computer Science},
  Vol.~\bibinfo{volume}{11423})}. \bibinfo{publisher}{Springer},
  \bibinfo{pages}{611--639}.
\newblock


\bibitem[\protect\citeauthoryear{Bernardi, Dardha, Gay, and Kouzapas}{Bernardi
  et~al\mbox{.}}{2014}]%
        {bernardidardha14}
\bibfield{author}{\bibinfo{person}{Giovanni Bernardi}, \bibinfo{person}{Ornela
  Dardha}, \bibinfo{person}{Simon~J. Gay}, {and} \bibinfo{person}{Dimitrios
  Kouzapas}.} \bibinfo{year}{2014}\natexlab{}.
\newblock \showarticletitle{On Duality Relations for Session Types}.
\newblock In \bibinfo{booktitle}{\emph{Trustworthy Global Computing}}.
  \bibinfo{publisher}{Springer Berlin Heidelberg}, \bibinfo{pages}{51--66}.
\newblock
\urldef\tempurl%
\url{https://doi.org/10.1007/978-3-662-45917-1_4}
\showDOI{\tempurl}


\bibitem[\protect\citeauthoryear{Bernardy, Boespflug, Newton, Jones, and
  Spiwack}{Bernardy et~al\mbox{.}}{2018}]%
        {bernardyboespflug18}
\bibfield{author}{\bibinfo{person}{Jean-Philippe Bernardy},
  \bibinfo{person}{Mathieu Boespflug}, \bibinfo{person}{Ryan~R. Newton},
  \bibinfo{person}{Simon~Peyton Jones}, {and} \bibinfo{person}{Arnaud
  Spiwack}.} \bibinfo{year}{2018}\natexlab{}.
\newblock \showarticletitle{Linear Haskell: practical linearity in a
  higher-order polymorphic language}.
\newblock \bibinfo{journal}{\emph{Proc.\ of {POPL}}}  \bibinfo{volume}{2}
  (\bibinfo{year}{2018}), \bibinfo{pages}{1--29}.
\newblock
\urldef\tempurl%
\url{https://doi.org/10.1145/3158093}
\showDOI{\tempurl}


\bibitem[\protect\citeauthoryear{Brady}{Brady}{2013}]%
        {brady13}
\bibfield{author}{\bibinfo{person}{Edwin Brady}.}
  \bibinfo{year}{2013}\natexlab{}.
\newblock \showarticletitle{Idris, a general-purpose dependently typed
  programming language: Design and implementation}.
\newblock \bibinfo{journal}{\emph{Journal of Functional Programming}}
  \bibinfo{volume}{23}, \bibinfo{number}{5} (\bibinfo{year}{2013}),
  \bibinfo{pages}{552–593}.
\newblock
\urldef\tempurl%
\url{https://doi.org/10.1017/S095679681300018X}
\showDOI{\tempurl}


\bibitem[\protect\citeauthoryear{Brady}{Brady}{2017}]%
        {brady17}
\bibfield{author}{\bibinfo{person}{Edwin Brady}.}
  \bibinfo{year}{2017}\natexlab{}.
\newblock \showarticletitle{{TYPE}-{DRIVEN} {DEVELOPMENT} {OF} {CONCURRENT}
  {COMMUNICATING} {SYSTEMS}}.
\newblock \bibinfo{journal}{\emph{Computer Science}} \bibinfo{volume}{18},
  \bibinfo{number}{3} (\bibinfo{year}{2017}), \bibinfo{pages}{219}.
\newblock
\urldef\tempurl%
\url{https://doi.org/10.7494/csci.2017.18.3.1413}
\showDOI{\tempurl}


\bibitem[\protect\citeauthoryear{Brand and Zafiropulo}{Brand and
  Zafiropulo}{1983}]%
        {Brand1983CFM}
\bibfield{author}{\bibinfo{person}{Daniel Brand} {and} \bibinfo{person}{Pitro
  Zafiropulo}.} \bibinfo{year}{1983}\natexlab{}.
\newblock \showarticletitle{On Communicating Finite-State Machines}.
\newblock \bibinfo{journal}{\emph{J. ACM}} \bibinfo{volume}{30},
  \bibinfo{number}{2} (\bibinfo{date}{April} \bibinfo{year}{1983}),
  \bibinfo{pages}{323--342}.
\newblock
\urldef\tempurl%
\url{https://doi.org/10.1145/322374.322380}
\showDOI{\tempurl}


\bibitem[\protect\citeauthoryear{Caires and Pfenning}{Caires and
  Pfenning}{2010}]%
        {CP10}
\bibfield{author}{\bibinfo{person}{Lu\'{\i}s Caires} {and}
  \bibinfo{person}{Frank Pfenning}.} \bibinfo{year}{2010}\natexlab{}.
\newblock \showarticletitle{Session Types as Intuitionistic Linear
  Propositions}. In \bibinfo{booktitle}{\emph{CONCUR}}
  \emph{(\bibinfo{series}{LNCS}, Vol.~\bibinfo{volume}{6269})}.
  \bibinfo{publisher}{Springer}, \bibinfo{pages}{222--236}.
\newblock
\urldef\tempurl%
\url{https://doi.org/10.1007/978-3-642-15375-4_16}
\showDOI{\tempurl}


\bibitem[\protect\citeauthoryear{Dardha and Gay}{Dardha and Gay}{2018}]%
        {dardhagay18}
\bibfield{author}{\bibinfo{person}{Ornela Dardha} {and}
  \bibinfo{person}{Simon~J. Gay}.} \bibinfo{year}{2018}\natexlab{}.
\newblock \showarticletitle{A New Linear Logic for Deadlock-Free Session-Typed
  Processes}. In \bibinfo{booktitle}{\emph{Proc.\ of {FoSSaCS}}}
  \emph{(\bibinfo{series}{LNCS}, Vol.~\bibinfo{volume}{10803})}.
  \bibinfo{publisher}{Springer}, \bibinfo{pages}{91--109}.
\newblock


\bibitem[\protect\citeauthoryear{Dardha, Giachino, and Sangiorgi}{Dardha
  et~al\mbox{.}}{2012}]%
        {dardhagiachino12}
\bibfield{author}{\bibinfo{person}{Ornela Dardha}, \bibinfo{person}{Elena
  Giachino}, {and} \bibinfo{person}{Davide Sangiorgi}.}
  \bibinfo{year}{2012}\natexlab{}.
\newblock \showarticletitle{Session types revisited}. In
  \bibinfo{booktitle}{\emph{Proc.\ of {PPDP}}}. \bibinfo{publisher}{ACM},
  \bibinfo{pages}{139--150}.
\newblock


\bibitem[\protect\citeauthoryear{Dardha, Giachino, and Sangiorgi}{Dardha
  et~al\mbox{.}}{2017}]%
        {DardhaGS17}
\bibfield{author}{\bibinfo{person}{Ornela Dardha}, \bibinfo{person}{Elena
  Giachino}, {and} \bibinfo{person}{Davide Sangiorgi}.}
  \bibinfo{year}{2017}\natexlab{}.
\newblock \showarticletitle{Session types revisited}.
\newblock \bibinfo{journal}{\emph{Inf. Comput.}}  \bibinfo{volume}{256}
  (\bibinfo{year}{2017}), \bibinfo{pages}{253--286}.
\newblock
\urldef\tempurl%
\url{https://doi.org/10.1016/j.ic.2017.06.002}
\showDOI{\tempurl}


\bibitem[\protect\citeauthoryear{Demangeon, Honda, Hu, Neykova, and
  Yoshida}{Demangeon et~al\mbox{.}}{2015}]%
        {DHHNY2015}
\bibfield{author}{\bibinfo{person}{Romain Demangeon}, \bibinfo{person}{Kohei
  Honda}, \bibinfo{person}{Raymond Hu}, \bibinfo{person}{Rumyana Neykova},
  {and} \bibinfo{person}{Nobuko Yoshida}.} \bibinfo{year}{2015}\natexlab{}.
\newblock \showarticletitle{Practical Interruptible Conversations: Distributed
  Dynamic Verification with Multiparty Session Types and {Python}}.
\newblock \bibinfo{journal}{\emph{Formal Methods in System Design}}
  (\bibinfo{year}{2015}).
\newblock
\urldef\tempurl%
\url{https://doi.org/10.1007/s10703-014-0218-8}
\showDOI{\tempurl}


\bibitem[\protect\citeauthoryear{Dezani-Ciancaglini, de'Liguoro, and
  Yoshida}{Dezani-Ciancaglini et~al\mbox{.}}{2009}]%
        {dezani-ciancagliniliguoro09progress}
\bibfield{author}{\bibinfo{person}{Mariangiola Dezani-Ciancaglini},
  \bibinfo{person}{Ugo de'Liguoro}, {and} \bibinfo{person}{Nobuko Yoshida}.}
  \bibinfo{year}{2009}\natexlab{}.
\newblock \showarticletitle{On Progress for Structured Communications}. In
  \bibinfo{booktitle}{\emph{Proc.\ of {TGC}}} \emph{(\bibinfo{series}{LNCS},
  Vol.~\bibinfo{volume}{4912})}. \bibinfo{publisher}{Springer},
  \bibinfo{pages}{257--275}.
\newblock


\bibitem[\protect\citeauthoryear{Dezani-Ciancaglini, Mostrous, Yoshida, and
  Drossopoulou}{Dezani-Ciancaglini et~al\mbox{.}}{2006}]%
        {dezani-ciancaglinimostrous06}
\bibfield{author}{\bibinfo{person}{Mariangiola Dezani-Ciancaglini},
  \bibinfo{person}{Dimitris Mostrous}, \bibinfo{person}{Nobuko Yoshida}, {and}
  \bibinfo{person}{Sophia Drossopoulou}.} \bibinfo{year}{2006}\natexlab{}.
\newblock \showarticletitle{Session Types for Object-Oriented Languages}. In
  \bibinfo{booktitle}{\emph{Proc.\ of {ECOOP}}} \emph{(\bibinfo{series}{LNCS},
  Vol.~\bibinfo{volume}{4067})}. \bibinfo{publisher}{Springer},
  \bibinfo{pages}{328--352}.
\newblock


\bibitem[\protect\citeauthoryear{Eisenberg, Vytiniotis, Jones, and
  Weirich}{Eisenberg et~al\mbox{.}}{2014}]%
        {eisenbergvytiniotis14}
\bibfield{author}{\bibinfo{person}{Richard~A. Eisenberg},
  \bibinfo{person}{Dimitrios Vytiniotis}, \bibinfo{person}{Simon~Peyton Jones},
  {and} \bibinfo{person}{Stephanie Weirich}.} \bibinfo{year}{2014}\natexlab{}.
\newblock \showarticletitle{Closed type families with overlapping equations}.
  In \bibinfo{booktitle}{\emph{Proceedings of the 41st {ACM} {SIGPLAN}-{SIGACT}
  Symposium on Principles of Programming Languages}}.
  \bibinfo{publisher}{{ACM}}.
\newblock
\urldef\tempurl%
\url{https://doi.org/10.1145/2535838.2535856}
\showDOI{\tempurl}


\bibitem[\protect\citeauthoryear{Filinski}{Filinski}{1994}]%
        {filinski94}
\bibfield{author}{\bibinfo{person}{Andrzej Filinski}.}
  \bibinfo{year}{1994}\natexlab{}.
\newblock \showarticletitle{Representing monads}. In
  \bibinfo{booktitle}{\emph{Proceedings of the 21st {ACM} {SIGPLAN}-{SIGACT}
  symposium on Principles of programming languages - {POPL}
  {\textquotesingle}94}}. \bibinfo{publisher}{{ACM} Press}.
\newblock
\urldef\tempurl%
\url{https://doi.org/10.1145/174675.178047}
\showDOI{\tempurl}


\bibitem[\protect\citeauthoryear{Fowler}{Fowler}{2016}]%
        {Fowler16}
\bibfield{author}{\bibinfo{person}{Simon Fowler}.}
  \bibinfo{year}{2016}\natexlab{}.
\newblock \showarticletitle{An {E}rlang Implementation of Multiparty Session
  Actors}. In \bibinfo{booktitle}{\emph{{ICE}}}.
\newblock
\urldef\tempurl%
\url{https://doi.org/10.4204/EPTCS.223.3}
\showDOI{\tempurl}


\bibitem[\protect\citeauthoryear{Fowler, Lindley, Morris, and Decova}{Fowler
  et~al\mbox{.}}{2019}]%
        {fowlerlindley19}
\bibfield{author}{\bibinfo{person}{Simon Fowler}, \bibinfo{person}{Sam
  Lindley}, \bibinfo{person}{J.~Garrett Morris}, {and}
  \bibinfo{person}{S\'{a}ra Decova}.} \bibinfo{year}{2019}\natexlab{}.
\newblock \showarticletitle{Exceptional Asynchronous Session Types: Session
  Types without Tiers}.
\newblock \bibinfo{journal}{\emph{Proc.\ of POPL}}  \bibinfo{volume}{3},
  Article \bibinfo{articleno}{28} (\bibinfo{year}{2019}),
  \bibinfo{numpages}{29}~pages.
\newblock
\urldef\tempurl%
\url{https://doi.org/10.1145/3290341}
\showDOI{\tempurl}


\bibitem[\protect\citeauthoryear{Gaboardi, ya~Katsumata, Orchard, Breuvart, and
  Uustalu}{Gaboardi et~al\mbox{.}}{2016}]%
        {gaboardikatsumata16}
\bibfield{author}{\bibinfo{person}{Marco Gaboardi}, \bibinfo{person}{Shin ya
  Katsumata}, \bibinfo{person}{Dominic Orchard}, \bibinfo{person}{Flavien
  Breuvart}, {and} \bibinfo{person}{Tarmo Uustalu}.}
  \bibinfo{year}{2016}\natexlab{}.
\newblock \showarticletitle{Combining effects and coeffects via grading}. In
  \bibinfo{booktitle}{\emph{Proceedings of the 21st {ACM} {SIGPLAN}
  International Conference on Functional Programming}}.
  \bibinfo{publisher}{{ACM}}.
\newblock
\urldef\tempurl%
\url{https://doi.org/10.1145/2951913.2951939}
\showDOI{\tempurl}


\bibitem[\protect\citeauthoryear{Gay, Thiemann, and Vasconcelos}{Gay
  et~al\mbox{.}}{2020}]%
        {gaythiemann20}
\bibfield{author}{\bibinfo{person}{Simon~J. Gay}, \bibinfo{person}{Peter
  Thiemann}, {and} \bibinfo{person}{Vasco~T. Vasconcelos}.}
  \bibinfo{year}{2020}\natexlab{}.
\newblock \showarticletitle{Duality of Session Types: The Final Cut}.
\newblock \bibinfo{journal}{\emph{Electronic Proceedings in Theoretical
  Computer Science}}  \bibinfo{volume}{314} (\bibinfo{date}{April}
  \bibinfo{year}{2020}), \bibinfo{pages}{23--33}.
\newblock
\urldef\tempurl%
\url{https://doi.org/10.4204/eptcs.314.3}
\showDOI{\tempurl}


\bibitem[\protect\citeauthoryear{Gay and Vasconcelos}{Gay and
  Vasconcelos}{2010}]%
        {gayvasconcelos10}
\bibfield{author}{\bibinfo{person}{Simon~J. Gay} {and}
  \bibinfo{person}{Vasco~T. Vasconcelos}.} \bibinfo{year}{2010}\natexlab{}.
\newblock \showarticletitle{Linear type theory for asynchronous session types}.
\newblock \bibinfo{journal}{\emph{Journal of Functional Programming}}
  \bibinfo{volume}{20}, \bibinfo{number}{1} (\bibinfo{year}{2010}),
  \bibinfo{pages}{19--50}.
\newblock


\bibitem[\protect\citeauthoryear{Gay and Vasconcelos}{Gay and
  Vasconcelos}{2012}]%
        {gayvasconcelos12}
\bibfield{author}{\bibinfo{person}{Simon~J. Gay} {and}
  \bibinfo{person}{Vasco~T. Vasconcelos}.} \bibinfo{year}{2012}\natexlab{}.
\newblock \showarticletitle{Linear type theory for asynchronous session types}.
\newblock \bibinfo{journal}{\emph{JFP}} \bibinfo{volume}{20},
  \bibinfo{number}{1} (\bibinfo{year}{2012}), \bibinfo{pages}{19--50}.
\newblock
\newblock
\shownote{Extended version of \cite{gayvasconcelos10}.}


\bibitem[\protect\citeauthoryear{Girard}{Girard}{1987}]%
        {girard87}
\bibfield{author}{\bibinfo{person}{Jean-Yves Girard}.}
  \bibinfo{year}{1987}\natexlab{}.
\newblock \showarticletitle{Linear Logic}.
\newblock \bibinfo{journal}{\emph{Theoretical Computer Science}}
  \bibinfo{volume}{50} (\bibinfo{year}{1987}), \bibinfo{pages}{1--102}.
\newblock


\bibitem[\protect\citeauthoryear{Honda}{Honda}{1993}]%
        {honda93}
\bibfield{author}{\bibinfo{person}{Kohei Honda}.}
  \bibinfo{year}{1993}\natexlab{}.
\newblock \showarticletitle{Types for Dyadic Interaction}. In
  \bibinfo{booktitle}{\emph{Proc.\ of {CONCUR}}} \emph{(\bibinfo{series}{LNCS},
  Vol.~\bibinfo{volume}{715})}. \bibinfo{publisher}{Springer},
  \bibinfo{pages}{509--523}.
\newblock


\bibitem[\protect\citeauthoryear{Honda, Vasconcelos, and Kubo}{Honda
  et~al\mbox{.}}{1998}]%
        {hondavasconcelos98}
\bibfield{author}{\bibinfo{person}{Kohei Honda},
  \bibinfo{person}{Vasco~Thudichum Vasconcelos}, {and} \bibinfo{person}{Makoto
  Kubo}.} \bibinfo{year}{1998}\natexlab{}.
\newblock \showarticletitle{Language Primitives and Type Discipline for
  Structured Communication-Based Programming}. In
  \bibinfo{booktitle}{\emph{Proc.\ of {ESOP}}} \emph{(\bibinfo{series}{LNCS},
  Vol.~\bibinfo{volume}{1381})}. \bibinfo{publisher}{Springer},
  \bibinfo{pages}{122--138}.
\newblock


\bibitem[\protect\citeauthoryear{Honda, Yoshida, and Carbone}{Honda
  et~al\mbox{.}}{2008}]%
        {hondayoshida08}
\bibfield{author}{\bibinfo{person}{Kohei Honda}, \bibinfo{person}{Nobuko
  Yoshida}, {and} \bibinfo{person}{Marco Carbone}.}
  \bibinfo{year}{2008}\natexlab{}.
\newblock \showarticletitle{Multiparty asynchronous session types}. In
  \bibinfo{booktitle}{\emph{Proc.\ of {POPL}}}, Vol.~\bibinfo{volume}{43(1)}.
  \bibinfo{publisher}{ACM}, \bibinfo{pages}{273--284}.
\newblock


\bibitem[\protect\citeauthoryear{Hu and Yoshida}{Hu and Yoshida}{2016}]%
        {HY16}
\bibfield{author}{\bibinfo{person}{Raymond Hu} {and} \bibinfo{person}{Nobuko
  Yoshida}.} \bibinfo{year}{2016}\natexlab{}.
\newblock \showarticletitle{Hybrid Session Verification Through Endpoint {API}
  Generation}. In \bibinfo{booktitle}{\emph{Proc.\ of FASE}}.
\newblock
\urldef\tempurl%
\url{https://doi.org/10.1007/978-3-662-49665-7_24}
\showDOI{\tempurl}


\bibitem[\protect\citeauthoryear{Imai, Yuen, and Agusa}{Imai
  et~al\mbox{.}}{2010}]%
        {imaiyuen10}
\bibfield{author}{\bibinfo{person}{Keigo Imai}, \bibinfo{person}{Shoji Yuen},
  {and} \bibinfo{person}{Kiyoshi Agusa}.} \bibinfo{year}{2010}\natexlab{}.
\newblock \showarticletitle{Session Type Inference in Haskell}. In
  \bibinfo{booktitle}{\emph{Proc.\ pf {PLACES}}}
  \emph{(\bibinfo{series}{{EPTCS}}, Vol.~\bibinfo{volume}{69})}.
  \bibinfo{pages}{74--91}.
\newblock
\urldef\tempurl%
\url{https://doi.org/10.4204/EPTCS.69.6}
\showDOI{\tempurl}


\bibitem[\protect\citeauthoryear{Jespersen, Munksgaard, and Larsen}{Jespersen
  et~al\mbox{.}}{2015}]%
        {JML15}
\bibfield{author}{\bibinfo{person}{Thomas Bracht~Laumann Jespersen},
  \bibinfo{person}{Philip Munksgaard}, {and} \bibinfo{person}{Ken~Friis
  Larsen}.} \bibinfo{year}{2015}\natexlab{}.
\newblock \showarticletitle{Session types for {Rust}}. In
  \bibinfo{booktitle}{\emph{Proc.\ of WGP@ICFP}}.
\newblock
\urldef\tempurl%
\url{https://doi.org/10.1145/2808098.2808100}
\showDOI{\tempurl}


\bibitem[\protect\citeauthoryear{Kobayashi}{Kobayashi}{2006}]%
        {kobayashi06}
\bibfield{author}{\bibinfo{person}{Naoki Kobayashi}.}
  \bibinfo{year}{2006}\natexlab{}.
\newblock \showarticletitle{A New Type System for Deadlock-Free Processes}. In
  \bibinfo{booktitle}{\emph{Proc.\ of {CONCUR}}} \emph{(\bibinfo{series}{LNCS},
  Vol.~\bibinfo{volume}{4137})}. \bibinfo{publisher}{Springer},
  \bibinfo{pages}{233--247}.
\newblock


\bibitem[\protect\citeauthoryear{Kobayashi, Pierce, and Turner}{Kobayashi
  et~al\mbox{.}}{1999}]%
        {KPT99}
\bibfield{author}{\bibinfo{person}{Naoki Kobayashi},
  \bibinfo{person}{Benjamin~C. Pierce}, {and} \bibinfo{person}{David~N.
  Turner}.} \bibinfo{year}{1999}\natexlab{}.
\newblock \showarticletitle{Linearity and the pi-calculus}.
\newblock \bibinfo{journal}{\emph{{ACM} Trans. Program. Lang. Syst.}}
  \bibinfo{volume}{21}, \bibinfo{number}{5} (\bibinfo{year}{1999}),
  \bibinfo{pages}{914--947}.
\newblock
\urldef\tempurl%
\url{https://doi.org/10.1145/330249.330251}
\showDOI{\tempurl}


\bibitem[\protect\citeauthoryear{Kokke}{Kokke}{2019}]%
        {kokke19}
\bibfield{author}{\bibinfo{person}{Wen Kokke}.}
  \bibinfo{year}{2019}\natexlab{}.
\newblock \showarticletitle{{Rusty Variation}: Deadlock-free Sessions with
  Failure in {Rust}}.
\newblock \bibinfo{journal}{\emph{{EPTCS}}}  \bibinfo{volume}{304}
  (\bibinfo{date}{Sept.} \bibinfo{year}{2019}), \bibinfo{pages}{48–60}.
\newblock
\showISSN{2075-2180}
\urldef\tempurl%
\url{https://doi.org/10.4204/eptcs.304.4}
\showDOI{\tempurl}
\newblock
\shownote{Renamed to Sesh.}


\bibitem[\protect\citeauthoryear{Kokke and Dardha}{Kokke and Dardha}{[n.d.]}]%
        {kokkedardha21}
\bibfield{author}{\bibinfo{person}{Wen Kokke} {and} \bibinfo{person}{Ornela
  Dardha}.} \bibinfo{year}{[n.d.]}\natexlab{}.
\newblock \bibinfo{title}{Prioritise the Best Variation}.
  (\bibinfo{year}{[n.\,d.]}).
\newblock
\urldef\tempurl%
\url{https://wen.works/public/drafts/pgv.pdf}
\showURL{%
\tempurl}
\newblock
\shownote{Forthcoming.}


\bibitem[\protect\citeauthoryear{Kouzapas, Dardha, Perera, and Gay}{Kouzapas
  et~al\mbox{.}}{2016}]%
        {KDPG16}
\bibfield{author}{\bibinfo{person}{Dimitrios Kouzapas}, \bibinfo{person}{Ornela
  Dardha}, \bibinfo{person}{Roly Perera}, {and} \bibinfo{person}{Simon~J.
  Gay}.} \bibinfo{year}{2016}\natexlab{}.
\newblock \showarticletitle{Typechecking protocols with {Mungo} and {StMungo}}.
  In \bibinfo{booktitle}{\emph{PPDP}}. \bibinfo{pages}{146--159}.
\newblock


\bibitem[\protect\citeauthoryear{Launchbury and Peyton~Jones}{Launchbury and
  Peyton~Jones}{1994}]%
        {launchburypeytonjones94}
\bibfield{author}{\bibinfo{person}{John Launchbury} {and}
  \bibinfo{person}{Simon~L. Peyton~Jones}.} \bibinfo{year}{1994}\natexlab{}.
\newblock \showarticletitle{Lazy Functional State Threads}. In
  \bibinfo{booktitle}{\emph{Proc. \ of {PLDI}}} (Orlando, Florida, USA).
  \bibinfo{publisher}{ACM}, \bibinfo{address}{New York, NY, USA},
  \bibinfo{pages}{24--35}.
\newblock
\showISBNx{089791662X}
\urldef\tempurl%
\url{https://doi.org/10.1145/178243.178246}
\showDOI{\tempurl}


\bibitem[\protect\citeauthoryear{Lindley and McBride}{Lindley and
  McBride}{2013}]%
        {lindleymcbride13}
\bibfield{author}{\bibinfo{person}{Sam Lindley} {and} \bibinfo{person}{Conor
  McBride}.} \bibinfo{year}{2013}\natexlab{}.
\newblock \showarticletitle{Hasochism}. In
  \bibinfo{booktitle}{\emph{Proceedings of the 2013 {ACM} {SIGPLAN} symposium
  on Haskell - Haskell {\textquotesingle}13}}. \bibinfo{publisher}{{ACM}
  Press}.
\newblock
\urldef\tempurl%
\url{https://doi.org/10.1145/2503778.2503786}
\showDOI{\tempurl}


\bibitem[\protect\citeauthoryear{Lindley and Morris}{Lindley and
  Morris}{2015}]%
        {lindleymorris15}
\bibfield{author}{\bibinfo{person}{Sam Lindley} {and}
  \bibinfo{person}{J.~Garrett Morris}.} \bibinfo{year}{2015}\natexlab{}.
\newblock \showarticletitle{A Semantics for Propositions as Sessions}. In
  \bibinfo{booktitle}{\emph{Proc.\ of {ESOP}}}. \bibinfo{pages}{560--584}.
\newblock


\bibitem[\protect\citeauthoryear{Lindley and Morris}{Lindley and
  Morris}{2016}]%
        {lindleymorris16}
\bibfield{author}{\bibinfo{person}{Sam Lindley} {and}
  \bibinfo{person}{J.~Garrett Morris}.} \bibinfo{year}{2016}\natexlab{}.
\newblock \showarticletitle{Embedding session types in Haskell}. In
  \bibinfo{booktitle}{\emph{Proc.\ of Haskell}}. \bibinfo{publisher}{{ACM}},
  \bibinfo{pages}{133--145}.
\newblock
\urldef\tempurl%
\url{https://doi.org/10.1145/2976002.2976018}
\showDOI{\tempurl}


\bibitem[\protect\citeauthoryear{Lindley and Morris}{Lindley and
  Morris}{2017}]%
        {LM17:fst}
\bibfield{author}{\bibinfo{person}{Sam Lindley} {and}
  \bibinfo{person}{J~Garrett Morris}.} \bibinfo{year}{2017}\natexlab{}.
\newblock \showarticletitle{Lightweight Functional Session Types}.
\newblock In \bibinfo{booktitle}{\emph{Behavioural Types: from Theory to
  Tools}}. \bibinfo{publisher}{River Publishers}, \bibinfo{pages}{265--286}.
\newblock


\bibitem[\protect\citeauthoryear{Lopez, Marques, Martins, Ng, Santos,
  Vasconcelos, and Yoshida}{Lopez et~al\mbox{.}}{2015}]%
        {LMMNSVY2015}
\bibfield{author}{\bibinfo{person}{Hugo~A. Lopez}, \bibinfo{person}{Eduardo
  R.~B. Marques}, \bibinfo{person}{Francisco Martins},
  \bibinfo{person}{Nicholas Ng}, \bibinfo{person}{Casar Santos},
  \bibinfo{person}{Vasco~Thudichum Vasconcelos}, {and} \bibinfo{person}{Nobuko
  Yoshida}.} \bibinfo{year}{2015}\natexlab{}.
\newblock \showarticletitle{Protocol-Based Verification of Message-Passing
  Parallel Programs}. In \bibinfo{booktitle}{\emph{OOPSLA}}.
\newblock
\urldef\tempurl%
\url{https://doi.org/10.1145/2814270.2814302}
\showDOI{\tempurl}


\bibitem[\protect\citeauthoryear{Neubauer and Thiemann}{Neubauer and
  Thiemann}{2004}]%
        {neubauerthiemann04}
\bibfield{author}{\bibinfo{person}{Matthias Neubauer} {and}
  \bibinfo{person}{Peter Thiemann}.} \bibinfo{year}{2004}\natexlab{}.
\newblock \showarticletitle{An Implementation of Session Types}. In
  \bibinfo{booktitle}{\emph{Proc.\ of {PADL}}} \emph{(\bibinfo{series}{Lecture
  Notes in Computer Science}, Vol.~\bibinfo{volume}{3057})}.
  \bibinfo{publisher}{Springer}, \bibinfo{pages}{56--70}.
\newblock
\urldef\tempurl%
\url{https://doi.org/10.1007/978-3-540-24836-1\_5}
\showDOI{\tempurl}


\bibitem[\protect\citeauthoryear{Neykova, Bocchi, and Yoshida}{Neykova
  et~al\mbox{.}}{2017}]%
        {NBY2017}
\bibfield{author}{\bibinfo{person}{Rumyana Neykova}, \bibinfo{person}{Laura
  Bocchi}, {and} \bibinfo{person}{Nobuko Yoshida}.}
  \bibinfo{year}{2017}\natexlab{}.
\newblock \showarticletitle{Timed Runtime Monitoring for Multiparty
  Conversations}.
\newblock \bibinfo{journal}{\emph{Formal Aspects of Computing}}
  (\bibinfo{year}{2017}).
\newblock
\urldef\tempurl%
\url{https://doi.org/10.1007/s00165-017-0420-8}
\showDOI{\tempurl}


\bibitem[\protect\citeauthoryear{Neykova and Yoshida}{Neykova and
  Yoshida}{2017a}]%
        {NY2017}
\bibfield{author}{\bibinfo{person}{Rumyana Neykova} {and}
  \bibinfo{person}{Nobuko Yoshida}.} \bibinfo{year}{2017}\natexlab{a}.
\newblock \showarticletitle{{Let It Recover: Multiparty Protocol-Induced
  Recovery}}. In \bibinfo{booktitle}{\emph{CC}}.
\newblock
\urldef\tempurl%
\url{https://doi.org/10.1145/3033019.3033031}
\showDOI{\tempurl}


\bibitem[\protect\citeauthoryear{Neykova and Yoshida}{Neykova and
  Yoshida}{2017b}]%
        {NY2017A}
\bibfield{author}{\bibinfo{person}{Rumyana Neykova} {and}
  \bibinfo{person}{Nobuko Yoshida}.} \bibinfo{year}{2017}\natexlab{b}.
\newblock \showarticletitle{{Multiparty Session Actors}}.
\newblock \bibinfo{journal}{\emph{{Logical Methods in Computer Science}}}
  \bibinfo{volume}{13}, \bibinfo{number}{1} (\bibinfo{date}{March}
  \bibinfo{year}{2017}).
\newblock
\urldef\tempurl%
\url{https://doi.org/10.23638/LMCS-13(1:17)2017}
\showDOI{\tempurl}


\bibitem[\protect\citeauthoryear{Orchard, Wadler, and Eades}{Orchard
  et~al\mbox{.}}{2020}]%
        {orchardwadler20}
\bibfield{author}{\bibinfo{person}{Dominic Orchard}, \bibinfo{person}{Philip
  Wadler}, {and} \bibinfo{person}{Harley Eades}.}
  \bibinfo{year}{2020}\natexlab{}.
\newblock \showarticletitle{Unifying graded and parameterised monads}.
\newblock \bibinfo{journal}{\emph{Electronic Proceedings in Theoretical
  Computer Science}}  \bibinfo{volume}{317} (\bibinfo{date}{May}
  \bibinfo{year}{2020}), \bibinfo{pages}{18--38}.
\newblock
\urldef\tempurl%
\url{https://doi.org/10.4204/eptcs.317.2}
\showDOI{\tempurl}


\bibitem[\protect\citeauthoryear{Orchard and Yoshida}{Orchard and
  Yoshida}{2017}]%
        {orchardyoshida17}
\bibfield{author}{\bibinfo{person}{Dominic Orchard} {and}
  \bibinfo{person}{Nobuko Yoshida}.} \bibinfo{year}{2017}\natexlab{}.
\newblock \showarticletitle{Session types with linearity in Haskell}.
\newblock \bibinfo{journal}{\emph{Behavioural Types: from Theory to Tools}}
  (\bibinfo{year}{2017}), \bibinfo{pages}{219}.
\newblock


\bibitem[\protect\citeauthoryear{Orchard and Yoshida}{Orchard and
  Yoshida}{2016}]%
        {orchardyoshida16}
\bibfield{author}{\bibinfo{person}{Dominic~A. Orchard} {and}
  \bibinfo{person}{Nobuko Yoshida}.} \bibinfo{year}{2016}\natexlab{}.
\newblock \showarticletitle{Effects as sessions, sessions as effects}. In
  \bibinfo{booktitle}{\emph{Proc.\ of {POPL}}}. \bibinfo{publisher}{{ACM}},
  \bibinfo{pages}{568--581}.
\newblock
\urldef\tempurl%
\url{https://doi.org/10.1145/2837614.2837634}
\showDOI{\tempurl}


\bibitem[\protect\citeauthoryear{Padovani}{Padovani}{2014}]%
        {padovani14}
\bibfield{author}{\bibinfo{person}{Luca Padovani}.}
  \bibinfo{year}{2014}\natexlab{}.
\newblock \showarticletitle{{Deadlock and Lock Freedom in the Linear
  $\pi$-Calculus}}. In \bibinfo{booktitle}{\emph{Proc.\ of {CSL-LICS}}}.
  \bibinfo{publisher}{ACM}, \bibinfo{pages}{72:1--72:10}.
\newblock


\bibitem[\protect\citeauthoryear{Padovani}{Padovani}{2017}]%
        {PadFuse}
\bibfield{author}{\bibinfo{person}{Luca Padovani}.}
  \bibinfo{year}{2017}\natexlab{}.
\newblock \showarticletitle{A simple library implementation of binary
  sessions}.
\newblock \bibinfo{journal}{\emph{Journal of Functional Programming}}
  \bibinfo{volume}{27} (\bibinfo{year}{2017}).
\newblock
\urldef\tempurl%
\url{https://doi.org/10.1017/S0956796816000289}
\showDOI{\tempurl}
\newblock
\shownote{Website:
  \url{http://www.di.unito.it/~padovani/Software/FuSe/FuSe.html}.}


\bibitem[\protect\citeauthoryear{Padovani and Novara}{Padovani and
  Novara}{2015}]%
        {padovaninovara15}
\bibfield{author}{\bibinfo{person}{Luca Padovani} {and} \bibinfo{person}{Luca
  Novara}.} \bibinfo{year}{2015}\natexlab{}.
\newblock \showarticletitle{{Types for Deadlock-Free Higher-Order Programs}}.
  In \bibinfo{booktitle}{\emph{Proc.\ of {FORTE}}}
  \emph{(\bibinfo{series}{LNCS}, Vol.~\bibinfo{volume}{9039})}.
  \bibinfo{publisher}{Springer}, \bibinfo{pages}{3--18}.
\newblock


\bibitem[\protect\citeauthoryear{{Peyton Jones}, Gordon, and Finne}{{Peyton
  Jones} et~al\mbox{.}}{[n.d.]}]%
        {peytonjonesgordon96}
\bibfield{author}{\bibinfo{person}{Simon~L. {Peyton Jones}},
  \bibinfo{person}{Andrew~D. Gordon}, {and} \bibinfo{person}{Sigbj{\o}rn
  Finne}.} \bibinfo{year}{[n.d.]}\natexlab{}.
\newblock \showarticletitle{Concurrent Haskell}. In
  \bibinfo{booktitle}{\emph{Proc.\ of {POPL}}}. \bibinfo{publisher}{{ACM}},
  \bibinfo{pages}{295--308}.
\newblock
\urldef\tempurl%
\url{https://doi.org/10.1145/237721.237794}
\showDOI{\tempurl}


\bibitem[\protect\citeauthoryear{Polakow}{Polakow}{2015}]%
        {polakow15}
\bibfield{author}{\bibinfo{person}{Jeff Polakow}.}
  \bibinfo{year}{2015}\natexlab{}.
\newblock \showarticletitle{Embedding a full linear Lambda calculus in
  Haskell}. In \bibinfo{booktitle}{\emph{Proc/ of the Symposium on Haskell}}.
  \bibinfo{publisher}{{ACM}}.
\newblock
\urldef\tempurl%
\url{https://doi.org/10.1145/2804302.2804309}
\showDOI{\tempurl}


\bibitem[\protect\citeauthoryear{Pucella and Tov}{Pucella and Tov}{2008}]%
        {pucellatov08}
\bibfield{author}{\bibinfo{person}{Riccardo Pucella} {and}
  \bibinfo{person}{Jesse~A. Tov}.} \bibinfo{year}{2008}\natexlab{}.
\newblock \showarticletitle{Haskell session types with (almost) no class}. In
  \bibinfo{booktitle}{\emph{Proc.\ of Haskell}}. \bibinfo{publisher}{{ACM}}.
\newblock
\urldef\tempurl%
\url{https://doi.org/10.1145/1411286.1411290}
\showDOI{\tempurl}


\bibitem[\protect\citeauthoryear{Sackman and Eisenbach}{Sackman and
  Eisenbach}{2008}]%
        {sackmaneisenbach08}
\bibfield{author}{\bibinfo{person}{Matthew Sackman} {and}
  \bibinfo{person}{Susan Eisenbach}.} \bibinfo{year}{2008}\natexlab{}.
\newblock \showarticletitle{Session Types in Haskell Updating Message Passing
  for the 21st Century}.
\newblock  (\bibinfo{date}{01} \bibinfo{year}{2008}).
\newblock


\bibitem[\protect\citeauthoryear{Sangiorgi and Walker}{Sangiorgi and
  Walker}{2001}]%
        {sangiorgiwalker01}
\bibfield{author}{\bibinfo{person}{Davide Sangiorgi} {and}
  \bibinfo{person}{David Walker}.} \bibinfo{year}{2001}\natexlab{}.
\newblock \bibinfo{booktitle}{\emph{The $\pi$-calculus: a Theory of Mobile
  Processes}}.
\newblock \bibinfo{publisher}{Cambridge University Press}.
\newblock


\bibitem[\protect\citeauthoryear{Scalas, Dardha, Hu, and Yoshida}{Scalas
  et~al\mbox{.}}{2017}]%
        {Scalas2017}
\bibfield{author}{\bibinfo{person}{Alceste Scalas}, \bibinfo{person}{Ornela
  Dardha}, \bibinfo{person}{Raymond Hu}, {and} \bibinfo{person}{Nobuko
  Yoshida}.} \bibinfo{year}{2017}\natexlab{}.
\newblock \showarticletitle{A Linear Decomposition of Multiparty Sessions for
  Safe Distributed Programming}. In \bibinfo{booktitle}{\emph{Proc.\ of
  {ECOOP}}} \emph{(\bibinfo{series}{LIPIcs}, Vol.~\bibinfo{volume}{74})}.
  \bibinfo{publisher}{Schloss Dagstuhl - Leibniz-Zentrum f{\"{u}}r Informatik},
  \bibinfo{pages}{24:1--24:31}.
\newblock
\urldef\tempurl%
\url{https://doi.org/10.4230/LIPIcs.ECOOP.2017.24}
\showDOI{\tempurl}


\bibitem[\protect\citeauthoryear{Scalas and Yoshida}{Scalas and
  Yoshida}{2016}]%
        {ScalasY16}
\bibfield{author}{\bibinfo{person}{Alceste Scalas} {and}
  \bibinfo{person}{Nobuko Yoshida}.} \bibinfo{year}{2016}\natexlab{}.
\newblock \showarticletitle{Lightweight Session Programming in Scala}. In
  \bibinfo{booktitle}{\emph{ECOOP}}.
\newblock
\urldef\tempurl%
\url{https://doi.org/10.4230/LIPIcs.ECOOP.2016.21}
\showDOI{\tempurl}


\bibitem[\protect\citeauthoryear{Sivaramakrishnan, Nagaraj, Ziarek, and
  Eugster}{Sivaramakrishnan et~al\mbox{.}}{2010}]%
        {SivaramakrishnanNZE10}
\bibfield{author}{\bibinfo{person}{K.~C. Sivaramakrishnan},
  \bibinfo{person}{Karthik Nagaraj}, \bibinfo{person}{Lukasz Ziarek}, {and}
  \bibinfo{person}{Patrick Eugster}.} \bibinfo{year}{2010}\natexlab{}.
\newblock \showarticletitle{Efficient Session Type Guided Distributed
  Interaction}. In \bibinfo{booktitle}{\emph{Proc.\ of {COORDINATION}}},
  Vol.~\bibinfo{volume}{6116}.
\newblock
\urldef\tempurl%
\url{https://doi.org/10.1007/978-3-642-13414-2_11}
\showDOI{\tempurl}


\bibitem[\protect\citeauthoryear{Takeuchi, Honda, and Kubo}{Takeuchi
  et~al\mbox{.}}{1994}]%
        {takeuchihonda94}
\bibfield{author}{\bibinfo{person}{Kaku Takeuchi}, \bibinfo{person}{Kohei
  Honda}, {and} \bibinfo{person}{Makoto Kubo}.}
  \bibinfo{year}{1994}\natexlab{}.
\newblock \showarticletitle{An Interaction-Based Language and its Typing
  System}. In \bibinfo{booktitle}{\emph{Proc.\ of {PARLE}}}
  \emph{(\bibinfo{series}{LNCS}, Vol.~\bibinfo{volume}{817})}.
  \bibinfo{publisher}{Springer}, \bibinfo{pages}{398--413}.
\newblock


\bibitem[\protect\citeauthoryear{Wadler}{Wadler}{2012}]%
        {wadler12}
\bibfield{author}{\bibinfo{person}{Philip Wadler}.}
  \bibinfo{year}{2012}\natexlab{}.
\newblock \showarticletitle{Propositions as sessions}. In
  \bibinfo{booktitle}{\emph{Proc.\ of {ICFP}}}. \bibinfo{pages}{273--286}.
\newblock


\bibitem[\protect\citeauthoryear{Wadler}{Wadler}{2014}]%
        {wadler14}
\bibfield{author}{\bibinfo{person}{Philip Wadler}.}
  \bibinfo{year}{2014}\natexlab{}.
\newblock \showarticletitle{Propositions as sessions}.
\newblock \bibinfo{journal}{\emph{Journal of Functional Programming}}
  \bibinfo{volume}{24}, \bibinfo{number}{2-3} (\bibinfo{date}{Jan.}
  \bibinfo{year}{2014}), \bibinfo{pages}{384--418}.
\newblock
\newblock
\shownote{Extended version of \cite{wadler12}.}


\bibitem[\protect\citeauthoryear{Wadler}{Wadler}{2015}]%
        {wadler15}
\bibfield{author}{\bibinfo{person}{Philip Wadler}.}
  \bibinfo{year}{2015}\natexlab{}.
\newblock \showarticletitle{Propositions as types}.
\newblock \bibinfo{journal}{\emph{Commun. ACM}} \bibinfo{volume}{58},
  \bibinfo{number}{12} (\bibinfo{year}{2015}), \bibinfo{pages}{75--84}.
\newblock


\end{thebibliography}

\end{document}